\documentclass[aps,prd,twocolumn,nopacs,nofootinbib]{revtex4-1}

\usepackage{amsfonts}
\usepackage{amsmath}
\usepackage{xcolor}
\usepackage{bm}
\usepackage{fancybox}
\usepackage{comment}
\usepackage{tipa}
\usepackage{longtable}
\definecolor{refs}{RGB}{245,156,74}
\usepackage[colorlinks=true,hyperfootnotes=true,citecolor=cyan]{hyperref}

\usepackage{enumitem}

\newcommand{\be}{\begin{equation}}
\newcommand{\ee}{\end{equation}}
\newcommand{\ba}{\begin{eqnarray}}
\newcommand{\ea}{\end{eqnarray}}
\newcommand{\bs}{\begin{subequations}}
\newcommand{\es}{\end{subequations}}

\newcommand{\diff}{\textrm{d}}

\newcommand{\lp}{\left(}
\newcommand{\rp}{\right)}
\newcommand{\lb}{\left[}
\newcommand{\rb}{\right]}

\newcommand{\nn}{\nonumber}

\newcommand{\1}{1$^\text{st}$}
\newcommand{\2}{2$^\text{nd}$}
\newcommand{\3}{3$^\text{rd}$}

\newcommand{\Hyp}{Z}

\newcommand{\josec}[1]{{\textcolor{orange}{ #1}}}

\begin{document}

\title{Energy and entropy in the Geometrical Trinity of gravity}

\author{D{\'e}bora Aguiar Gomes}
\affiliation{Departamento de F{\'i}sica - Universidade Federal do Cear{\'a}, Brazil}
\email{deboragomes@fisica.ufc.br}

\author{Jose Beltr{\'a}n Jim{\'e}nez}
\affiliation{Departamento de Física Fundamental and IUFFyM, Universidad de Salamanca, E-37008 Salamanca, Spain}
\email{jose.beltran@usal.es}

\author{Tomi S. Koivisto}
\address{Laboratory of Theoretical Physics, Institute of Physics, University of Tartu, W. Ostwaldi 1, 50411 Tartu, Estonia}
\address{National Institute of Chemical Physics and Biophysics, Rävala pst. 10, 10143 Tallinn, Estonia}
\email{tomi.koivisto@ut.ee}

\begin{abstract}

All energy is gravitational energy. That is the consequence of the equivalence
principle, according to which gravity is the universal interaction. The physical charges of this interaction have remained undisclosed, but the Advent
of the Geometrical Trinity opened a new approach to this foundational problem. Here it is shown to provide a background-independent unification of the previous, non-covariant approaches of Bergmann-Thomson, Cooperstock, Einstein, von Freud, Landau-Lifshitz, Papapetrou and Weinberg.
First, the Noether currents are derived for a generic Palatini theory of gravity coupled with generic matter fields, and then the canonical i.e. the unique charges are robustly derived and analysed, particularly in the metric teleparallel and the symmetric teleparallel versions of General Relativity. These results, and their application to black holes and gravitational waves, are new. 

\end{abstract}

\maketitle

\tableofcontents

\section{Introduction}
\label{introduction}

As it is well-known, there are 3 different but equivalent ``pictures'' of quantum mechanics. Each of these {\it formulations}
of quantum mechanics, {\`a} la Schr\"odinger, {\`a} la Heisenberg and {\`a} la Dirac, can provide the most convenient toolbox depending on the problem. On the other hand, at the foundational rather than technical level, contemplating the different {\it interpretations} of quantum mechanics, such as the Copenhagen or the many worlds interpretation, can give rise to new ideas about how to approach the remaining fundamental problems of physics.
Indeed, some approaches have led to {\it extensions} of the standard theory of quantum mechanics, in particular in regard to its relation with gravity that is still an unresolved issue. In the ideal case, such alternative extensions give rise to distinct predictions that predispose them to 
experimental falsification (e.g. \cite{penrose-roadtoreality-2005,Valentini:2008dq,Bose:2017nin}).

Analogously, there are 3 different but equivalent ``pictures'' of General Relativity (GR).
The Geometrical Trinity of Gravity \cite{BeltranJimenez:2019esp} is a
framework that presents a variety of alternatives at each of the 3 levels:
technical formulation, foundational interpretation and potential physical
extension that could eventually describe gravity according to the principles of quantum mechanics. Besides the standard
description of GR in terms of the metrical spacetime curvature, it is
possible to consider the equivalent dynamics as the workings of a relativistic force
field (torsion) or, in the most minimal case, to understand gravity solely through its equivalence with inertia, in terms of a gauge field (nonmetricity).
The torsion picture and the nonmetricity picture, called metric teleparallelism and
symmetric teleparallelism respectively, can also be combined and interpolated in 
a formulation of GR with a flat but otherwise generic affine connection \cite{BeltranJimenez:2019odq}.
As far as the dynamics is concerned, these alternatives are equivalent.
However, physically relevant differences may arise when considering the
Noether currents in gravitating systems (nota bene: that is, in any physical
systems!) because the currents are sensitive to the boundary terms in the
action. The action is a central 
tool in computations of the Noether currents corresponding to the symmetries of 
the theory, and in particular it is used to compute conserved charges with direct physical relevance, such as the energy and the entropy \cite{Komar:1958wp,Wald:1993nt}. Also, the 
action principle is of paramount importance for the quantum theory, both in canonical and 
in the path integral approaches. From this perspective, the Geometrical Trinity 
could be considered to represent 3 distinct theories of gravity. 

We recall that Einstein's theory of gravity already was a \1 order theory \cite{1916SPAW......1111E}, though this subtlety is often neglected, as the \2 order formulation due to Hilbert has been adopted as the standard action principle for GR. 
The problem with the Hilbert action is that
it is not well-defined in some situations, due to the \2 order derivatives of the metric, and then one is
forced to modify the theory by adding different boundary terms for different purposes, such as the Gibbons-Hawking-York term \cite{PhysRevLett.28.1082,PhysRevD.15.2752}. It is perhaps due to this
incompleteness of Hilbert's action, that it is a rather common misconception that the superpotentials
appearing in Noether \2 theorem are inherently ambiguous. This is not the case, as was recently clarified also in Ref. \cite{DeHaro:2021gdv}. We maintain that in a well-defined theory, the currents should be no more ambiguous than 
the field equations {in a sense that will become clear in the body of this work.}\footnote{In fact, as we shall find, in canonical gauge theories these 2 aspects of a theory are equivalent \cite{Koivisto:2022nar}.} In his original articles, Einstein proposed a definition of gravitational energy-momentum \cite{1916SPAW......1111E} which was met with criticism due {to} its lack of covariance \cite{article}.
However, Einstein's so-called ``pseudotensor'' is, at least according to some convention, a well-defined geometrical object, to wit a section of the \1 jet bundle, and more importantly, its behaviour under coordinate transformations properly describes the physical property of the gravitational field. Due to the equivalence principle, the energy-momentum vanishes for an observer whose acceleration exactly cancels gravitation, and for example a rotating observer sees gravitational energy-momentum where a static observer does not. Yet, on the other hand, there is also a point in requiring the coordinate-independence of physical quantities, and in this sense both Einstein and his critics were right. Einstein's action did have a problem, its ``pseudoinvariance'' i.e. invariance only up to a boundary term. It is the consequence of this that the energy-momentum pseudotensor, though its form is unique, turns out to give coordinate-dependent results for the physical charges.

A canonical reconciliation exists, though was introduced only recently \cite{BeltranJimenez:2017tkd}. From a modern perspective, the choice of coordinates, importance of which Einstein emphasised in his non-covariant formulation, translates into the fixing of the gauge in the new version of GR. A covariant definition of gravitational energy-momentum requires the introduction of some ``background structure'', with respect to which the theory can be ``covariantised'', such as an auxiliary reference metric \cite{PhysRev.57.147,10.2307/20488488} or an auxiliary reference connection \cite{Tomboulis:2017fim,Feng:2021lfa}, which for specific solutions are of course asymptotically naturally provided by the sufficiently symmetric boundary conditions for the dynamical fields \cite{Arnowitt:1959ah,Deser:2019acl}. The new resolution is based instead on a reformulation of gravity as a translation gauge theory \cite{BeltranJimenez:2017tkd}. The required extra field is provided by the translation gauge potential that is a fundamental ingredient of the theory on par with the metric. The resolution is canonical due to the flatness of the connection\footnote{It has been later proposed that the flatness a.k.a. teleparallelism could be the macroscopic consequence of the Planck mass $m_P$ being the mass of the gravitational connection \cite{Koivisto:2021jdy}. In this article we do not consider such possible extension beyond the Planck scale, but restrict to the regime where the theory is dynamically equivalent to GR.}: the gauge field generated purely by a coordinate transformation is obviously the minimal structure required to restore coordinate invariance\footnote{In other words, it is the (minimal) St\"uckelbergisation of the coordinate invariance \cite{BeltranJimenez:2022azb}.}. The resulting theory was called the Coincident GR since in the unitary gauge it coincides with Einstein's original formulation \cite{BeltranJimenez:2017tkd,Koivisto:2018aip}. The principle of relativity posits inertial frames, wherein the physical laws assume their standard, coordinate-independent form. We proposed the
criterion that {\it in an inertial frame, the divergence of the
gravitational field is the material current} \cite{Jimenez:2019yyx}\footnote{This agrees with Cooperstock's hypothesis \cite{cooperstock} that physical results are obtained 
(in the coordinate-dependent language) by imposing the vanishing of the Einstein pseudotensor. That the local energy-momentum should be solely due to matter is consistent with many alternative proposals to define gravitational energy-momentum, beginning with the most immediate responses to Einstein from T. Levi-Civita, F. Klein and H. A. Lorentz etc \cite{Duerr:2019nar}, and continuing up to various current proposals, e.g. \cite{Nikolic:2014kga,Wu:2016tzh,Aoki:2020nzm,Aoki:2022gez}.}. Thus, the inertial frame might be considered as the realisation of the relativity principle that perhaps addresses the long-standing criticisms of GR, \1 emphatically voiced by Kretschmann \cite{pittphilsci380}.


 

This motivates us to study the Noether currents in the 3 \1 order formulations of GR of the Geometrical Trinity. It is the task of this article to undertake the \1 robust
computations of these currents, with most focus at the two more unexplored, teleparallel corners of the Geometrical Trinity. Therein we will indeed find genuinely new insights to energy, momentum, and entropy associated with matter and with spacetime itself. For the same motivations as in Ref.\cite{BeltranJimenez:2018vdo}, in this article we will use exclusively the tensor language (Palatini formalism, introduced in Section \ref{formalism}). Noether charges in the Geometrical Trinity have already been derived in the exterior algebra parlance (metric-affine formalism) \cite{Jimenez:2021nup}. Also, though we derive the generic form of the currents in 
Section \ref{current}, in the article we will restrict the focus on theories (dynamically) 
equivalent to GR.\footnote{Various aspects of modifications of these theories have been considered in the recent literature \cite{Bahamonde:2021gfp,BeltranJimenez:2019tme,Dimakis:2021gby,Ayuso:2021vtj,Vignolo:2021frk,Albuquerque:2022eac,Capozziello:2022vyd,Ferreira:2022jcd,Capozziello:2022wgl,Hu:2022anq}, and the energy and the entropy might be further interesting aspects to consider in the future \cite{Mandal:2020lyq,Sharma:2021egn,Banerjee:2021mqk,Koussour:2022ycn,Koussour:2022jss}.} Diffeomorphism Noether charges of GR in its curved and in its teleparallel pictures will be investigated in Section \ref{palatini} and \ref{teleparallelism}, respectively. The interpretation, and relation to previous literature, is discussed in Section \ref{potentials}. The result is applied to black hole spacetime in Section \ref{energy-momentum} and to gravitational waves in Section \ref{gravitationalwaves}. In Section 
\ref{entropysection} we consider different prescriptions of entropy, and their relation to Wald's derivation. Section \ref{conclusions} concludes the article.   





\section{Formalism}
\label{formalism}

In this Section, which can be readily skipped by experts, we introduce notation for the tensor formalism a.k.a the Palatini formalism of gravity. We intend to be as systematic in the notation as possible (for example, since $T^\alpha{}_{\mu\nu}$ is the torsion tensor, let $T_\alpha$ be its trace and $t_\alpha{}^{\mu\nu}$ be its conjugate tensor, and then $\mathfrak{T}^\alpha{}_{\mu\nu}$, $\mathfrak{T}_\alpha$ and $\mathfrak{t}_\alpha{}^{\mu\nu}$ the respective tensor densities, instead of using a different letter for each of these quantities). Signaturewise, we follow the mostly positive convention.

\subsection{Connection}
\label{connection}

The covariant derivative $\nabla_\mu$ acts on tensor indices as follows
\begin{subequations}
\ba \label{covariants}
\nabla_\mu V^\alpha & = & V^\alpha_{\phantom{\alpha},\mu} + \Gamma^\alpha_{\phantom{\alpha}\mu\lambda}V^\lambda\,, \\
\nabla_\mu V_\alpha & = & V_{{\alpha},\mu} - \Gamma^\lambda_{\phantom{\alpha}\mu\alpha}V_\lambda\,.
\ea
\end{subequations}
The action of the commutator then defines the two tensorial properties of the connection
\bs
\label{geometry}
\ba
\big[\nabla_\mu,\nabla_\nu\big]V^\alpha & = & {R}^\alpha_{\phantom{\alpha}\beta\mu\nu}V^\beta - {T}^\beta_{\phantom{\beta}\mu\nu}\nabla_\beta V^\alpha\,, \\
\big[\nabla_\mu,\nabla_\nu\big]V_\alpha & = & -{R}^\beta_{\phantom{\beta}\alpha\mu\nu}V_\beta - {T}^\beta_{\phantom{\beta}\mu\nu}\nabla_\beta V_\alpha\,,
\ea
\es
the curvature and the torsion,
\begin{subequations}
\ba \label{riemann}
{R}^\alpha_{\phantom{\alpha}\beta\mu\nu} & = & 
2\partial_{[\mu} \Gamma^\alpha_{\phantom{\alpha}\nu]\beta}
+ 2\Gamma^\alpha_{\phantom{\alpha}[\mu\lvert\lambda\rvert}\Gamma^\lambda_{\phantom{\lambda}\nu]\beta}\,,  \\
T^\alpha{}_{\mu\nu} & = & 2\Gamma^\alpha{}_{[\mu\nu]}\,. 
\ea
\end{subequations}
By construction, these tensors obey the Bianchi identities
\begin{subequations}
\label{Bianchi}
\ba
R^\alpha_{\phantom{\alpha}\beta(\mu\nu)}&= & 0\,, \\
R^\mu_{\phantom{\mu}[\alpha\beta\gamma]} - \nabla_{[\alpha}T^\mu_{\phantom{\mu}\beta\gamma]} + T^\nu_{\phantom{\nu}[\alpha\beta}T^\mu_{\phantom{\mu}\gamma]\nu} & = & 0\,, \label{Bianchi2}\\
\nabla_{[\alpha}R^{\mu}_{\phantom{\mu}\lvert\nu\rvert\beta\gamma]} - T^\lambda_{\phantom{\lambda}[\alpha\beta}R^{\mu}_{\phantom{\mu}\vert\nu\rvert\gamma]\lambda} & = & 0\,,
\ea
\end{subequations}
that follow from (\ref{geometry}). We can further define the contractions 
$R_{\mu\nu} = R^\alpha{}_{\mu\alpha\nu} = -R^\alpha{}_{\mu\alpha\nu}$ called the curvature trace, $F_{\mu\nu} = R^\alpha{}_{\alpha\mu\nu}$ called
the homothetic curvature, and $T_\mu = T^\alpha{}_{\mu\alpha}=-T^\alpha{}_{\alpha\mu}$ called the torsion trace. All these quantities require only the connection $\Gamma^\alpha{}_{\mu\nu}$ for their definition. 

\subsection{Metric}
\label{metric}

If the manifold is endowed also with a metric tensor $g_{\mu\nu}$ (i.e. the manifold is Riemannian), we can introduce its Christoffel symbols, 
\be \label{christoffel}
\mathring{\Gamma}^\alpha{}_{\mu\nu}= -\frac{1}{2}g^{\alpha\beta}g_{\mu\nu,\beta} + g^{\alpha\beta}g_{\beta(\mu,\nu)}\,.  
\ee 
This is the unique connection $\mathring{\nabla}_\alpha$ that has no torsion and satisfies $\mathring{\nabla}_\alpha g_{\mu\nu} =0$.
Its curvature $\mathring{R}^\alpha{}_{\mu\beta\nu}$ is called the Riemann tensor, the trace $\mathring{R}_{\mu\nu}$
the Ricci tensor and the contraction $\mathring{R} = g^{\mu\nu}\mathring{R}_{\mu\nu}$ the Ricci scalar. The 
covariant derivative $Q_{\alpha\mu\nu}=\nabla_\alpha g_{\mu\nu} = -g_{\mu\rho} g_{\nu\sigma}\nabla_\alpha g^{\rho\sigma}$ does not vanish
in general. This tensor has 2 independent traces that we denote as $Q_\alpha=Q_{\alpha\mu}{}^{\mu}$ and $\tilde{Q}^\mu=Q_\alpha{}^{\mu\alpha}$.
The connection coefficients of $\nabla_\alpha$ can be decomposed wrt (with respect to) the metric $g_{\mu\nu}$ as
\be
\Gamma^\alpha{}_{\mu\nu} = \mathring{\Gamma}^\alpha{}_{\mu\nu} + K^\alpha{}_{\mu\nu} + L^\alpha{}_{\mu\nu}\,,
\ee
where the 2 new tensors are derived from the torsion and from the metric-incompatibility, respectively, as 
\begin{subequations}
\ba
K^\alpha{}_{\mu\nu} & = & \frac{1}{2}T^\alpha{}_{\mu\nu} - T_{(\mu\nu)}{}^\alpha\,, \\
L^\alpha{}_{\mu\nu}  & = & \frac{1}{2}Q^\alpha{}_{\mu\nu} - Q_{(\mu\nu)}{}^\alpha\,. 
\ea
\end{subequations}
These have the traces
\bs
\ba
K^\alpha{}_{\mu\alpha} & = & 0\,, \\
K^\alpha{}_{\alpha\mu} & = & K_\mu{}^\alpha{}_\alpha = T_\mu\,, \\
L^\alpha{}_{\mu\alpha} & = & L^\alpha{}_{\alpha\mu} = -\frac{1}{2}Q_\mu\,, \\ L_\mu{}^\alpha{}_\alpha & = & \frac{1}{2}Q_\mu - \tilde{Q}_\mu\,.
\ea
\es
The Bianchi identities (\ref{Bianchi}) are then supplemented with the identity for the non-metric curvature,
\be
\nabla_{[\mu}Q_{\nu]}{}^{\alpha\beta} = R^{(\alpha\beta)}{}_{\mu\nu} + \frac{1}{2}T^\lambda{}_{\mu\nu}Q_\lambda{}^{\alpha\beta}\,. \label{bianchi3}
\ee
It follows that 
\begin{subequations}
\ba
\nabla_{[\mu}Q_{\nu]} & = & F_{\mu\nu} + \frac{1}{2}T^\lambda{}_{\mu\nu}Q_\lambda\,, \label{bianchi3a} \\
\nabla_{[\mu}\tilde{Q}_{\nu]} & = &  Q_{[\mu\nu]\alpha}\tilde{Q}^\alpha + \frac{1}{2}\lp R_{[\mu\nu]} - \tilde{R}_{[\mu\nu]}\rp \nn \\ 
 & - & \frac{1}{2}T_{\alpha\beta[\mu}Q^{\alpha\beta}{}_{\nu]}\,,
\label{bianchi3b}
\ea
\end{subequations}
where $\tilde{R}^{\mu}{}_{\nu} = g^{\alpha\beta} R^{\mu}{}_{\alpha\nu\beta}$. 
This tensor, as well as the scalars $R = g^{\mu\nu}R_{\mu\nu}$ and $\tilde{R} = g^{\mu\nu}\tilde{R}_{\mu\nu}$, can be only defined wrt a given metric. (We will see that in symmetric teleparallelism, the Weyl 1-form $Q_\mu$ is pure gauge and can always be eliminated, even globally. In the curvature formulation, the same is true due to the projective invariance of the theory to be discussed in \ref{pinvariance}.)

\subsection{Lagrangian}
\label{lagrangian}

The lagrangian of the theory {is assumed to} depend only upon the 2 fields and their \1 derivatives,
\begin{subequations}
\be
L_G = L_G(g^{\mu\nu},g^{\mu\nu}{}_{,\lambda},\Gamma^\alpha{}_{\mu\nu},\Gamma^\alpha{}_{\mu\nu,\lambda})\,. 
\ee
If it is also assumed that the lagrangian is a scalar, it follows that it can always be written is terms of the tensors defined above, as
\be
L_G = L_G(g^{\mu\nu},Q_{\lambda}{}^{\mu\nu},T^\alpha{}_{\mu\nu},R^\alpha{}_{\mu\lambda\nu})\,. 
\ee
\end{subequations}
We may then define the conjugate tensors,
\begin{subequations}
\ba
q^\alpha{}_{\mu\nu} & = &  \frac{\partial L_G}{\partial Q_\alpha{}^{\mu\nu}}\,, \\ \label{qconju}
t_\alpha{}^{\mu\nu} & = &  \frac{\partial L_G}{\partial T^\alpha{}_{\mu\nu}}\,, \\ \label{tconju}
r_\alpha{}^{\beta\mu\nu} & = &  \frac{\partial L_G}{\partial R^\alpha{}_{\beta\mu\nu}}\,. \label{rconju}
\ea
\end{subequations}
For our purposes, it is important to include matter fields into the study. Thus, we also introduce arbitrary source fields $\psi$ described by the lagrangian $L_M$,
\be
L_M = L_M(g_{\mu\nu},\psi,\nabla_\alpha\psi)\,.
\ee
The dependence of $L_M$ upon the gravitational variables defines the energy-momentum tensor and the hypermomentum tensor of matter as
 \begin{subequations}
\ba
T_{\mu\nu} & = &  L_M g_{\mu\nu} -\frac{2\partial {L}_M}{\partial g^{\mu\nu}}\,, \label{hilbert} \\
\Hyp_\alpha{}^{\mu\nu} & = &  -\frac{\partial {L}_M}{\partial \Gamma^\alpha{}_{\mu\nu}}\,,
\ea
\end{subequations}
respectively. All physical fields contribute to energy-momentum. To our knowledge, there are 2 possible fundamental sources of hypermomentum, spinor fields in Einstein-Cartan-Kibble and related theories \cite{Kibble:1961ba,Trautman:2006fp}, and gauge fields in the iso-Khronon \1 order formulation of Yang-Mills theory \cite{Gallagher:2022kvv}.   
The total lagrangian is $L=L_G+L_M$, and the action integral of the coupled gravity-matter system is given as
\be
I = \int\diff^n x\sqrt{-\mathfrak{g}} L = \int \diff^n x \mathfrak{L}\,.
\ee
We have denoted the scalar density $\mathfrak{L}=\sqrt{-\mathfrak{g}}L$ and to be systematic, we will adopt this same notation for any tensor density, and thus in the following 
we might conveniently refer to $\mathfrak{T}^\alpha{}_{\mu\nu} = \sqrt{-\mathfrak{g}}{T}^\alpha{}_{\mu\nu}$, $\mathfrak{t}_\alpha{}^{\mu\nu} = \sqrt{-\mathfrak{g}}{t}_\alpha{}^{\mu\nu}$, etc., etc. For generality, we can also work in $n$ dimensions since relaxing $n=4$ is mostly inconsequential. 

The Euler-Lagrange variation of $I$ wrt the metric and the connection imply the 2 sets of field equations, respectively $\mathfrak{E}_{\mu\nu} =0$ and $\mathfrak{E}_\alpha{}^{\mu\nu} = 0$, where the 2 variations are
\begin{subequations}
\ba
 \mathfrak{E}_{\mu\nu} & = & - \frac{1}{2}\mathfrak{T}_{\mu\nu} + \lp \nabla_\alpha+T_\alpha\rp \mathfrak{q}^\alpha{}_{\mu\nu}+  \frac{\partial\mathfrak{L}_G}{\partial g^{\mu\nu}}\,, \label{geom1} \\
 \frac{1}{2}\mathfrak{E}_\alpha{}^{\mu\nu} & = & -\frac{1}{2}\mathfrak{\Hyp}_\alpha{}^{\mu\nu} + \lp \nabla_\beta+T_\beta\rp \mathfrak{r}_\alpha{}^{\nu\mu\beta}  +  \frac{1}{2}T^\mu{}_{\beta\gamma}\mathfrak{r}_\alpha{}^{\nu\beta\gamma} \nn \\  
 & + &  \mathfrak{t}_\alpha{}^{\mu\nu} - \mathfrak{q}^{\mu\nu}{}_\alpha\,, \label{ceom1}
\ea
\end{subequations}
or, in terms of tensors instead of tensor densities, 
\begin{subequations}
\ba
 {E}_{\mu\nu} & = & - \frac{1}{2}{T}_{\mu\nu} + \hat{\nabla}_\alpha q^\alpha{}_{\mu\nu}
  +    \frac{\partial{L_G}}{\partial g^{\mu\nu}} - \frac{1}{2}L_G g_{\mu\nu}\,, \quad  \quad  \label{geom2}\\
 \frac{1}{2}{E}_\alpha{}^{\mu\nu} & = & - \frac{1}{2}\Hyp_\alpha{}^{\mu\nu} + \hat{\nabla}_\beta r_\alpha{}^{\nu\mu\beta} +  \frac{1}{2}T^\mu{}_{\beta\gamma}r_\alpha{}^{\nu\beta\gamma}  \nn \\   & + & 
  t_\alpha{}^{\mu\nu} - q^{\mu\nu}{}_\alpha\,, \label{ceom2}
\ea
\end{subequations}
where we defined
\be
\hat{\nabla}_\mu = \nabla_\mu + T_\mu + \frac{1}{2}Q_\alpha\,. \label{hatnabla}
\ee
The usefulness of this covariant derivative symbol stems from partial integrations, since we have, for an arbitrary vector $X^\mu$
\bs
\be
\partial_\mu\mathfrak{X}^\mu = \sqrt{-\mathfrak{g}}\mathring{\nabla}_\mu X^\mu = \sqrt{-\mathfrak{g}}\hat{\nabla}_\mu X^\mu\,.
\ee
In case of an antisymmetric tensor $X^{[\mu\nu]}$,
\ba
\partial_\mu\mathfrak{X}^{[\mu\nu]} & = & \sqrt{-\mathfrak{g}}\mathring{\nabla}_\mu X^{[\mu\nu]} \nn \\
& = & \sqrt{-\mathfrak{g}}\lp \hat{\nabla}_\mu X^{[\mu\nu]} - \frac{1}{2}T^\nu{}_{\alpha\beta}X^{\alpha\beta}\rp
\nn \\
& = & \nabla_\mu\mathfrak{X}^{[\mu\nu]} - \frac{1}{2}T^\nu{}_{\alpha\beta}\mathfrak{X}^{[\alpha\beta]}
+ T_\mu\mathfrak{X}^{[\mu\nu]}\,. \label{antisymmetrictensor}
\ea
\es
The equations of motion $E_M=0$ for the matter fields $\psi$ are given by the Euler-Lagrange variation
\be
E_M = \frac{\partial L_M}{\partial\psi} - \hat{\nabla}_\mu \frac{\partial L_M}{\partial\nabla_\mu\psi}\,.
\ee
In the case that $\psi$ has any gauge charge, it is important to note that we consider $\nabla_\mu$ to be  the total covariant derivative. Only then $E_M$, as defined above, is fully covariant. 
In the following we will need to look at the Euler-Lagrange variations 
more carefully, and in particular, keep track of the total derivatives.  

\section{Current}
\label{current}

Now we are ready to take a step from notational to the more physical aspects of the Palatini-formulated theories. The transformation of interest is the most general coordinate transformation or, equivalently, the active diffeomorphisms. We review it in \ref{diffeomorphism} for the gravitational fields, i.e. the metric and the connection. We also discuss the properly implemented diffeomorphism of the matter fields, which is a nontrivial, and perhaps yet partly a nonresolved issue when gauge symmetries are involved. In subsection \ref{bianchi} we review the Bianchi identities, and finally in \ref{noethercharges} the Noether currents from a generic symmetry transformation. We emphasise the fundamentally quasilocal definition of physical charges. The local formula is a point-wise idealisation that is not always applicable. 

\subsection{Diffeomorphism}
\label{diffeomorphism}

The variation of the fields under an infinitesimal diffeomorphism 
\begin{subequations}
\label{morf}
\be
x^\alpha \rightarrow x^\alpha + v^\alpha\,,
\ee 
is given by (minus) their Lie derivative along the vector $v^\alpha$. The variation of the metric is therefore 
\ba
\delta_v g_{\mu\nu} & = & - g_{\mu\nu,\alpha}v^\alpha - 2g_{\alpha(\mu}  v^\alpha{}_{,\nu)}  \label{gmorf} \\
& = & -2\mathring{\nabla}_{(\mu}v_{\nu)}  \nn \\
& = & -2g_{\alpha(\mu}\nabla_{\nu)}v^\alpha + \lp 2T_{(\mu\nu)\alpha}-Q_{\alpha\mu\nu}\rp v^\alpha \nn \\
& = & - 2\nabla_{(\mu}v_{\nu)} + 2\lp T_{(\mu\nu)\alpha} - L_{\alpha\mu\nu}\rp v^\alpha\,. \nn
\ea
In the 1{}$^{\text{st}}$ line we wrote the variation in terms of the partial derivative $\partial_\alpha$, in the 2{}$^{\text{nd}}$ line
in terms of the metric connection $\mathring{\nabla}_\alpha$, and in the 3{}$^{\text{rd}}$ line in terms of the independent connection $\nabla_\alpha$. 
We will write the corresponding 3 expressions for the variation of the connection in the same order, 
\ba
\delta_v \Gamma^\alpha{}_{\mu\nu} & = & -v^\alpha{}_{,\mu\nu} +\Gamma^\alpha{}_{\mu\nu} v^\lambda{}_{,\lambda}  -  
\Gamma^\alpha{}_{\mu\lambda}v^\lambda{}_{,\nu}   \nn\\ & - & \Gamma^\alpha{}_{\lambda\nu}v^\lambda{}_{,\mu} - \Gamma^\alpha{}_{\mu\nu,\lambda}v^\lambda  \label{cmorf} \\   
& = & -\mathring{\nabla}_\mu\mathring{\nabla}_\nu v^\alpha - T^\alpha{}_{\lambda\nu}\mathring{\nabla}_\mu v^\lambda \nn \\
 &- & N^\alpha{}_{\mu\lambda}\mathring{\nabla}_\nu v^\lambda +  N^\lambda{}_{\mu\nu}\mathring{\nabla}_\lambda v^\alpha \nn \\
& - & 
\lp \mathring{\nabla}_\mu\bar{N}^\alpha{}_{\nu\lambda}+N^\alpha{}_{\mu\beta}\bar{N}^\beta{}_{\nu\lambda} - N^\beta{}_{\mu\nu}\bar{N}^\alpha{}_{\beta\lambda}\rp v^\lambda \nn \\
& = & -\nabla_\mu\nabla_\nu v^\alpha - \nabla_\mu\lp T^\alpha{}_{\beta\nu}v^\beta\rp - R^\alpha{}_{\nu\beta\mu}v^\beta\,. \nn
\ea
\end{subequations}
In the 2{}$^{\text{nd}}$ equality above we used the short-hands
\begin{subequations}
\ba
N^\alpha{}_{\mu\nu} = K^\alpha{}_{\mu\nu} + L^\alpha{}_{\mu\nu}\,, \\
\bar{N}^\alpha{}_{\mu\nu} = N^\alpha{}_{\mu\nu} - T^\alpha{}_{\mu\nu}\,.
\ea
\end{subequations}
The variation of a scalar field $\psi$ is simply 
\be
\delta_v \psi = -v^\alpha\partial_\alpha\psi\,.
\ee
We will however need to take into account generic matter and gauge fields, since our purpose is to elucidate the universality of gravitation: {\it all} energy is gravitational energy, in the sense that it should be computed from the gravitational fields $g_{\mu\nu}$,  $\Gamma^\alpha{}_{\mu\nu}$ and not from the source fields $\psi$. This will become clear in the course of the proceedings.  

Thus, in the following, the symbol $\psi$ can represent an arbitrary multiplet or whatever indexed collection $\psi=\{\psi_A\}_{A=1,2,\dots}$ of fields though we omit the indices to ease notation. Then it is important to take into account the gauge symmetries of the field theory under consideration, be it for matter fields or for gauge fields or both. As it is well-known, did one use the 
so-called ``canonical'' translation, the resulting Noether current would {not} in general coincide with the Hilbert energy-momentum tensor (\ref{hilbert}) but instead one obtains a so-called ``canonical Noether tensor''. In our view, it would be more appropriate to call the resulting current the pseudocanonical pseudotensor, since 
neither is it obtained gauge-covariantly wrt the internal symmetry nor is it a proper tensor wrt coordinate transformations. 

Instead, let us illustrate the more gauge-theoretical approach with the example of the  electromagnetic field $A_\mu$. We then take $v^\alpha$ to be electromagnetically charged, and we take $\nabla_\mu$ to be also electromagnetically charged. 
To begin, note that since $A_\mu$ is a 1-form, its ``canonical'' translation is given as
\be \label{wronglie}
\tilde{\delta}_v A_\mu =-(\mathcal{L}_v A)_\mu= -v^\alpha A_{\mu,\alpha} - v^\alpha{}_{,\mu} A_\alpha\,.   
\ee
In the case of the electromagnetic lagrangian, $L_M = -A_{\mu\nu}A^{\mu\nu}/4$, where $A_{\mu\nu}=2A_{[\nu,\mu]}$, (\ref{wronglie}) leads to the pseudocanonical pseudotensor,
\be \label{wrongAemt}
\tilde{T}^\mu{}_\nu = A^{\mu\alpha}A_{\alpha,\nu} - \frac{1}{4}\delta^\mu_\nu A^{\alpha\beta}A_{\alpha\beta}\,,
\ee
when restricting to constant translations, $v^\alpha{}_{,\mu}=0$. We should rather restrict to covariantly constant translations as follows. \1ly, we consider the gauge-covariant diffeomorphism,
\be \label{electriclie}
{\delta}_v A_\mu = -v^\alpha \nabla_\alpha A_{\mu} - \lp\nabla_\mu v^\alpha{}\rp A_\alpha\,. \ee
Yet, this transformation is ambiguous, since we assumed that $v^\alpha$ is electrically charged but have not specified the charges. We can fix the charge of each of the 4 components of $v^\alpha$ by requiring that the space-time scalar $v^\alpha A_\alpha$ is covariantly constant, i.e. $\nabla_\mu(v^\alpha A_\alpha) = 0$. This allows to rewrite the (\ref{electriclie}) as
\be \label{electriclieC}
{\delta}_v A_\mu = -2v^\alpha \nabla_{[\alpha}A_{\mu]} = -v^\alpha A_{\alpha\mu}\,,   
\ee
and we obtain  
\be \label{Aemt}
{T}^\mu{}_\nu = A^{\mu\alpha}A_{\nu\alpha} - \frac{1}{4}\delta^\mu_\nu A^{\alpha\beta}A_{\alpha\beta}\,.
\ee
This is the Noether current that is equivalent to the Hilbert energy-momentum tensor (\ref{hilbert}) computed from the electromagnetic lagrangian $L_M$. In our view, this should be called the canonical (and thus, the unique) electromagnetic energy-momentum current. The derivation can be immediately generalised to non-Abelian gauge fields. 

The difference between the currents (\ref{wrongAemt}) and (\ref{Aemt}) is a term that vanishes on-shell or, more generally, a superpotential term. It has been argued, originally by Kijowski, that both currents can correspond to a Hamiltonian for the electromagnetic field, depending upon what boundary conditions are appropriate to choose in a specific physical situation \cite{Chrusciel:1983dz}. However, taking into account gravity, it is always the energy-momentum (\ref{Aemt}) that sources the gravitational field, and it is in this sense we argue for a uniquely defined current. Boundary conditions applicable to a specific physical situation may allow to reduce the gauge-invariant current into an expression such as (\ref{wrongAemt}).
 
As another caveat, strictly speaking the unified treatment of all matter, gauge and gravitational fields using the universal covariant $\nabla_\mu$ works
according to the minimal coupling principle for the symmetric, but not for the metric teleparallel geometry. The complete clarification of the minimal coupling in generic metric-affine geometry has been carried out \cite{BeltranJimenez:2020sih}, but the gauge-covariant and proper canonical Noether currents of the matter sector would deserve a separate study\footnote{For related discussions on Lie derivatives in gauge theories, see e.g. \cite{Ortin:2002qb,Obukhov:2007se,Elgood:2020svt,Baker:2021hly,Kourkoulou:2022ajr}. 
Concerning the gauge formulation of gravity, we have checked that both the ``canonical'' and the covariant Lie derivatives \cite{Obukhov:2007se,Sauro:2022chz} (see Eq.(8) of Ref.\cite{Jimenez:2021nup} or section 2.5.2 of Ref.\cite{JimenezCano:2021rlu}) consistently lead to (\ref{morf}) via the tetrad postulate.}. In this article we shall keep track of the possible difference between the proper canonical energy-momentum currents and the pseudocanonical pseudotensors by writing down the symbol $\Delta T^\mu{}_\nu$ for this difference, though in our unified gauge theory framework it should follow consistently from \1 principles that  $\Delta T^\mu{}_\nu=0$.

\subsection{Bianchi identity}
\label{bianchi}

The variation of the gravitational action wrt  the inverse metric, $\delta g^{\mu\nu} = -g^{\mu\alpha}g^{\nu\alpha}\delta g_{\alpha\beta}$,  is 
\begin{subequations}
\ba
\delta_g I & = & \int\diff^n x \lb -\nabla_\alpha\lp \mathfrak{q}^\alpha{}_{\mu\nu}\delta g^{\mu\nu}\rp + \lp {\mathfrak{E}}_{\mu\nu} - T_\alpha \mathfrak{q}^\alpha{}_{\mu\nu}\rp\delta g^{\mu\nu} \rb \nn \\
& = & \int\diff^4 x\sqrt{-\mathfrak{g}} \Big[ -\mathring{\nabla}_\alpha\lp q^\alpha{}_{\mu\nu}\delta g^{\mu\nu}\rp  + E_{\mu\nu}\delta g^{\mu\nu} \Big]\,. \label{gvary}
\ea
In the 1$^{\text{st}}$ equality the expression $\mathfrak{E}_{\mu\nu}$ vanishes on-shell according to (\ref{geom1}), and in the 2$^{\text{nd}}$ equality we have on-shell $E_{\mu\nu}=0$ according to (\ref{geom2}). In the same way, we obtain for the variation 
wrt the connection
\ba
\delta_{\Gamma} I 
& = & \int\diff^n x\sqrt{-\mathfrak{g}} \left[   \mathring{\nabla}_\beta\lp 2{r}_\alpha{}^{\nu\beta\mu}\delta \Gamma^\alpha{}_{\mu\nu}\rp 
 +   E_\alpha{}^{\mu\nu}\delta\Gamma^\alpha{}_{\mu\nu} \right]\,. \label{cvary} \quad\quad 
\ea
\end{subequations}
Now consider the case of a diffeomorphism (\ref{morf}). Then (\ref{gvary}) becomes
\begin{subequations}
\label{dvary}
\ba
\delta_{g} I & = & 2 \int\diff^n x \sqrt{-\mathfrak{g}}\Big[ \mathring{\nabla}_\alpha\lp - {q}^\alpha{}_{\mu\nu}\mathring{\nabla}^\mu v^\nu + {E}^\alpha{}_\nu v^\nu\rp \nn \\
&  + & \mathring{\nabla}_\mu {E}^\mu{}_\nu v^\mu\Big]\,. \quad 
\ea
The variation (\ref{cvary}) of the connection under the diffeomorphism (\ref{cmorf}) results in
\ba
\delta_{\Gamma} I 
& = & \int\diff^n x\sqrt{-\mathfrak{g}} \Big[   \mathring{\nabla}_\beta\Big( 2{r}_\alpha{}^{\nu\beta\mu}\delta \Gamma^\alpha{}_{\mu\nu} - E_\alpha{}^{\beta\nu}T^\alpha{}_{\mu\nu} v^\mu \nn \\
& - & E_\alpha{}^{\beta\nu}\nabla_\nu v^\alpha + \hat{\nabla}_\mu E_\alpha{}^{\mu\beta}v^\alpha\Big) -\hat{\nabla}_\nu\hat{\nabla}_\mu E_\alpha{}^{\mu\nu} v^\alpha  \nn \\
& + & T^\beta{}_{\alpha\nu} \hat{\nabla}_\mu E_\beta{}^{\mu\nu}v^\alpha + R^\beta{}_{\nu\mu\alpha}E_\beta{}^{\mu\nu}v^\alpha\Big]\,.
\ea
Since we have included also matter sources $\psi$, we have to yet take into account the variation due to the matter fields,
\be
\delta_\psi I = \int\diff^n x\lb \mathfrak{E}_M \delta_v\psi +\partial_\mu\lp \frac{\partial \mathfrak{L}_M}{\partial\nabla_\mu\psi}\delta_v \psi\rp\rb\,.
\ee
Finally, we note that since we restrict to diffeomorphism invariant theories and, thus, $L$ must be a scalar, the total variation of the action reduces to the simple expression
\be
\delta_v I ={-\int\diff^nx\mathcal{L}_v (\sqrt{-\mathfrak{g}} \,L) }= -\int\diff^n x\sqrt{-\mathfrak{g}}\mathring{\nabla}_\mu\lp Lv^\mu\rp\,. 
\ee
\end{subequations}
Summing up the results (\ref{dvary}),  we arrive at the alternative expression for the total variation of the action,
\be
\delta_v I =  \int\diff^n x\Big[ \mathfrak{E}_\mu v^\mu + \mathfrak{E}_M\delta_v \psi + \partial_\mu \lp \mathfrak{J}^\mu - \mathfrak{L}v^\mu\rp\Big]\,,
\ee
where
\begin{subequations}
\ba
E_\mu & = & -2\mathring{\nabla}_\nu E^\nu{}_\mu - \hat{\nabla}_\beta  \hat{\nabla}_\alpha E_\mu{}^{\alpha\beta} \nn \\
&+ & T^\beta{}_{\mu\gamma}\hat{\nabla}_\alpha E_\beta{}^{\alpha\gamma} + R^\alpha{}_{\nu\beta\mu}E_\alpha{}^{\beta\nu}\,, \label{bianchi1}
\ea
and 
\ba
J^\mu & = & -2{q}^\mu{}_{\alpha\beta}\mathring{\nabla}^\alpha v^\beta + 2 {E}^\mu{}_\nu v^\nu \nn \\
& - & 2r_\alpha{}^{\nu\mu\beta}\Big[  \nabla_\beta\nabla_\nu v^\alpha + \nabla_\beta\lp T^\alpha{}_{\gamma\nu}v^\gamma\rp + R^\alpha{}_{\nu\gamma\beta}v^\gamma\Big] \nn \\
& - &  E_\alpha{}^{\mu\nu}\nabla_\nu v^\alpha + \hat{\nabla}_\nu E_\alpha{}^{\nu\mu}v^\alpha + T^\alpha{}_{\nu\beta}E_\alpha{}^{\mu\nu}v^\beta \nn  \\
& + & Lv^\mu +  \frac{\partial L_M}{\partial\nabla_\mu\psi}\delta_v\psi\,. \label{gnoether}
\ea
\end{subequations}
Assuming the vanishing of the total derivative term at the boundary $\partial$, i.e. $(J^\mu - Lv^\mu)\mid_\partial = 0$, we obtain the Noether identity 
\be
E_\mu = E_M\delta_\mu\psi\,. \label{nidentity}
\ee
In vacuum, or when the matter fields are on-shell, $E_M=0$, this reduces to what is known as the generalised Bianchi identity, $E_\mu = 0$ \cite{Magnano:1993bd,Koivisto:2005yk}.
This guarantees the conservation of the current density $\mathfrak{J}^\mu$,
\be
\mathring{\nabla}_\mu J^\mu = 0 \quad \text{i.e.} \quad \partial_\mu \mathfrak{J}^\mu = 0\,. \label{nconservation}
\ee
This is the Noether current and its conservation is simply the consequence of the diffeomorphism invariance of $I$.

The expression (\ref{gnoether}) can be rewritten in the standard form that makes it explicit that the Noether
currents from the \2 theorem (for gauge symmetries, such as the diffeomorphism symmetry) vanish on shell if the 
surface terms are dropped away. The standard form of the diffeomorphism Noether current in metric-affine gravity has been derived previously, both in the exterior calculus formulation in the language of differential forms \cite{Jimenez:2021nup} and in the Palatini formulation in the tensor language of linear algebra \cite{Obukhov:2015eha}. The result is
\bs
\ba
J^\mu & = & -v^\nu\lp 2E^\mu{}_\nu - \nabla_\alpha E_\nu{}^{\mu\alpha} - T^\beta{}_{\nu\alpha}E_\beta{}^{\mu\alpha}\rp  
\nn \\
& - &  \lp\nabla_\alpha v^\nu\rp E_\nu{}^{\mu\alpha} - 
\frac{\partial L_M}{\partial\nabla_\mu \psi}\delta_v\psi + \mathring{\nabla}_\nu J^{\nu\mu}\,,
\ea
where the surface term is
\be
J^{\mu\nu} = 2 v^\alpha t_\alpha{}^{\mu\nu} + 2\lp \nabla_\beta v^\alpha 
+ T^\alpha{}_{\lambda\beta}v^\lambda\rp r_\alpha{}^{\beta\mu\nu}\,.
\ee
\es
The antisymmetry $J^{\mu\nu}=J^{[\mu\nu]}$ renders manifest on-shell conservation of the current, $\mathring{\nabla}_\mu J^\mu \approx \mathring{\nabla}_\mu \mathring{\nabla}_\nu J^{\mu\nu} = 2\mathring{R}_{\mu\nu}J^{\mu\nu}=0$.  

\subsection{Noether charges}
\label{noethercharges}

Considering some arbitrary transformation $\delta$, the canonical Noether procedure leads, in the case of the invariance $\delta I =0$ of the action $I=I_G+I_M$, to the 
off-shell condition
\ba
0 & = & E_{\mu\nu}\delta g^{\mu\nu} + E_\alpha{}^{\mu\nu}\delta\Gamma^\alpha{}_{\mu\nu} + E_M\delta\psi \nn \\
& + &  \mathring{\nabla}_\alpha\Big( -q^\alpha{}_{\mu\nu}\delta g^{\mu\nu} +  2{r}_\beta{}^{\nu\alpha\mu}\delta \Gamma^\beta{}_{\mu\nu} \nn\\
& + &  L\delta x^\alpha + \frac{\partial L_M}{\partial\nabla_\alpha\psi}\delta\psi \Big)\,. \label{off-shell}
\ea
On-shell, the 1$^{\text{st}}$ line vanishes identically, and the expression in parenthesis is the canonical conserved Noether current.
In the case of diffeomorphisms $\delta =\delta_v$, we obtain the on-shell conserved current from (\ref{gnoether}) simply by setting $E^\mu{}_\nu = E_\alpha{}^{\mu\nu}=0$. 

There exists, at least locally\footnote{In the language of differential forms, this is because every closed form is locally exact.}, an antisymmetric 2$^{\text{nd}}$ rank tensor $J^{\mu\nu}=J^{[\mu\nu]}$ s.t. $J^\mu = \mathring{\nabla}_\nu J^{\mu\nu}$, then the Noether charge $C$ in a volume
$\mathcal{V}$ can be expressed as
\begin{subequations}
\label{integral}
\be \label{integral1}
C = \int_{\mathcal{V}} \diff^{n-1}\sigma_\mu \mathfrak{J}^\mu = \frac{1}{2}\oint_{\partial \mathcal{V}} \diff^{n-2}\sigma_{\mu\nu} \mathfrak{J}^{\mu\nu}\,. 
\ee
Here $\diff^{n-1}\sigma_\mu$ is understood as the infinitesimal element of the $(n-1)$-dimensional volume $\mathcal{V}$ (with the convention that the normal is pointing to the future), and 
$\diff^{n-1}\sigma_{\mu\nu}$ as the infinitesimal element of the $(n-2)$-dimensional hypersurface $\partial\mathcal{V}$ that encloses the volume $\mathcal{V}$. Another useful expression
is given by 1+3 decomposition, and considering the integration over a spatial volume $\mathcal{V}$,
\be \label{integral2}
C = \int_{\mathcal{V}} \diff^{n-1} x \mathfrak{J}^0 = \oint_{\partial \mathcal{V}} \diff^{n-2}x  \mathfrak{J}^{0 i}n_i\,,
\ee
\end{subequations}
where $n^i$ is the (outward pointing) unit normal to the surface $\partial \mathcal{V}$.
(We employ the somewhat clumsy notation for the volume and the surface elements in (\ref{integral}), the reason being that in this article we deliberately refrain from using the language of differential forms.)

We should stress that it is the \2 form in both of the equations (\ref{integral}) that is the fundamental
definition of the charge. 
In particular, physical charges are fundamentally quasilocal. That term is misleading if one takes it to mean that the charges would be defined only asymptotically, or that the quasilocal charge otherwise would not describe physics truly locally. It is simply a technical term meaning that the expression is a surface integral rather than a volume integral. For concreteness, let us consider the electromagnetic field $A_{\alpha\beta}$, sourced by a current ${J}^\alpha$ in 4 dimensions $n=4$. The electromagnetic charge can be sometimes written
as
\bs
\be \label{localcharge}
q = \int_{\mathcal{V}} \diff^3 x \mathfrak{J}^0\,.
\ee
That the current ${J}^\alpha$ is conserved, is dictated by gauge invariance.\footnote{{A clarification might be in order here. Certainly, the gauge symmetry, being an unphysical redundancy cannot strictly dictate the (on-shell) conservation of the physical current, with the understanding that physical charges can be used to classify physically different configurations. However, if we have a gauge field with a certain gauge symmetry, the so-called reducibility parameters of the gauge symmetry (i.e., those transformations that leave the gauge field invariant) provide the possible global transformations in the matter sector to which the gauge field can couple. These global symmetries are in fact responsible for the existence of the conserved currents and it is in this sense that we can say that the current conservation is dictated by gauge symmetry. Obviously, we can have conserved currents without associated gauge symmetries like the axial current for massless fermions.}} However, the charge $q$ need not be constant in time if there is flux in or out of the volume $\mathcal{V}$. This is all well known. The point we are stressing is that (\ref{localcharge}) may often fail to give the correct answer \cite{Ortin:2015hya}. Instead, using the Maxwell equation and the Stokes theorem suggests that
\be \label{qlocalcharge}
q = \frac{1}{2}\oint_{\partial \mathcal{V}} \diff^{2}\sigma_{\mu\nu}\sqrt{-\mathfrak{g}} A^{\mu\nu}\,. 
\ee
\es
This is the more fundamental definition of charge\footnote{{In fact, one can take a step further and {\it define} charges as the sources of electric fields and this would include situations without matter where charges would arise from topology. This more fundamental interpretation can be extended also to the gravitational case where we could have empty spacetimes with energy and momentum arising from non-trivial topologies.}}, and because it is a surface integral, it is called quasilocal. It is not meaningful to demand the density $\mathfrak{J}^0$ to remain valid point-wise i.e. at {\it infinitely} small scales, mathematically one can only consider an {\it infinitesimal} surface elements. Physically, of course even that is too much. The atoms of spacetime should presumably be considered to be closed surfaces with dimensions of the Planck scale at most.  

A long-standing problem was to define the charges of the gravitational field, energy and momentum. If we had a gravitational analogy to the electromagnetic field, it would have to be of the form $\mathfrak{h}^{\mu\nu}{}_\alpha$, i.e. have a component for each spacetime dimension labelled by an extra index. 
Again, the charge should be a quasilocal surface integral (now over 4 fluxes), rather than a volume integral over a source current (now with 4 components). Depending on the topology, fluxes may exist with nontrivial charges, even when there are no sources. Thus, the fundamental definition of the energy-momentum would have to be of the form
\be \label{qlocalchargeg}
C_\alpha = \frac{1}{2}\oint_{\partial \mathcal{V}} \diff^{2}\sigma_{\mu\nu} \mathfrak{h}^{\mu\nu}{}_\alpha\,. 
\ee
However, the Noether theorem does not yield a current of this form in the standard (Hilbert) formulation of GR, unlike in electromagnetism of course. 
We will review the calculation (using the equivalent Palatini formulation)
in the next section \ref{palatini}.
Many different forms for $\mathfrak{h}^{\mu\nu}{}_\alpha$ have been suggested, each with their own motivations, benefits and problems. A universal problem was perhaps the most obvious one, that there was no unique definition. Also highly elaborate mathematico-geometrical constructions had been developed as alternatives, without necessarily any connection with the Noether theorems \cite{Szabados:2009eka}. We could only very cursorily review some aspects of some of these developments in the introduction \ref{introduction}, but in section \ref{potentials} we will present several examples of the proposed gravitational field excitations, also called superpotentials in this context \cite{PhysRev.89.263}, found in the vast literature on the topic.

\section{Palatini gravity}
\label{palatini}

The \1 order formulation of the Hilbert action is sometimes called the Einstein-Palatini theory, and it is just a guise of the well-known Einstein-Cartan-Kibble-Sciama theory. When ignoring hypermomentum, the independent connection becomes dynamically equivalent to the Levi-Civita connection, up to projective invariance which we study in detail below. Plugging back the solution results in the standard Hilbert's \2 order purely metric formulation. Thus it is not surprising that the Noether current in the Einstein-Palatini theory is the same as in the Hilbert theory and given by the Komar expression that does not conform to (\ref{qlocalcharge}).

\subsection{Projective invariance}
\label{pinvariance}

The Einstein-Palatini lagrangian is
\be \label{epalatini}
{L}_G = \frac{m_P^2}{2}R\,. 
\ee
Then
\be \label{grconst}
r_\alpha{}^{\beta\mu\nu} =\frac{m_P^2}{2}g^{\beta[\nu}\delta^{\mu]}_\alpha\,,  
\ee
and (\ref{ceom2}) can be written as
\ba \label{pceom}
2m_P^{-2} E_\alpha{}^{\mu\nu} & = & T^\mu{}_\alpha{}^\nu +Q_\alpha{}^{\mu\nu} +\lp T^\nu+\frac{1}{2}Q^\nu - \tilde{Q}^\nu\rp\delta^\mu_\alpha \nn \\
& - & \lp T_\alpha+\frac{1}{2}Q_\alpha\rp g^{\mu\nu} - 2m_P^{-2} \Hyp_\alpha{}^{\mu\nu}\,. 
\ea
On-shell, the 3 traces of this equation give
\begin{subequations}
\label{htraces}
\ba
m_P^{-2} \Hyp_{\alpha}{}^{\mu\alpha} & = & 0\,, \\
m_P^{-2} \Hyp_{\alpha}{}^{\alpha\nu} & = & \lp n-2\rp T^\nu +\lp n-1\rp\lp  \frac{1}{2} Q^\nu -\tilde{Q}^\nu\rp\,, \quad \\
m_P^{-2} \Hyp_{\alpha}{}^{\mu}{}_{\mu} & = & -\lp n-2\rp T_\alpha - \frac{1}{2}\lp n-3\rp Q_\alpha   - \tilde{Q}_\alpha\,. 
\ea
\end{subequations}
These equations reflect the projective invariance of the action (\ref{epalatini}). The projective transformation generated by the 1-form $A_\mu$ acts on the connection as
\begin{subequations}
\label{projective}
\be
\delta_A \Gamma^\alpha{}_{\mu\nu} = A_\mu\delta^\alpha_\nu\,,
\ee
and, consequently, the 3 traces are transformed as
\ba
\delta_A T_\mu & = & \lp n-1\rp A_\mu\,, \\
\delta_A Q_\mu & = & -2n A_\mu\,, \\
\delta_A \tilde{Q}_\mu & = & -2A_\mu\,.
\ea
\end{subequations}
For traceless hypermomentum, the solution to the set of 3 equations (\ref{htraces}) is given by precisely such a 
projective relation (\ref{projective}) between the 3 traces. For vanishing hypermomentum, the general solution for
$E_\alpha{}^{\mu\nu}=0$ in (\ref{pceom}) is the connection $\Gamma^\alpha{}_{\mu\nu} =  \mathring{\Gamma}^\alpha{}_{\mu\nu}
+  A_\mu\delta^\alpha_\nu$ with an arbitrary $A_\mu$. This is a special type of vector distortion \cite{BeltranJimenez:2015pnp,BeltranJimenez:2016wxw,BeltranJimenez:2017vop}.
Thus, 1$^\text{st}$ order GR is a projectively invariant theory. The metric field equations (\ref{geom2}) can be expressed as
\be \label{efe}
R_{(\mu\nu)}-\frac{1}{2}R  g_{\mu\nu}= \mathring{R}_{\mu\nu}-\frac{1}{2}\mathring{R}g_{\mu\nu} = m_P^{-2}T_{\mu\nu}\,,
\ee
where in the \2 equality we used the connection field equation. We see that (\ref{efe}) is unaffected by the projective transformation, because $\delta_A R_{\mu\nu} = 2\nabla_{[\mu}A_{\nu]}$. 
Clearly, requiring either metricity, or requiring vanishing torsion fixes the gauge to 
$\Gamma^\alpha{}_{\mu\nu} =  \mathring{\Gamma}^\alpha{}_{\mu\nu}$ but the dynamics are unaffected by the choice of $A_\mu$.
The significance of the projective transformation of the affine 
connection\footnote{The projective transformation of the GL connection is a
Weyl rescaling of the affine conection \cite{Koivisto:2018aip,Iosifidis:2018zwo}.} is that it leaves the autoparallel paths invariant (path is the image of a curve, independent
of its parameterisation). Consider the equation for the autoparallel $x(t)$ wrt an arbitrary connection
\be \label{geodesic}
\ddot{x}^\alpha + \Gamma^\alpha{}_{\mu\nu}\dot{x}^\mu\dot{x}^\nu = 0\,.
\ee 
The autoparallel is called a geodesic when $\Gamma^\alpha{}_{\mu\nu} = \mathring{\Gamma}^\alpha{}_{\mu\nu}$. 
Under the reparameterisation
\begin{subequations}
\label{repara}
\ba
x^\alpha(\tau) & = & x^\alpha(\tau(t))\,, \\
\dot{x}^\alpha & = & \frac{\diff x^\alpha}{\diff \tau}\dot{\tau}\,, \\
\ddot{x}^\alpha & = & \frac{\diff^2 x^\alpha}{\diff \tau^2}\dot{\tau}^2 +  \frac{\diff x^\alpha}{\diff \tau}\ddot{\tau}\,,
\ea
we should also transform the connection,
\be
\Gamma^\alpha{}_{\mu\nu} = \Gamma^\alpha{}_{\mu\nu}  + \delta_A  \Gamma^\alpha{}_{\mu\nu}\,, \label{atransf}
\ee
\end{subequations}
so that we can rewrite the autoparallel equation (\ref{geodesic})
\be \label{geodesic2}
\frac{\diff^2 x^\alpha}{\diff \tau^2} + \Gamma^\alpha{}_{\mu\nu}\frac{\diff x^\mu}{\diff \tau}\frac{\diff x^\nu}{\diff \tau} + \lp A_\mu \frac{\diff x^\mu}{\diff \tau}+ \frac{\ddot{\tau}}{\dot{\tau}^2}\rp \frac{\diff x^\alpha}{\diff \tau} = 0\,.
\ee 
Unless $\ddot{\tau} =0$, in order to recover form invariance of the autoparallel equation, we should set the projection $A_\mu$ in (\ref{atransf}) according to the reparameterisation as 
\be
\frac{1}{\dot{\tau}}=\exp\left[\int\frac{\diff x^\mu }{\diff \tau}A_\mu\diff t\right]\,.
\ee
We have paid some attention to the $\delta_A$-invariance of Einstein-Palatini gravity 
since it may be useful to clarify that it plays a different role in 
the currents wrt the connection symmetries in teleparallelism which we will consider in the following section \ref{teleparallelism}. 

In particular, the projective symmetry of the Einstein-Palatini theory is a  trivial symmetry. From the expression (\ref{off-shell}), we see that the Noether current corresponding
to the symmetry (\ref{projective}) is just $J_A^\alpha = 2r_\mu{}^{\mu\alpha\nu}A_\nu$, and for the case (\ref{epalatini}) this expression vanishes. In such a case the symmetry is called trivial. Now, the triviality of the projective invariance reflects the geometrical fact that a particle's trajectory in 4 dimensions is the same physical object, a {\it path}, regardless of whether it is parameterised by the proper time of the particle, the 4$^{\text{th}}$ coordinate which may or not coincide with the former, or indeed parameterised by an arbitrary $\tau$. 


\subsection{Noether current}

We now consider the generalised Noether current (\ref{gnoether}) due to a diffeomorphism {generated} by $v^\mu$. We obtain, using (\ref{grconst}) and $t_\alpha{}^{\mu\nu}=q^\alpha{}_{\mu\nu}=0$,
\begin{subequations}
\ba
J^\mu & = & \frac{m_P^2}{2}\Big[ g^{\mu\nu}\nabla_\alpha\nabla_\nu v^\alpha-\delta^\mu_\alpha\Box v^\alpha + g^{\mu\nu}\nabla_\alpha\lp T^\alpha{}_{\gamma\nu} v^\gamma\rp \nn \\  
& - & \nabla^\nu\lp T^\mu{}_{\gamma\nu} v^\gamma\rp - R^\mu{}_\nu v^\nu - \tilde{R}^\mu{}_\nu v^\nu - Rv^\mu\Big] \nn \\
& + & \frac{\partial L_M}{\partial\nabla_\mu\psi}\delta_v \psi + L_Mv^\mu\,.
\ea
In the \3 line appears the canonical energy-momentum pseudotensor of matter. 
By rearranging the terms we get
\ba
J^\mu & = & \frac{m_P^2}{2}\Big[ 2\nabla_\alpha \lp \nabla^{[\mu}v^{\alpha]} + T^{[\mu\alpha]}{}_\nu v^\nu\rp \nn \\
& + & Q_\alpha{}^{\mu\nu}\lp \nabla_\nu v^\alpha + T^\alpha{}_{\gamma\nu}v^\gamma\rp  +  \Delta R^\mu{}_\nu v^\nu\Big]  \nn \\
& + & \Delta T^\mu{}_\nu v^\nu \,,\quad\quad \label{grcurrent}
\ea
\end{subequations}
where we used the field equation (\ref{efe}) and the shorthand $\Delta R_{\mu\nu} = 2 \mathring{R}_{\mu\nu}-R_{\mu\nu} - \tilde{R}_{\mu\nu}$, and
\be \label{difference}
\Delta T^\mu{}_\nu v^\nu = \frac{\partial L_M}{\partial\nabla_\mu\psi}\delta_v \psi - 2\frac{\partial L_M}{\partial g^{\mu\nu}}v^\nu\,,
\ee
where, for scalar matter, $\Delta T^\mu{}_\nu$ is the difference between the matter energy-momentum tensor and the pseudocanonical pseudotensor. (As discussed in \ref{diffeomorphism}, in the proper canonical formulation $\delta_v \psi$ should be such that $\Delta T^\mu{}_\nu v^\nu=0$.)

If there is no hypermomentum, $R_{\mu\nu} = \mathring{R}_{\mu\nu} + 2\mathring{\nabla}_{[\mu}A_{\nu]}$ 
and then $\Delta R_{\mu\nu} =0$. The current (\ref{grcurrent}) simplifies to
\begin{subequations}
\be \label{komar}
J^\mu = m_P^2\mathring{\nabla}_\alpha \lp \mathring{\nabla}^{[\mu}v^{\alpha]}\rp + \Delta T^\mu{}_\nu v^\nu\,.  
\ee 
The Noether potential can be identified as
\be \label{komar2}
J_G^{\mu\nu}  = m_P^2  \mathring{\nabla}^{[\mu}v^{\nu]}\,. \ee
This is the standard result in GR, the Komar superpotential \cite{Komar:1958wp}. 
  
It is also well known that the use of the Komar superpotential for the definition of energy does not always yield the desired results. In particular, the black hole energy appears to be only half of the Schwarzschild mass if computed from (\ref{komar}) in the standard coordinates. One cannot simply renormalise the expression, since then the angular momentum of the Kerr solution, computed in the Boyer-Linquist coordinates, becomes twice too 
large \cite{Bak:1993us}. Another drawback of the expression is that it gives an energy to the radiating asymptotic solution of Bondi {\it et al.} \cite{Bondi:1962px} that does not coincide with the Bondi mass. Rather than the standard energy-momentum current $\sim T^\mu{}_\nu$, the Komar superpotential corresponds to the trace-corrected $\sim T^\mu{}_\nu -\frac{1}{2}T\delta^\mu{}_\nu$. In fact, we will show in section \ref{gravitationalwaves} that it reduces asymptotically to the ADM (Arnowitt-Deser-Misner \cite{Arnowitt:1959ah,witten1962gravitation}) energy expression only under special circumstances (for pedagogic discussion see Ref.\cite{Jaramillo:2010ay}, and for the relation of the Komar superpotential and the Hamiltonian see e.g. \cite{Chrusciel:1983dz,Wald:1993nt}). 

It is interesting to note it has been proposed already some time ago that the problems with the Komar superpotential could be dealt with by invoking an additional vector field, which would enter the superpotential precisely as the projective vector distortion $A^\mu$ {\it if} our final result (\ref{komar2}) could be modified to
\be \label{komar3}
\tilde{J}_G^{\mu\nu} = m_P^2  \lp \mathring{\nabla}^{[\mu}v^{\nu]} + A^{[\mu}v^{\nu]}\rp\,.
\ee
\end{subequations}
This would result {\it if} we could set $\mathring{\nabla}_\mu \rightarrow {\nabla}_\mu$ in (\ref{komar2}). In 1985, Katz suggested to consider a flat reference metric as the suitable background structure that could be used to fix the results in situations when they were not in agreement with more succesfull superpotentials \cite{Katz_1985}. The suggestion was inspired by the Rosen's original bimetric construction \cite{ROSEN19631}, and the vector field $A^\mu$ in that case was identified as 1 of the 10 Killing vectors of the flat reference metric. The ``formal tricks of an artificially introduced flat background'' \cite{Katz_1985} are rather  non-minimal and admittedly rather ad hoc method of setting straight the anomalous factors in Komar expressions. However, the method produces the desired results.
A main result of this article will be that there is no need to postulate a reference metric or resort to any other artificial device to provide the required vector field extension (\ref{komar3}) of the Komar superpotential (\ref{komar2}), but instead it is built into geometrical foundation of the theory of coincident GR. 

In the next section \ref{teleparallelism}, we show that the canonical resolution of the energy problem emerges naturally in the teleparallel formulations, wherein the field equations can be understood as the statement that the current from Noether's \1 theorem
equals the current from Noether's \2 theorem, i.e. the divergence of the superpotential equals the sum of the metric and the matter energy-momentum tensors \cite{Koivisto:2022nar}.   
In the present case this seems not to be possible.
We could formally arrive at the expression for a superpotential
\be
{h}_K^{\mu\nu}{}_\alpha = m_P^2\mathring{\nabla}^{[\mu}\delta^{\nu]}_{\hat{\alpha}}\,,
\ee
where the hat upon the index signifies that the index is hidden from the covariant derivative operator. The inertial current becomes, again formally,
\be
t^\mu_{K\nu} = \frac{m_P^2}{2}\lb \lp \mathring{R}-\mathring{\Box}\rp\delta^\mu_{\hat{\nu}} + 
\lp 2\mathring{\nabla}^\mu\mathring{\nabla}_\alpha - \mathring{\nabla}_\alpha\mathring{\nabla}^\mu\rp\delta^\alpha_{\hat{\nu}}\rb\,. 
\ee
In the teleparallel pictures of GR, the inertial current i.e. the metric energy-momentum tensor is instead \1 order in the derivatives of the metric, and there is gauge freedom which allows to set the tensor to vanish, i.e. to choose an inertial frame. This freedom is absent in the curvature picture of GR, due to the triviality of the projective invariance.   

\section{Teleparallelism}
\label{teleparallelism}


Teleparallelism is defined by the flatness condition $R^\alpha{}_{\mu\beta\nu}=0$, in which case the connection takes the form of a pure gauge transformation of the trivial connection \cite{Schouten1954,BeltranJimenez:2019odq}
\be
\Gamma^\alpha{}_{\mu\nu} = (\Lambda^{-1})^\alpha{}_\beta\partial_\mu \Lambda^\beta{}_\nu\,, \label{lambda}
\ee
with $\Lambda^\alpha{}_\mu\in\text{GL}(4,\mathbb{R})$. We may then consider the invertible matrix $\Lambda^\alpha{}_\beta$ as the dynamical variable associated to the connection. Therefore, we need to reconsider the
variations wrt the connections, since now they are restricted to be of the form
\be
\delta \Gamma^\alpha{}_{\mu\nu} = \nabla_\mu\lb  (\Lambda^{-1})^\alpha{}_\beta\delta\Lambda^\beta{}_\nu\rb\,.  
\ee
The consequence is that the connection field equation changes to 
\be
\hat{\nabla}_\mu E_\alpha{}^{\mu\nu} = 0\,.
\ee
Thus, the vanishing of $E_\alpha{}^{\mu\nu}$ given in (\ref{ceom2}) is no longer guaranteed on-shell. However, since the derivation of the Noether identity (\ref{nidentity}) in \ref{bianchi} 
did not assume the field equations to hold, but only that the connection transforms under diffeomorphisms as in (\ref{cmorf}) as it does regardless of having curvature or
not, the generalised Bianchi identity $E_\mu=0$ remains valid for the $E_\mu$ given in (\ref{bianchi1}), and the conservation $\partial_\mu J^\mu =0$ still holds for the 
generalised  Noether current $J^\mu$ given in (\ref{gnoether}). Therefore, in teleparallelism the 1$^{\text{st}}$ and the 3$^{\text{rd}}$ terms on the 3$^{\text{rd}}$ line of
(\ref{gnoether}) have to be taken into account even on-shell. This is consistent with the result from the canonical Noether procedure, when we note that\footnote{{It might be convenient to stress that $\Lambda^\alpha{}_\mu$ transforms under diffeomorphisms as a set of 4 1-forms (contravariant vectors) so that the $\Gamma^\alpha{}_{\mu\nu}$ transforms as a connection.}}
\ba
\delta_v \Lambda^\mu{}_\nu&=&-(\mathcal{L}_v\Lambda)^\mu{}_\nu =-v^\lambda\partial_\lambda \Lambda^\mu{}_\nu-\partial_\nu v^\lambda \Lambda^\mu{}_\lambda\nn\\
&=&-\Lambda^\mu{}_\alpha\Big( \nabla_\nu v^\alpha + T^\alpha{}_{\beta\nu}v^\beta\Big)\,,
\ea
and take 1 further step from (\ref{off-shell}) to rewrite the off-shell identity 
\begin{subequations}
\ba
0 & = & E_{\mu\nu}\delta g^{\mu\nu} - (\Lambda^{-1})^\alpha{}_\beta\hat{\nabla}_\mu E_\alpha{}^{\mu\nu}\delta \Lambda^\beta{}_\nu + E_M\delta\psi \nn \\
& + &  \mathring{\nabla}_\mu\Big[ \hat{\nabla}_\nu E_\alpha{}^{\nu\mu}v^\alpha  -   2{q}^\mu{}_{\alpha\beta}\mathring{\nabla}^\alpha v^\beta -  E_\alpha{}^{\mu\nu}\nabla_\nu v^\alpha  \nn \\
 & + &  T^\alpha{}_{\nu\beta}E_\alpha{}^{\mu\nu}v^\beta
  +     Lv^\mu + \frac{\partial L_M}{\partial\nabla_\mu\psi} \delta_v\psi\Big]\,, \label{off-shell_tele}
\ea
where the expression in square brackets is the Noether current in teleparallelism. It is conserved on-shell, wherein only the \1 term vanishes identically. 
Indeed, we note that this is almost identical to the off-shell conserved current appearing in (\ref{gnoether}),
\ba
J^\mu & = & 2 {E}^\mu{}_\nu v^\nu + \hat{\nabla}_\nu E_\alpha{}^{\nu\mu}v^\alpha  - 2{q}^\mu{}_{\alpha\beta}\mathring{\nabla}^\alpha v^\beta - E_\alpha{}^{\mu\nu}\nabla_\nu v^\alpha \nn \\
& + &  T^\alpha{}_{\nu\beta}E_\alpha{}^{\mu\nu}v^\beta 
 + Lv^\mu +  \frac{\partial L_M}{\partial\nabla_\mu\psi}\delta_v\psi\,, \label{telenoether}
\ea
apart from just the 1$^{\text{st}}$ term. 

\subsection{Metric teleparallelism}
\label{metrictele}

At this point, we restrict to the metric teleparallel framework.
{By writing then open the 2$^{\text{nd}}$ term above}, we obtain the current in the form
\ba
J^\mu & = & \lb \hat{\nabla}_\nu \lp 2t_\alpha{}^{\nu\mu} - Z_\alpha{}^{\nu\mu}\rp 
 +  2g_{\nu\alpha} \frac{\partial L}{\partial g^{\mu\nu}}\rb v^\alpha \nn  \\
& - & E_\alpha{}^{\mu\nu}\nabla_\nu v^\alpha + T^\alpha{}_{\nu\beta}E_\alpha{}^{\mu\nu}v^\beta\,.  \label{telenoether2}
\ea 
\end{subequations}
The variations then reduce to 
\begin{subequations}
\ba
\mathfrak{E}_{\mu\nu} & = & -\frac{1}{2}\mathfrak{T}_{\mu\nu} + \frac{\partial\mathfrak{L}_G}{\partial g^{\mu\nu}}\,, \\
\mathfrak{E}_\alpha{}^{\mu\nu} & = & -\mathfrak{\Hyp}_\alpha{}^{\mu\nu} + 2\mathfrak{t}_\alpha{}^{\mu\nu}\,.
\ea
\end{subequations}
For further simplicity we specify that the action has a quadratic form 
\begin{subequations}
\be
\mathfrak{L}_G = \frac{1}{2}\mathfrak{t}_\alpha{}^{\mu\nu} {T}^\alpha{}_{\mu\nu}\,,  
\ee
such that
\be
 \frac{\partial{L}_G}{\partial g^{\mu\nu}} = -\frac{1}{2}t_{(\mu}{}^{\alpha\beta} T_{\nu)\alpha\beta} + t_{\alpha\beta(\mu}T^{\alpha\beta}{}_{\nu)}\,.
\ee
\end{subequations}
Then it is not difficult to see that (\ref{telenoether2}) simplifies to
\ba
J^\mu & = &  \mathring{\nabla}_\nu \Big[ \lp 2t_\alpha{}^{\nu\mu} - \Hyp_\alpha{}^{\nu\mu}\rp v^\alpha\Big] \nn \\ 
 & - & \frac{1}{2}\Hyp_{(\mu}{}^{\alpha\beta} T_{\nu)\alpha\beta} + \Hyp_{\alpha\beta(\mu}T^{\alpha\beta}{}_{\nu)} + \Delta T^\mu{}_\nu v^\nu\,.
\label{tcurrent}
\ea
The 2$^{\text{nd}}$ line takes into account the hypermomentum and the part defined in (\ref{difference}). The gravitational superpotential
\be \label{metricsuperpotential}
J_G^{\mu\nu} = -2t_\alpha{}^{\mu\nu}v^\alpha\,,
\ee
is the one expected from many previous studies of metric teleparallelism\footnote{The currents derived in the lagrangian (or hamiltonian) approach are sometimes contrasted with the expression deduced directly from the field equations \cite{Maluf:2005kn}. The latter perspective is discussed in Section \ref{potentials}, and we shall conclude that both ways lead to the same result. We could remark that if it is argued that the lagrangian-based approach is ambiguous since one may add total derivatives to the action (ambiguity 2) as in \ref{boundary}), so are the equations of motion-based approach ambiguous since {one may as well add superpotentials with vanishing divergence to the equations of motion} (ambiguity 1) as in \ref{boundary}).} \cite{nla.cat-vn2925903,Maluf:1994ji,deAndrade:2000kr,Maluf:2002zc,Emtsova:2019moq}. In particular, this expression is known to reduce (at the leading order in Riemann normal coordinates, with a judicious choice of frame) to the Bel-Robinson expression for the gravitational energy \cite{So:2008kr} which has physically desirable properties.  

Metric teleparallel theory is conventionally formulated in terms of tetrads \cite{Aldrovandi:2013wha} rather than using the Palatini formalism \cite{BeltranJimenez:2018vdo}.\footnote{{Both formulations are however equivalent when working with the inertial connection $\Lambda^\alpha{}_\mu$} that plays the role of the tetrad once the metric-compatibility constraint is solved, see e.g. \cite{BeltranJimenez:2019odq,BeltranJimenez:2019nns}.} The idea that the tetrad formulation could help with the pseudotensorial ambiguity of the superpotentials goes back at least to M{\o}ller's work in the 1960's \cite{MOLLER1961118}. However, the coordinate ambiguity is only translated to the Lorentz frame ambiguity in the tetrad formulation, and the same problem is of course seen in the Palatini form of the result (\ref{metricsuperpotential}), which depends upon the chosen solution for the 2 fields $\Lambda^\mu{}_\nu$ and $g_{\mu\nu}$, even though the superpotential is covariant wrt simultaneous transformations of the field. Here we need the definition of the general-relativistic inertial frame \cite{Jimenez:2019yyx} that fixes, given 1 of the 2 fields, the other one. We shall now review some previous calculations in the literature, and see that they will not give the correct energy charge for the black hole (except asymptotically). Then we will redo the calculation in an inertial frame. To our knowledge, that is the \1 exact derivation of the black hole energy in metric teleparallelism. (Perhaps, it indeed can be understood in terms of a Newtonian potential energy in the force interpretation of gravity compatible with metric teleparallelism \cite{Knox2011-KNONTA,Read:2018ogw}).

Let us consider the following spherically symmetric metric:
\be \label{sphericalmetric}
 \diff s^2 = -A^2\diff t^2 + B^{-2}\diff r^2 + r^2\sin^2\theta\diff \phi^2 + r^2\diff\theta^2\,.
\ee
In the gauge corresponding to a diagonal tetrad, the non-vanishing components of $t_\alpha{}^{\mu\nu}=t_\alpha{}^{[\mu\nu]}$ are 
\begin{subequations}
\label{st21}
\ba
t_0{}^{0r} & = & -m_P^2\frac{2B^2}{r}\,, \\
t_0{}^{0\theta} & = & t_r{}^{r\theta} = -m_P^2\frac{\cot\theta}{r^2}\,, \\
t_\theta{}^{r\theta} & = & t_\phi{}^{r\phi} = m_P^2 B^2\lp \frac{1}{r} +  \frac{A'}{A}\rp\,.
\ea
\end{subequations}
Computing the Noether charge with respect to the time-like Killing vector $v^\mu=\delta^\mu_0$ gives $C_v = 8\pi m_P^2 r B^2$. 
This result diverges as $r \rightarrow \infty$. This is interpreted, in the standard context of metric teleparallelism \cite{Aldrovandi:2013wha}, to be due to inertial forces. To eliminate such forces, the gauge freedom 
is used to Lorentz-rotate the connection. The finiteness of the action has been considered as a criterion that distinguishes the absence of inertial forces \cite{Obukhov:2006sk,Lucas:2009nq,Krssak:2015rqa,Krssak:2015lba,Krssak:2018ywd}. A solution with finite action was reported by Kr{\u s}{\u s}{\'a}k {\it et al} \cite{Krssak:2018ywd} to be associated with the superpotential (again we write only the non-vanishing components
and recall the antisymmetry)
 \begin{subequations}
 \label{st1}
\ba
t_0{}^{0r} & = & -\frac{2m_P^2}{r}\lp B - B^2\rp\,, \\
t_\theta{}^{r\theta} & = & t_\phi{}^{r\phi} = m_P^2 B^2\lp \frac{1}{r} +  \frac{A'}{A}\rp - m_P^2 \frac{B}{r} \,.
\ea
\end{subequations}
Now the charge is finite, and equal to the mass of the black hole when the integration limit is extrapolated to the asymptotic infinity $r \rightarrow \infty$. 

Brown and York introduced regularisations in the context of the energy problem in GR, in their derivation of a famous quasilocal energy expression \cite{Brown:1992br}. Before Kr{\u s}{\u s}{\'a}k {\it et al}, e.g. Maluf {\it et al} had used regularisations in the context of metric teleparallel energy expression \cite{Maluf:2005sr,Ulhoa:2010wv}. In fact, 1 of the prescriptions adopted in metric teleparallelism is claimed to yield the same energy as the Brown-York expression \cite{Castello-Branco:2013iza}. For the Schwarzschild black hole with mass $m_S$, the expression implies that the energy contained inside the horizon is $2m_S$, and outside the horizon, the whole spacetime is filled with negative energy energy density such that asymptotically the total energy converges to $m_S$. This probably could be interpreted to describe something that could be related to some effective energy concept, but it is not the canonical energy charge we are concerned with in this article.

More recently, Emtsova and Toporensky also introduced a 
different regularisation scheme, using the metric connection (\ref{christoffel}) to normalise the spin connection \cite{Emtsova:2019moq,Emtsova:2020gxa}. 
They considered the Schwarzschild black hole 
\be
A=B=\sqrt{1-\frac{m_S}{4\pi m_P^2}}\,,
\ee
and reported the regularised spin connection 
\begin{subequations}
\label{st22}
\ba
t_0{}^{0r} & = & -\frac{2m_P^2}{r}\lp 1-A\rp - \frac{m_S}{4\pi r^2}\,, \\
t_\theta{}^{r\theta} & = & t_\phi{}^{r\phi} = -\frac{m_P^2}{r}\lp 1- A \rp + \frac{m_S}{4\pi r^2}\,.
\ea
\end{subequations}
This gives the correct value $C_0=m_S$, but again only when integrating the current over the whole  spacetime. The scheme based on the {comparison metric connection}, as well as the scheme based on the finiteness of the action (the scheme which could be also called holographic renormalisation \cite{Krssak:2015lba,Krssak:2018ywd}), are both regularisations of only the total value of the energy i.e. the global mass within the spacetime manifold.

The novel proposition is that the energy measured {\it locally} by a physical observer can be determined in an inertial frame \cite{Jimenez:2019yyx}. The inertial frame is characterised by the vanishing of the energy-momentum associated with the metric field (or equivalently, the tetrad). This has not been investigated in metric teleparallelism (see however Ref. \cite{So:2008kr}, where the Riemann normal coordinates are considered). Even though we argued in Ref. \cite{Jimenez:2021nup} that the results should agree with computations in symmetric teleparallelism, it is  quite interesting to check explicitly the inertial frame hypothesis for the case of metric teleparallelism. A freely falling frame can be obtained by switching (\ref{sphericalmetric}) to the Lema{\^i}tre coordinates:
\be \label{lemaitre}
A = 1\,, \quad B^2 = \frac{m_S}{4\pi m_P^2 r}\,. 
\ee
The Lema{\^i}tre frame has been considered previously in metric teleparallelism \cite{Maluf:2007qq} and the components of the superpotential have been computed for the regularised spin connection \cite{Emtsova:2021ehh}, 
 \begin{subequations}
 \label{lemaitret}
\ba
t_0{}^{0r} & = & -\frac{m_S}{4\pi r^2} = 4t_\phi{}^{r\phi}= 4t_\theta{}^{r\theta}\,, \\
t_r{}^{0r} & = & \frac{m_P^2}{r}B = 4t_\phi{}^{0\phi}= 4t_\theta{}^{0\theta}\,.
\ea
\end{subequations}
According to the Ref. \cite{Emtsova:2021ehh}, the gravitational energy-momentum charges are vanishing in the freely falling frame. Our interpretation is quite different as we maintain that the gravitational energy momentum current should {\it always} vanish in an inertial frame and it is only in such a frame that we may compute the physically meaningful result for the gravitational charges. (Recall that we insist that not the local integral of the form (\ref{localcharge}) but the so-called quasilocal integral of the form (\ref{qlocalcharge}) should be regarded as the fundamental definition of the charge - in the case at hand this makes all the difference.) Indeed, now $C_0=m_S$ in the Lema{\^i}tre frame, as we readily compute from (\ref{lemaitret}). The crucial point here is that we obtain $C_0=m_S$ regardless of which surface we choose to compute the flux (as long as it contains the singularity), in contrast to the previous prescriptions such as (\ref{st1}) and (\ref{st22}) which, more or less accidentally, happen to give the correct result when $r$ is taken to infinity\footnote{The many regularisation schemes used in the literature could be said belong to the
locality class $N$ according to the respective corrections behaving as $r^{-N}$ at large $r$. The inertial frame regularisation would then have $N=\infty$.}. Of course, it is intuitively clear that energy is well defined in Minkowski space, and when such is available as the asymptotic limit of an otherwise arbitrary spacetime the limiting expression is unambiguous and its expression is known as the ADM energy. However, at any finite $r$ all previous calculations in metric teleparallelism we are aware of, fail to give the correct answer.  

In conclusion, we have proposed the definition of energy in metric teleparallel gravity and, at least for the Schwarzschild geometry, confirmed the validity of the definition. This vindicates the idea originally put forward by Christian M{\o}ller in 1961 \cite{Moller:233632}.

\subsection{Symmetric teleparallelism}
\label{1lmformalism}

We shall consider the quadratic action
\be
\mathfrak{L}_G = \frac{1}{2}\mathfrak{q}^\alpha{}_{\mu\nu}Q_\alpha{}^{\mu\nu} + \mathfrak{l}_\alpha{}^{\mu\nu} T^\alpha{}_{\mu\nu}\,, \label{Cquadratic}
\ee
where the symmetricity constraint is imposed by the Lagrange multiplier 
$\mathfrak{l}_\alpha{}^{\mu\nu}=\mathfrak{l}_\alpha{}^{[\mu\nu]}$. The field equations for the Lagrange multiplier, the metric and the connection are respectively given by
\bs
\ba
\frac{\delta\mathfrak{L_G}}{\delta\mathfrak{l}_\alpha{}^{\mu\nu}} & = & {T}^\alpha{}_{\mu\nu}\,, \\
2\mathfrak{E}^\mu{}_\nu & = & -\mathfrak{T}^\mu{}_\nu + 2\nabla_\alpha \mathfrak{q}^{\alpha\mu}{}_\nu + 2\mathfrak{q}^{\alpha\beta}{}_\nu Q_{\alpha\beta}{}^\mu \nn \\
& + & 2\sqrt{\mathfrak{-g}}\frac{\partial L_G}{\partial g^{\mu\nu}}
- \mathfrak{L}_G\delta^\mu_\nu\,, \label{Cmetric} \\
\mathfrak{E}_\alpha{}^{\mu\nu} & = & 2\mathfrak{l}_\alpha{}^{\mu\nu} 
- 2\mathfrak{q}^{\mu\nu}{}_\alpha + \mathfrak{Z}_\alpha{}^{\mu\nu}\,.
\label{connectionvariation}
\ea
\es
We have massaged the metric field equation (\ref{geom1}) in a form (\ref{Cmetric}) that will be helpful in the following. Our aim is again to derive the Noether potential.
To that aim, we begin by setting the teleparallel current (\ref{telenoether}) on shell,
\be \label{Ctelenoether}
J^\mu = Lv^\mu 
+ \frac{\partial L_M}{\partial \nabla_\mu\psi}\delta_v \psi-2q^\mu{}_{\alpha\beta}\mathring{\nabla}^\alpha v^\alpha -
E_\alpha{}^{\mu\nu}\nabla_\nu v^\alpha\,. 
\ee
Consider the connection field equation $\nabla_\mu \mathfrak{E}_\alpha{}^{\mu\nu} = 0$, where $\mathfrak{E}_\alpha{}^{\mu\nu}$ is given by (\ref{connectionvariation}). At this point, we just assume
that there is a solution $\mathfrak{l}_\alpha{}^{\mu\nu}=\frac{1}{2}\mathfrak{h}^{\mu\nu}{}_\alpha$ s.t.
\be
\nabla_\mu  \mathfrak{h}^{\mu\nu}{}_\alpha = 
\nabla_\mu\lp 2\mathfrak{q}^{\mu\nu}{}_\alpha - \mathfrak{Z}_\alpha{}^{\mu\nu}\rp\,. \label{assumedsolution}
\ee
We shall return to discuss this solution in details later. Now, plugging the
connection variation (\ref{connectionvariation}) and the assumed solution
(\ref{assumedsolution}) into the current (\ref{Ctelenoether}) it becomes
\bs
\ba
J^\mu & = & Lv^\mu 
+ \frac{\partial L_M}{\partial \nabla_\mu\psi}\delta_v \psi - q^\mu{}_{\alpha\beta}Q_\nu{}^{\alpha\beta}v^\nu \nn \\
& - & \lp h^{\mu\nu}{}_\alpha + Z_\alpha{}^{\mu\nu}\rp \nabla_\nu v^\alpha\,. 
\ea
Now we just use the (assumed) solution (\ref{assumedsolution}) to replace the derivative of the non-metricity conjugate in the metric field equation
(\ref{Cmetric}) and plug it into the above, 
\be \label{Ccurrent}
\mathfrak{J}^\mu = \nabla_\alpha\lp \mathfrak{h}^{\alpha\mu}{}_\nu v^\nu\rp 
+ \lp \nabla_\alpha\mathfrak{Z}_\nu{}^{\alpha\mu}\rp v^\nu -
\mathfrak{Z}_\nu{}^{\mu\alpha}\nabla_\alpha v^\nu + \Delta T^\mu{}_\nu\,.
\ee
\es
Apart from the hypermomentum, the current is precisely\footnote{Like in the metric teleparallel case (\ref{tcurrent}), there are additional hypermomentum terms which do not appear in the canonical form. Because the hypothetical hypermomentum is introduced as just a parameterisation of a possible connection-dependence of the lagrangian $L_M$, we don't have the required fundamental equations to determine the structure of the $Z_\alpha{}^{\mu\nu}$-terms. For example, if the
equations implied that $\nabla_\alpha \mathfrak{Z}_\nu{}^{\alpha\mu} = 
-\nabla_\alpha \mathfrak{Z}_\nu{}^{\mu\alpha}$, the current could be put to the canonical form. A conjecture is that a consistent realisation of hypermomentum should decouple from the canonical current.} the one derived in the metric-affine gauge formulation \cite{Jimenez:2021nup}, and deduced
earlier from the premetric axioms \cite{Koivisto:2021jdy}.

This completes the derivations of the Noether currents in the Palatini formulation of the Geometrical Trinity. 

\subsection{Alternative formulations}
\label{alternatives}

It may be worthwhile to clarify the formulation-dependence of the Noether currents. Therefore we shall derive some non-canonical currents that result in some alternative formulations of symmetric teleparallel gravity.
The ambiguity of the currents in different formulations can be regarded as 
a strength rather than a weakness of the Noether formalism. It allows to distinguish between different formulations of the action principle giving rise to equivalent dynamics.
That we don't find exactly the right current in the tensor formulation (with 2 Lagrange multipliers in \ref{lmformalism} or with 0 Lagrange multipliers in \ref{icformalism}), indicates that the gauge formulation used in Ref. \cite{Jimenez:2021nup} is a more fundamental representation of the theory (whose proper translation to tensor language is the formulation in
\ref{1lmformalism}). Defects that can be detected already at the classical level hint that the formulation may encounter serious obstacles in the quantum theory. It is well known that the path integral can depend upon the choice of fundamental variables. 

\subsubsection{Lagrange multiplier formalism}
\label{lmformalism}

To drop some inessential baggage from the derivation, in this subsection we 1) set hypermomentum to zero $Z_\alpha{}^{\mu\nu}=0$ and 2) consider on-shell relations, i.e. set $E_{\mu\nu}=E_\alpha{}^{\mu\nu}=0$, without explicitly reminding about it. 
In the Lagrange multiplier formulation \cite{BeltranJimenez:2017tkd}, we vary 
\be
\mathfrak{L}_G = \frac{1}{2}\mathfrak{q}^\alpha{}_{\mu\nu}Q_\alpha{}^{\mu\nu}
+ \mathfrak{l}_\alpha{}^{\mu\nu}T^\alpha{}_{\mu\nu} + \mathfrak{l}_\alpha{}^{\beta\mu\nu}R^\alpha{}_{\beta\mu\nu}\,, \label{altquadratic}
\ee
wrt a general connection. The multipliers have the antisymmetry wrt their 2 last indices,
\bs
\ba
\mathfrak{l}_\alpha{}^{(\mu\nu)} & = & 0\,, \label{altsym1} \\
\mathfrak{l}_\alpha{}^{\beta(\mu\nu)} & = & 0\,. \label{altsym2}
\ea
\es
We then obtain the equations of motion
\bs
\ba
T^\alpha{}_{\mu\nu} & = & 0\,, \label{alteom1} \\
R^\alpha{}_{\beta\mu\nu} & = & 0\,, \label{alteom2} \\
\mathfrak{T}_{\mu\nu} &= & 2\nabla_\alpha\mathfrak{q}^\alpha{}_{\mu\nu} + 2\frac{\partial\mathfrak{L}_G}{\partial g^{\mu\nu}}\,, \label{alteom3} \\
\nabla_\beta\mathfrak{l}_\alpha{}^{\nu\mu\beta} & = & \mathfrak{q}^{\mu\nu}{}_\alpha -\mathfrak{l}_\alpha{}^{\mu\nu}\,, \label{alteom4}
\ea
\es
where the \1 and the \2 equations are of course the 2 new equations of motion resulting from the variation wrt the Lagrange multipliers. Now, instead of the 
teleparallel current (\ref{telenoether2}) we should go back to the general current (\ref{gnoether}), which now reduces to
\be
\mathfrak{J}^\mu = -2\mathfrak{q}^\mu{}_{\alpha\beta}\mathring{\nabla}^\alpha v^\beta + 2\mathfrak{l}_\alpha{}^{\nu\beta\mu}\nabla_\beta\nabla_\nu v^\alpha + \mathfrak{L}v^\mu + \frac{\partial \mathfrak{L}_M}{\partial \nabla_\mu\psi}\delta_v\psi\,. \label{altcurrent1}
\ee
Let us consider the \2 term
\ba
\mathfrak{l}_\alpha{}^{\nu\beta\mu}\nabla_\beta\nabla_\nu v^\alpha
 &=& \nabla_\beta\lp \mathfrak{l}_\alpha{}^{\nu\beta\mu}\nabla_\nu v^\alpha \rp + \lp\nabla_\beta\mathfrak{l}_\alpha{}^{\nu\mu\beta}\rp\nabla_\nu v^\alpha 
 \nn \\
 &=& \nabla_\beta\lp \mathfrak{l}_\alpha{}^{\nu\beta\mu}\nabla_\nu v^\alpha \rp \nn \\ & + & 
 \lp\mathfrak{q}^{\mu\nu}{}_\alpha - \mathfrak{l}_{\alpha}{}^{\mu\nu} \rp\nabla_\nu v^\alpha\,. 
 \ea
In the \1 equality we used (\ref{altsym2}) and in the \2 we used (\ref{alteom4}).
Plugging this into the (\ref{altcurrent1}) gives
\ba
\mathfrak{J}^\mu & = & 2\mathfrak{q}^{\mu\nu}{}_{\alpha}\lp{\nabla}_\alpha v^\beta - \mathring{\nabla}_\alpha v^\beta\rp - 2\mathfrak{l}_\alpha{}^{\mu\nu}\nabla_\nu v^\alpha \nn \\
& + & 2\nabla_\beta\lp \mathfrak{l}_\alpha{}^{\nu\beta\mu}\nabla_\nu v^\alpha \rp
 +   \mathfrak{L} v^\mu + \frac{\partial \mathfrak{L}_M}{\partial \nabla_\mu}\delta_v\psi \nn \\
& = & 2\mathfrak{q}^\mu{}_{\alpha\beta}L^{\beta\alpha}{}_\lambda v^\lambda 
- 2\nabla_\nu\lp \mathfrak{l}_\alpha{}^{\mu\nu} v^\alpha\rp 
+ 2 \lp\nabla_\nu\mathfrak{l}_\alpha{}^{\mu\nu}\rp v^\alpha \nn \\
 & + & 2\nabla_\beta\lp \mathfrak{l}_\alpha{}^{\nu\beta\mu}\nabla_\nu v^\alpha \rp
 +  \mathfrak{L}v^\mu + \frac{\partial \mathfrak{L}_M}{\partial \nabla_\mu}\delta_v\psi\,.
 \label{altcurrent2}
\ea
Taking the divergence of (\ref{alteom4}) yields, due to the combination of symmetry (\ref{commutativity}) and the antisymmetry (\ref{altsym2}) which cancels the RHS,
\be 
\nabla_\mu\mathfrak{l}_\alpha{}^{\mu\nu} = \nabla_\mu\mathfrak{q}^{\mu\nu}{}_\alpha\,. 
\ee
This can be recognised as the ``remarkable relation'' Eq.(16) of Ref.\cite{Koivisto:2021jdy}.
We shall encounter this relation in a different guise in the next section \ref{potentials}, and we'll see that it allows us to identify the Lagrange multiplier field
\be \label{remarkablesolution}
\mathfrak{l}_\alpha{}^{\mu\nu} = \frac{1}{2}\mathfrak{h}^{\mu\nu}{}_\alpha\,,
\ee
as what is called the gravitational excitation $\mathfrak{h}^{\mu\nu}{}_\alpha$. From (\ref{alteom3}) we get that, using also (\ref{thevariation}),    
\be
2\nabla_\nu\mathfrak{l}_\alpha{}^{\mu\nu} =-\lp \mathfrak{T}^{\mu}{}_{\alpha}   + \delta^\mu_\alpha\mathfrak{L}_G+ 2\mathfrak{q}^\mu{}_{\nu\beta}L^{\nu\beta}{}_\alpha\rp\,.
\ee
Using this in (\ref{altcurrent2}) it assumes the form
\be
\mathfrak{J}^\mu  =  2\nabla_\nu\lp \mathfrak{l}_\alpha{}^{\beta\nu\mu}\nabla_\beta v^\alpha
- \mathfrak{l}_\alpha{}^{\mu\nu}v^\alpha \rp + \Delta T^\mu{}_\nu v^\nu\,. 
\ee
The Lagrange multiplier $\mathfrak{l}_\alpha{}^{\beta\nu\mu}$ cannot be fully determined by the field equations, since it possesses gauge symmetries, as detailed in Section IV.B.1 of Ref.\cite{BeltranJimenez:2018vdo}. However, 
the essentially unique solution has the same form (\ref{grconst}) as in the derivation of the Komar superpotential, as it will again be shown in detail in the next section 
\ref{potentials}. Putting all these results together, the final result for the on-shell current reads
\bs
\be
\mathfrak{J}^\mu  = \nabla_\nu\lp m_P^2\sqrt{-\mathfrak{g}}\nabla^{[\mu}v^{\nu]}  - \mathfrak{h}^{\mu\nu}{}_\alpha v^\alpha\rp  + \Delta T^\mu{}_\nu v^\nu\,. 
\ee
We may separate the 2 contributions to the charge tensor,
\be 
\mathfrak{J}^\mu  = \nabla_\nu\lp J^{\mu\nu}_{\text{can}} + J^{\mu\nu}_{\text{non}} \rp  + \Delta T^\mu{}_\nu v^\nu\,. 
\ee
\es
where
\bs
\ba
J^{\mu\nu}_{\text{can}} & = & -{h}^{\mu\nu}{}_\alpha v^\alpha\,, \\
J^{\mu\nu}_{\text{non}} & = & m_P^2\nabla^{[\mu}v^{\nu]} = 
J^{\mu\nu}_{K} - m_P^2 Q^{[\mu\nu]}{}_\alpha v^\alpha\,.
\ea
\es
The canonical piece $J^{\mu\nu}_{\text{can}}$ is given by the
excitation tensor. The \2 piece $J^{\mu\nu}_{\text{non}}$ since, like the standard Komar superpotential, it features a derivative of the parameter $v^\alpha$. Regarding the Komar superpotential, we shall write yet 1 more equivalent expression, 
\be \label{komar_improved}
J^{\mu\nu}  =  m_P^2\lp \mathring{\nabla}^{[\mu}v^{\nu]} - Q^{[\mu}v^{\nu]} + \tilde{Q}^{[\mu}v^{\nu]}\rp\,,
\ee
which was also used in Ref.\cite{Heisenberg:2022nvs}. Notably, this has the form (\ref{komar3}), where now $A^\mu = -Q^\mu + \tilde{Q}^\mu$.

The non-canonical piece $J^{\mu\nu}_{\text{non}}$ is absent in the Noether current obtained in the metric-affine gauge formulation of the theory \cite{Jimenez:2021nup}. Our current view is that this piece is an unphysical artefact of translating to theory into the tensor (Palatini) formalism. In all known fundamental theories, the charges are linear in the parameter and do not depend upon the derivatives of the parameter. In the context of the premetric programme \cite{Hehl:2016glb,Koivisto:2021jdy}, incorporating such a derivative-dependence would require some new axiom the physical motivation of which seems unclear. (Perhaps, an inhomogeneous response to the transformation parameter could be interpreted in terms of an excitation gauge potential instead of the usual excitation tensor.)

\subsubsection{Inertial connection formalism}
\label{icformalism}

It can be illustrative to demonstrate the formulation-dependent property of the Noether
currents with another example. In the derivation above (and in the derivation in Ref.\cite{Jimenez:2021nup}) the action principle was formulated in terms of Lagrange multipliers. The suggested interpretation is the origin of GR in its symmetric teleparallel guise in a symmetry-broken massive gauge theory. This effective theory is teleparallel i.e. the connection is flat at observable scales because the Planck mass is the mass of the connection. A less fundamental formulation is obtained by solving the constraint equations (\ref{alteom1},\ref{alteom2}) and plugging the solution back into the action. This way one obtains Golovnev's ``inertial connection'' formulation of a teleparallel theory \cite{Golovnev:2017dox,Golovnev:2021omn}. Even though the (classical) dynamics in the inertial connection formulation are completely equivalent \cite{Koivisto:2018aip} (for the detailed and comprehensive study, see \cite{Hohmann:2021fpr}), the Noether currents are sensitive to the action principle since they are affected by the boundary terms in the action. 

According to (\ref{alteom1},\ref{alteom2}), in symmetric teleparallelism both the curvature $R^\alpha{}_{\mu\beta\nu}=0$ and the torsion $T^\alpha{}_{\mu\nu}=0$ vanish.
Because of the former, the connection has to be given by the matrix $\Lambda^\alpha{}_\beta$ in (\ref{lambda}), and because of the latter, the 
matrix can always be given by the more special form \cite{Schouten1954,BeltranJimenez:2017tkd}
\be
\Lambda^\alpha{}_\beta = \partial_\beta\xi^\alpha\,. \label{xi}
\ee
The {functions} $\xi^\alpha$ may thus be regarded as the dynamical variables. As a consequence, not even $\hat{\nabla}_\mu E_\alpha{}^{\mu\nu}$ can be 
guaranteed to vanish on-shell, but instead the equation of motion for the connection assumes the form 
\be
\hat{\nabla}_{\nu}\hat{\nabla}_\mu E_\alpha{}^{\mu\nu} = -\hat{\nabla}_{\nu}\hat{\nabla}_\mu\lp 2q^{\mu\nu}{}_\alpha + \Hyp_\alpha{}^{\mu\nu} \rp = 0 \,.
\ee
It is often more convenient to deal with tensor densities than tensors in symmetric teleparallelism, since, as we see from (\ref{geometry}) the 
covariant derivative $\nabla_\mu$ is commutative, but the $\hat{\nabla}_\mu$ defined in (\ref{hatnabla}) is not,
\begin{subequations}
\ba
{}[\nabla_\mu,\nabla_\nu] & = & 0\,, \label{commutativity} \\
{}[\hat{\nabla}_\mu,\hat{\nabla}_\nu]  & = & 2Q_{[\mu}\nabla_{\nu]}\,.
\ea
\end{subequations}
To arrive at the 2$^{\text{nd}}$ identity we recalled (\ref{bianchi3b}). Taking into account the symmetrisation inherited from the commutativity (\ref{commutativity}), the current (\ref{gnoether}) is
\begin{subequations}
\ba
\mathfrak{J}^\mu & = & 2\mathfrak{q}^{(\mu\nu)}{}_\alpha{\nabla}_\nu v^\alpha - 2\mathfrak{q}^{\mu\nu}{}_\alpha\mathring{\nabla}_\nu v^\alpha - 2\nabla_\nu\mathfrak{q}^{(\mu\nu)}{}_\alpha v^\alpha \nn \\
 & + &   \mathfrak{L}_G v^\mu + 2\lp \mathfrak{E}^\mu{}_\nu  + \frac{1}{2}\mathfrak{T}^\mu{}_\nu\rp v^\nu + \mathfrak{J}_M^\mu\,,  \label{qnoether}
\ea
where the piece due to matter sources is
\be
 \mathfrak{J}_M^\mu  =  \nabla_\nu\mathfrak{\Hyp}_\alpha{}^{(\mu\nu)} v^\alpha - \mathfrak{\Hyp}_\alpha{}^{(\mu\nu)}\nabla_\nu v^\alpha + \Delta T^\mu{}_\nu v^\nu\,.  \label{mcurrent}
\ee
\end{subequations}
conserved. The off-shell current can be written as
\be
\mathfrak{J}^\mu = -2\mathfrak{q}^{[\mu\nu]}{}_\alpha{}\nabla_\nu v^\alpha + 2\mathfrak{\tilde{w}}^\mu{}_\nu v^\nu + \mathfrak{J}_M^\mu\,,
\ee
where we abbreviate $\mathfrak{\tilde{w}} = \mathfrak{w} - \nabla_\alpha\mathfrak{q}^{[\alpha\mu]}{}_\nu$,  
\ba
\mathfrak{\tilde{w}}^\mu{}_\nu & = &  \mathfrak{q}^{\mu\alpha}{}_\beta L^\beta{}_{\alpha\nu} - \nabla_\alpha\mathfrak{q}^{(\mu\alpha)}{}_\nu + \mathfrak{E}^\mu{}_\nu+ \frac{1}{2}\lp \mathfrak{T}^\mu{}_\nu + \mathfrak{L}_G\delta^\mu_\nu\rp \quad \nn \\
 & = & \mathfrak{q}^{\mu\alpha}{}_\beta L^\beta{}_{\alpha\nu} + \mathfrak{q}^\alpha{}_{\beta\nu} Q_\alpha{}^{\beta\mu}- \nabla_\alpha\mathfrak{q}^{[\mu\alpha]}{}_\nu + \mathfrak{g}^{\mu\lambda}\frac{\partial L_G}{\partial g^{\lambda\nu}}\,. \quad \nn
\ea
For the  2$^{\text{nd}}$ form we used the identity
\ba
\mathfrak{E}^\mu{}_\nu + \frac{1}{2}\mathfrak{T}^\mu{}_\nu & = &  \nabla_\alpha\mathfrak{q}^{\alpha\mu}{}_{\nu} + \mathfrak{q}^\alpha{}_{\beta\nu} Q_\alpha{}^{\beta\mu} 
+  g^{\mu\lambda}\frac{\partial \mathfrak{L}_G}{\partial g^{\lambda\nu}}\,. \quad
\ea
Finally, we obtain the off-shell conserved current in the desired form,
\be \label{desired}
\mathfrak{J}^\mu = -2\nabla_\nu\lp \mathfrak{q}^{[\mu\nu]}{}_\alpha{} v^\alpha \rp + 2\mathfrak{{w}}^\mu{}_\nu v^\nu +  \mathfrak{J}_M^\mu\,.
\ee
where we again abbreviate
\be \label{w}
\mathfrak{{w}}^\mu{}_\nu = \mathfrak{q}^{\mu\alpha}{}_\beta L^\beta{}_{\alpha\nu} + \mathfrak{q}^\alpha{}_{\beta\nu} Q_\alpha{}^{\beta\mu} + \mathfrak{g}^{\mu\lambda}\frac{\partial L_G}{\partial g^{\lambda\nu}}\,. 
\ee
From (\ref{antisymmetrictensor}), for any antisymmetric tensor density $\mathfrak{a}^{\mu\nu}=\mathfrak{a}^{[\mu\nu]}$ we have $\nabla_\mu \mathfrak{a}^{\mu\nu} = \mathring{\nabla}_\mu \mathfrak{a}^{\mu\nu} 
= \partial_\mu \mathfrak{a}^{\mu\nu}$. 
Let us consider a quadratic theory
\be
\mathfrak{L}_G = \frac{1}{2}\mathfrak{q}^\alpha{}_{\mu\nu}Q_\alpha{}^{\mu\nu}\,, \label{quadratic}
\ee
to verify explicitly that $\mathfrak{w}^\mu{}_\nu=0$. The field equations (\ref{geom2}) are
\ba
\mathfrak{E}^\mu{}_\nu = \nabla_\alpha\mathfrak{q}^{\alpha\mu}{}_{\nu} - \frac{1}{2}\delta^\mu_\nu\mathfrak{L}_G + \frac{1}{2}\mathfrak{q}^{\mu}{}_{\alpha\beta}Q_{\nu}{}^{\alpha\beta} - \frac{1}{2}\mathfrak{T}^\mu{}_\nu\,. \label{steom} 
\ea
We need not suppose the field equations, but only to calculate from (\ref{quadratic}) the variation 
\ba
\mathfrak{g}^{\mu\lambda}\frac{\partial L_G}{\partial{g}^{\lambda\nu}} & = & -\mathfrak{q}^{\alpha}{}_{\beta\nu}Q_\alpha{}^{\beta\mu} +\frac{1}{2}\mathfrak{q}^\mu{}_{\alpha\beta}Q_{\nu}{}^{\alpha\beta} \nn \\
& = &  -\mathfrak{q}^{\alpha}{}_{\beta\nu}Q_\alpha{}^{\beta\mu} - \mathfrak{q}^\mu{}_{\alpha\beta}L^{\alpha\beta}{}_\nu\,, \label{thevariation}
\ea
which plugged into (\ref{w}) gives
\be
\mathfrak{{w}}^\mu{}_\nu =  0\,,
\ee
simplifying the result (\ref{desired}) to
\be \label{theresult}
\mathfrak{J}^\mu = -2\nabla_\nu\lp \mathfrak{q}^{[\mu\nu]}{}_\alpha{} v^\alpha \rp +  \mathfrak{J}_M^\mu\,.
\ee
As expected, (\ref{theresult}) does not match with what was found earlier. In the inertial connection formulation, the action would have to be adjusted with boundary terms in order 
to obtain Noether currents representing the physical energy-momentum currents. 
In fact, the inertial connection implied by (\ref{xi}), when plugged into the action (\ref{quadratic}), makes it a \2 order action. We are thus faced with the same issue as with the Hilbert's formulation of GR, and would be forced to modify the formulation with boundary terms, in order to render the variational problem mathematically well defined\footnote{The singular property of the GR equivalent is that these higher derivatives only enter as total derivatives, so this technicality represents a less fatal obstruction. In the generic case, however, the presence of higher derivatives of $\xi^\alpha$ is undesirable as it is prone to introduce Ostrogradski instabilities. This is of course associated to the common problem of breaking diffeomorphisms for a massless spin-2 particle. In metric teleparallelism, there may not be explicit higher derivatives, but the analogous pathologies appear associated with the breaking of Lorentz invariance. For some discussions on these problems of modified teleparallel gravity, see e.g. \cite{Koivisto:2018loq,BeltranJimenez:2019nns,BeltranJimenez:2020fvy,BeltranJimenez:2021auj,Hu:2022anq}}. 

We summarise the formulation-dependence of the Noether currents in table \ref{formulationdependence}. For completeness, we have mentioned also the possibility of hybrid formulation, wherein one a priori restricts to symmetric connections and varies the action
\be
\mathfrak{L}_G = \frac{1}{2}\mathfrak{q}^\alpha{}_{\mu\nu}Q_\alpha{}^{\mu\nu} + \mathfrak{l}_\alpha{}^{\beta\mu\nu}R_\alpha{}^{\beta\mu\nu}\,. \label{quadraticH}
\ee
However, this does not lead to dynamical equivalence with GR. The field equations,
\bs
\ba
R^\alpha{}_{\beta\mu\nu} & = & 0\,, \label{Heom1} \\
\mathfrak{T}_{\mu\nu} &= & 2\nabla_\alpha\mathfrak{q}^\alpha{}_{\mu\nu} + 2\frac{\partial\mathfrak{L}_G}{\partial g^{\mu\nu}}\,, \label{Heom2} \\
\nabla_\beta\mathfrak{l}_\alpha{}^{\nu\mu\beta} & = & \mathfrak{q}^{\mu\nu}{}_\alpha\,, \label{Heom3}
\ea
\es
imply now that the sum of material and inertial currents must vanish since (\ref{Heom3})
forces $\nabla_\mu \mathfrak{q}^{\mu\nu}{}_\alpha = 0$.

\begin{center}
\begin{table}
\begin{tabular}{|c| c| c| }
\hline
 space of connections & Noether potential $J^{\mu\nu}$  \\ \hline
 all & $\nabla^{[\mu}v^{\nu]} - h^{\mu\nu}{}_\alpha v^\alpha$  \\   \hline
 flat & $-h^{\mu\nu}{}_\alpha v^\alpha$  \\   \hline
 symmetric & inequivalent to GR    \\   \hline
  flat \& symmetric & $-2q^{[\mu\nu]}{}_\alpha v^\alpha$  \\ \hline
\end{tabular}
\caption{The formulation-dependence of the Noether current in symmetric teleparallelism. The canonical form of the current is obtained only when varying a flat connection and imposing its symmetricity with a Lagrange multiplier. \label{formulationdependence}}
\end{table}
\end{center}

\section{Potentials}
\label{potentials}

In this section we shall discuss the interpretation of the results, and in particular we shall put the canonical Noether potential $\mathfrak{h}^{\mu\nu}{}_\alpha$ in the contexts of previous discussions in the literature. In \ref{gravi1} we will recall how the inertial energy-momentum current appears in the field equations. In \ref{gravi2} the explicit form of the tensor density $\mathfrak{h}^{\mu\nu}{}_\alpha$ is finally uncovered. In the terminology of the premetric programme 
\cite{hehl2012foundations}, it is called the (kinetic) excitation (tensor density). We naively anticipated this object already at (\ref{qlocalchargeg}), and it is also the bottom line of the premetric reasoning. In \ref{boundary} we review 
previous derivations of gravitational energy-momentum, and conclude that they are consistent with $\mathfrak{h}^{\mu\nu}{}_\alpha$, though fundamentally more ambiguous. In particular, we list a handful of most serious suggestions of superpotentials in the literature, and show that each of them generalises to precisely  $\mathfrak{h}^{\mu\nu}{}_\alpha$ (when promoted by a certain minimal coupling principle into a proper tensor). 

\subsection{Inertial current}
\label{gravi1}
 
Consider the special theory of relativity in a manifestly diffeomorphism invariant form by using the symmetric teleparallel covariant derivative $\nabla_\mu$. In particular, the energy-momentum conservation is given in terms of the  symmetric energy-momentum tensor $T^\mu{}_\nu$ as $\nabla_\mu T^\mu{}_\nu =0$, which reduces in the concident gauge \cite{BeltranJimenez:2022azb} to $\partial_\mu T^\mu{}_\nu =0$.
However, in GR (and its modifications \cite{Koivisto:2005yk}) the conservation equation is $\mathring{\nabla}_\mu  T^\mu{}_\nu =0$, which can be written as
\be
\nabla_\mu \mathfrak{T}^\mu{}_\nu = \frac{1}{2}Q_{\nu\alpha\beta}\mathfrak{T}^{\alpha\beta}\,. \label{covcons}
\ee
Einstein formulated the conservation law as a divergence that includes the energy-momentum $G^\mu{}_\nu$ of the gravitational field \cite{1916SPAW......1111E}. In the symmetric teleparallel formulation, $G^\mu{}_\nu$ can
be written as a tensor, and Einstein's conservation law can be made manifestly covariant,
\be
\nabla_\mu\lp \mathfrak{T}^\mu{}_\nu + \mathfrak{G}^\mu{}_\nu\rp = 0\,. \label{einstein}
\ee
(Note that $G^\mu{}_\nu$ does not denote the Einstein tensor $\mathring{R}^\mu{}_\nu -\frac{1}{2}\delta^\mu_\nu \mathring{R}$, but so to speak, the covariantised form of the pseudotensor that depends only upon the 1$^{\text{st}}$ derivatives of the metric.) Now, let us use the field equations (\ref{geom1}) in (\ref{covcons}) to obtain
\ba
\nabla_\mu \mathfrak{T}^\mu{}_\nu &  = & Q_{\nu}{}^{\alpha\beta}\lp \nabla_\mu \mathfrak{q}^\mu{}_{\alpha\beta} + \frac{\partial\mathfrak{L}_G}{\partial g^{\alpha\beta}}\rp \nn \\
& = & \nabla_\mu\lp Q_{\nu}{}^{\alpha\beta}\mathfrak{q}^\mu{}_{\alpha\beta} - \delta^\mu_\nu \mathfrak{L}_G + 2\nabla_\alpha\mathfrak{q}^{\mu\alpha}{}_\nu\rp\,. \ea
where in the \2 equality we have used the identity (\ref{bianchi3}). We can now identify the gravitational energy momentum tensor in (\ref{einstein}) as
\be
\mathfrak{G}^\mu{}_\nu = \delta^\mu_\nu \mathfrak{L}_G-Q_{\nu}{}^{\alpha\beta}\mathfrak{q}^\mu{}_{\alpha\beta}\,, \label{pseudo}
\ee
since
\be
\nabla_\mu\lp \mathfrak{T}^\mu{}_\nu + \mathfrak{G}^\mu{}_\nu\rp = 2\nabla_\mu\nabla_\alpha \mathfrak{q}^{\mu\alpha}{}_\nu\,. \label{einstein2}
\ee
The expression (\ref{pseudo}) reduces to the canonical Noether current of the metric field with respect to the symmetry under global translations. In other words, it reduces to the pseudocanonical 
energy-momentum pseudotensor in the coincident gauge. The RHS of (\ref{einstein2}) represents the energy-momentum conservation equation of the connection. 

In the theory of coincident GR, the remarkable fact is that the RHS of (\ref{einstein2}) vanishes identically. The field equations (\ref{steom}) state that we always have on-shell
\be
2\nabla_\alpha{}\mathfrak{q}^{\alpha\mu}{}_\nu =  \mathfrak{T}^\mu{}_\nu + \mathfrak{G}^\mu{}_\nu\,. \label{steom2}
\ee
Nevertheless, the divergence of the LHS and RHS vanish identically only in the coincident GR. In that case $2\mathfrak{q}^{\alpha\mu}{}_\nu$ reduces to the Tolman's energy-momentum
complex \cite{Tolman1934-TOLRTA} in the coincident gauge. Our version of the theory is covariant and the equation (\ref{steom2}) holds for arbitrary $L_G$. The tensor $2\mathfrak{q}^{\alpha\mu}{}_\nu$, the non-metricity conjugate, could thus also be called the generalised Tolman tensor. We will see later that neither the generalised Tolman tensor, nor its antisymmetrisation in (\ref{theresult}) that we obtained as the Noether potential in the inertial connection formulation, provides a viable superpotential to describe the quasilocal energy charges.

At this point, we may state the definition of an inertial frame as that for which $G^\mu{}_\nu=0$. The local metric current induced by the diffeomorphism $\delta_v$ is $G^\mu{}_\nu v^\nu$, and the claim is quite simply that any local energy or momentum associated with the metric field (equivalently in the tetrad formulation, associated with the tetrad field) is not physical in the sense of being detectable by an inertial observer. It is often quoted from page 422 of Ref.\cite{Misner:1973prb} that ``Anybody who looks for a magic formula for 'local gravitational energy-momentum' is looking for the right answer to the wrong question''. Our approach is in agreement, as it excludes local gravitational energy-momentum, and we do not have to regard the coordinate-invariant expression of this fact, $G^\mu{}_\nu=0$, as a ``magic formula''. However, we maintain that energy can be and should be defined precisely, and that this is not prevented by gravitation being equivalent to inertia - 
contrary to what is perhaps the most common view on energy (for philosophical discussions see e.g. \cite{Pitts:2009km,Duerr:2019mqb}). 

\subsection{Kinetic excitation}
\label{gravi2}

The premetric formalism is a framework wherein the form of the theory is derived from conserved charges  \cite{hehl2012foundations}. One does not assume a priori a metric, nor a lagrangian for the theory, but begins with more primitive concepts whose generic relations are deduced from elementary geometric identities  \cite{Hehl:2016glb}. This way, one ends up with 2 fundamental equations, 1 of them
corresponding to the dynamical equation of motion (the inhomogeneous field equation), the other to a Bianchi identity (the homogeneous field equation). It should be understood that the premetric formalism is not any particular theory, but it provides a language to discuss, construct and analyse generic theories of physics with conserved charges as the fundamental starting point. 

In the case that one begins with the conservation of energy-momentum charges, the \1 fundamental equation takes the form \cite{Koivisto:2021jdy}
\be
\nabla_\alpha{}\tilde{\mathfrak{h}}^{\alpha\mu}{}_\nu =  \tilde{\mathfrak{T}}^\mu{}_\nu + \tilde{\mathfrak{G}}^\mu{}_\nu\,,\label{steom3}
\ee
where $\tilde{\mathfrak{h}}^{\alpha\mu}{}_\nu=\tilde{\mathfrak{h}}^{[\alpha\mu]}{}_\nu$ is the kinetic excitation tensor density, $\tilde{\mathfrak{T}}^\mu{}_\nu$ is the source current, and $ \tilde{\mathfrak{G}}^\mu{}_\nu$ is determined by the mass excitation tensor density that we denote $\tilde{\mathfrak{q}}^\mu{}_{\alpha\beta}$. The \2 fundamental equation is now understood as the expression of the equivalence principle: there is no force of gravity, therefore the affine geometry is symmetric and teleparallel. 

The dynamics of the theory remain undetermined before one specifies the form of the 2 tensor densities, the kinetic and the mass excitation tensors densities, by stating what is called {\it the constitutive law} of the premetric theory. If a local and linear constitutive law is postulated, it has 5632 independent components. If it is assumed that the constitutive law is reversible (which is a necessary condition for the existence of an action principle for the theory), there are 3712 independent components. If, on the other hand, the constitutive law is assumed to be metric, but not necessarily have a lagrangian formulation, there are only 14 distinct possibilities.  However, requiring the metrical, the lagrangian, and the parity-invariant properties of the constitutive law fixes it uniquely (up to an overall normalisation) \cite{Koivisto:2021jdy}. 

Thus, a result of the general premetric analysis is that there exists the unique pair of tensor densities 
$\tilde{\mathfrak{q}}^\mu{}_{\alpha\beta}=\tilde{\mathfrak{q}}^\mu{}_{(\alpha\beta)}$ and $\tilde{\mathfrak{h}}^{\mu\nu}{}_\alpha=\tilde{\mathfrak{h}}^{[\mu\nu]}{}_\alpha$ such that
\be
\nabla_\alpha \mathfrak{h}^{\alpha\mu}{}_{\nu} = 2\nabla_\alpha \mathfrak{q}^{\alpha\mu}{}_{\nu}\,. \label{complex}
\ee 
(The factor of $2$ is just a convention.) The unique theory has the mass excitation $\tilde{\mathfrak{q}}^\mu{}_{\alpha\beta} = \mathfrak{q}^\mu{}_{\alpha\beta}$ given by the so-called non-metricity conjugate (i.e. the generalised Tolman tensor) of the Coincident GR  \cite{BeltranJimenez:2017tkd}, 
\begin{subequations}
\label{mainpotentials}
\ba
\frac{4}{m_{P}^2}q^{\mu\nu}{}_\alpha & = & -Q^{\mu\nu}{}_\alpha + Q^{\nu\mu}{}_\alpha + Q_\alpha{}^{\mu\nu}  - \frac{1}{2}g^{\mu\nu}{Q}_\alpha  \nn \\
&  - & \frac{1}{2}\delta^\mu_\alpha Q^\nu +  \lp Q^\mu-\tilde{Q}^\mu\rp\delta^\nu_\alpha\,, \label{ptensor}
\ea
and the kinetic excitation given by $\tilde{\mathfrak{h}}^{\mu\nu}{}_\alpha={\mathfrak{h}}^{[\mu\nu]}{}_\alpha$ where
\be \label{complex2}
\frac{1}{m_P^2}h^{\mu\nu}{}_\alpha =  -Q^{[\mu\nu]}{}_\alpha - \delta^{[\mu}_\alpha Q^{\nu]} + \delta^{[\mu}_\alpha \tilde{Q}^{\nu]}\,.
\ee
The kinetic excitation tensor (\ref{complex2}) turns out to be the covariant version of the ``Einstein energy-momentum complex''. In the following subsection \ref{boundary} we will explain its precise relation to various pseudotensor superpotentials that have been considered in the literature. 
Then, it will be useful to note that we can write (\ref{complex2}) also as 
\be \label{complex0}
\mathfrak{h}^{\mu\nu}{}_\alpha = m_P^{2}\frac{g_{\alpha\beta}}{\sqrt{-\mathfrak{g}}}\nabla_\lambda\lp\mathfrak{g}^{\lambda[\mu}\mathfrak{g}^{\nu]\beta}\rp\,. \ee
\end{subequations}
Since the relations (\ref{complex}) and (\ref{mainpotentials}) are of paramount importance to our discussions, let us show explicitly that:
\ba
\frac{1}{m_P^2}({T}^\mu{}_\nu &+& {G}^\mu{}_\nu)  =  \frac{2}{m_P^2\sqrt{-\mathfrak{g}}}\nabla_\alpha\left(\mathfrak{q}^{\alpha\mu}{}_{\nu}\right)  \nn \\ &=&  \frac{2}{m_P^2}\hat{\nabla}_\alpha q^{\alpha\mu}{}_\nu \nonumber \\
 & = & -\hat{\nabla}_\alpha L^{\alpha\mu}{}_\nu + \frac{1}{2}\hat{\nabla}_\alpha\lp Q^\alpha-\tilde{Q}^\alpha\rp\delta^\mu_\nu
 \nn \\ & - & \frac{1}{4}\lb\hat{\nabla}_\nu Q^\mu + \hat{\nabla}_\alpha \lp g^{\alpha\mu}Q_\nu\rp\rb \nonumber \\
 & = & \hat{\nabla}_\alpha Q^{[\mu\alpha]}{}_\nu + \frac{1}{2}\hat{\nabla}_\alpha\lp Q^\alpha - \tilde{Q}^\alpha\rp\delta^\mu_\nu  
 \nn \\ & + & 
 \frac{1}{2}\hat{\nabla}_\nu\lp Q^\mu - \tilde{Q}^\mu\rp \nonumber \\
 & = & \hat{\nabla}_\alpha\lb Q^{[\mu\alpha]}{}_\nu + \lp Q^{[\alpha} - \tilde{Q}^{[\alpha}\rp\delta^{\mu]}_\nu\rb \nn \\
 & = & \frac{1}{\sqrt{-\mathfrak{g}}}\nabla_\alpha\Big[ \sqrt{-\mathfrak{g}}g_{\nu\beta}\Big( Q_\lambda{}^{\beta[\alpha}g^{\mu]\lambda} \nn \\ & + &  Q_\lambda{}^{\lambda[\mu}g^{\alpha]\beta} - Q_\lambda g^{\beta[\alpha}g^{\mu]\lambda}\Big)\Big]  \nn \\
 & = & \frac{1}{\sqrt{-\mathfrak{g}}}\nabla_\alpha\lb \frac{g_{\nu\beta}}{\sqrt{-\mathfrak{g}}}\nabla_\lambda\lp -g g^{\beta[\alpha}g^{\mu]\lambda}\rp\rb \nn \\
 & = &  \frac{1}{m_P^2\sqrt{-\mathfrak{g}}}\nabla_\alpha \mathfrak{h}^{\alpha\mu}{}_{\nu}\,. \label{complex22}
\ea
The curvature Lagrange multiplier is now easily solved from (\ref{alteom4}) which now reads, using (\ref{remarkablesolution}) and then (\ref{mainpotentials}) 
\ba
\frac{1}{\sqrt{-\mathfrak{g}}}\nabla_\beta\mathfrak{l}_\alpha{}^{\nu\mu\beta} & = & q^{\mu\nu}{}_\alpha - l_\alpha{}^{\mu\nu} 
= q^{\mu\nu}{}_\alpha - \frac{1}{2}h^{\mu\nu}{}_\alpha
\nn \\ & = &
\frac{m_P^2}{4}\Big(Q_\alpha{}^{\mu\nu} - \frac{1}{2}g^{\mu\nu}Q_\alpha \nn \\
  & + &   \frac{1}{2}\delta^\mu_\alpha Q^\nu - \delta^\mu_\alpha\tilde{Q}^\nu\Big)\,.
\ea
We find the solution 
\be \label{rsolution}
l_\alpha{}^{\nu\mu\beta} = \frac{m_P^2}{2}g^{\nu[\beta}\delta^{\mu]}_\alpha\,,
\ee
which could have been guessed since it is (up to a constant prefactor) the only available non-derivative tensor compatible with the symmetry of $l_\alpha{}^{\nu\mu\beta}$. 
Note, in particular, that since $2\nabla_\alpha \mathfrak{q}^{[\alpha\beta]}{}_\alpha \neq \nabla_\alpha \mathfrak{h}^{\alpha\beta}{}_\alpha$, the kinetic excitation cannot be 
given by the Noether potential in (\ref{theresult}). For the antisymmetrised tensor in the Noether potential we obtain from (\ref{ptensor})
\be 
\frac{4}{m_{P}^2}q^{[\mu\nu]}{}_\alpha  =  -2Q^{[\mu\nu]}{}_\alpha - \frac{3}{2}\delta^{[\mu}_\alpha Q^{\nu]} + \delta^{[\mu}_\alpha \tilde{Q}^{\nu]}\,. \label{consti2}
\ee
Now, we can put the field equation (\ref{steom}) into the fundamental premetric form (\ref{steom3}) as
\be
\nabla_\alpha{}\mathfrak{h}^{\alpha\mu}{}_\nu =  {\mathfrak{T}}^\mu{}_\nu + {\mathfrak{G}}^\mu{}_\nu\,,\label{steom4}
\ee
where ${\mathfrak{G}}^\mu{}_\nu$ is given by (\ref{pseudo}) with the specific form (\ref{ptensor}). 
We may now reconsider the current (\ref{Ccurrent}). Neglecting the hypothetical hypermomentum contribution and using the premetric form of the field equation (\ref{steom4}), we have 
\be
J^\mu = \lp T^\mu{}_\nu + G^\mu{}_\nu\rp v^\nu + h^{\alpha\mu}{}_\nu\nabla_\alpha v^\nu\,.
\ee
Thus, we see that the Noether current indeed reduces, in its local form, to the expected energy-momentum current, plus possible effects due to a covariantly non-constant transformation parameter $v^\mu$. 

Alternative superpotentials have been introduced also. For example, M{\o}ller proposed the energy-momentum complex
\be \label{moeller}
\mathfrak{h}_M^{\mu\nu}{}_\alpha = 2 m_P^2 Q^{[\mu\nu]}{}_\alpha\,,  
\ee
due to its improved covariance properties \cite{MOLLER1958347}. However, as we have emphasised, the premetric axioms are only compatible with (\ref{complex2}) together
with (\ref{ptensor}) \cite{Koivisto:2021jdy}. In section \ref{energy-momentum} we will show that the correct form of the Noether potential can be decided also ``empirically'', since the different prescriptions yield (contrary to some previous wisdom in the literature) inequivalent results, already in the case of a black hole. Before that, we will review various alternative superpotentials and clarify their relation to the canonical excitation tensor $h^{\mu\nu}{}_\alpha$.

\subsection{Superpotentials}
\label{boundary}

The inertial energy-momentum tensor (\ref{pseudo}) is the covariant version of the pseudocanonical Noether current of the metric field. 
We will now show that the excitation tensor (\ref{complex2}) is the covariant version of many superpotential pseudotensors that have been previously introduced in the literature. The uniqueness of these results should not be taken to be compromised by the formulation-dependence of the Noether currents, as summarised in table \ref{formulationdependence}.


Rather, it reflects the well-known ambiguity in the choice of superpotentials. 1) One can add an arbitrary $\mathfrak{h}^{\alpha\beta\gamma}{}_\mu = \mathfrak{h}^{[\alpha\beta\gamma]}{}_\mu$ antisymmetric in the \1 3 indices to $\mathfrak{h}^{\alpha\mu}{}_\nu$ such that $\mathfrak{h}^{\alpha\mu}{}_\nu + \nabla_\beta\mathfrak{h}^{\beta\alpha\mu}{}_\nu$ still satisfies (\ref{complex}). Such a term exists,
but this would not be consistent with the parity-invariance of the constitutive law. 2) One may add a boundary term determined by some $B^\mu$ to the lagrangian $L$ such that one considers instead $L + \mathring{\nabla}_\alpha B^\alpha$. This would change the current $J^\mu$ and consequently the superpotential $J^{\mu\nu}$. However, in the present case the only available vectors $B^\mu$ would be (at least) of the \1 derivative order, and therefore the boundary term would make the $L$ a higher derivative theory. At a more fundamental level, we consider the action principle as
(a part of) the definition of the theory. Adding a boundary term to the action would change the theory {according to our criterion that the action is fundamental on its own and not just through its capability of producing field equations.} 3) The so called $Y$-ambiguity enters when performing partial integrations in the derivation of the symplectic potential. Sort of ``integration constants'' may in principle be added to the total derivative terms. Though we shall not attempt to make this argument robustly in the present study, in practice it is rather obvious that there is, at least in the cases studied in this article, a canonical definition of the ``definite'' partial integration. 

In conclusion, the formulation-dependence of the currents provides a criterion to distinguish the more fundamental formulation of the action principle. In terms of the metric-affine gauge fields (frame, connection and metric), we should vary the generic connection (and set it symmetric teleparallel with Lagrange multipliers). In terms of the Palatini fields (spacetime metric and affine connection), we should vary the teleparallel connection (and set it symmetric with a Lagrange multiplier).

As Wald had explained, an integral (\ref{integral}) of the potential $J^{\mu\nu}$ can be used for the computation of the entropy associated with a horizon as a charge \cite{Wald:1993nt,Iyer:1994ys}. Crucial to this derivation is to take into account the possible difference between the canonical energy and  the integral of the Noether potential $J^{\mu\nu}$ at infinity. In general, a Hamiltonian for the dynamics generated by $\delta_v$ does exist iff one can find a suitable (not necessarily diffeomorphism covariant) boundary term, which can be computed from the symplectic potential. An algorithm, involving many subtleties, exists for this procedure. The Hamiltonian corresponding to asymptotic time translations may be generated by the current $J^\mu$ only if its integral yields the correct ADM energy. A purpose of the discussions of this section was to clarify that, in our formalism, the potential $\tilde{h}^{\alpha\mu}{}_\nu$ can correspond to the canonical energy-momentum current only if the equation of motion relates the divergence of the potential to the canonical gravitational energy momentum according to (\ref{steom3}).  For example, neither (\ref{consti2}) nor (\ref{moeller}) has this property, which is unique to the canonical excitation (\ref{complex0}). On the other hand, as seen from (\ref{steom2}), (\ref{ptensor}) has the property but it does not have the index symmetry deduced from the premetric axioms, or equivalently, in Wald's formalism, $\mathfrak{q}^{\alpha\mu}{}_\nu v^\nu$ would not be the components of a charge 2-form.  
Thus, the conjecture is that in an inertial frame the  $\mathfrak{h}^{\alpha\mu}{}_\nu v^\nu$ is both the unique Hamiltonian that plays the role of the generator of time translations and the unique Noether charge that plays the role of the conserved energy. Therefore, it should directly yield the physical energy and the entropy according to Wald's derivation. 


More generally, a given superpotential $\tilde{h}^{\alpha\mu}{}_\nu$ corresponds to a given inertial (metric) energy-momentum tensor. One can check that each divergence is non-vanishing,
\begin{subequations}
\label{pderivatives}
\ba
\nabla_\mu \mathfrak{Q}^{[\mu\nu]}{}_\alpha & = & \sqrt{-\mathfrak{g}}\lp \nabla_\mu {Q}^{[\mu\nu]}{}_\alpha -\frac{1}{2}Q_\mu {Q}^{[\mu\nu]}{}_\alpha\rp\,, \\
\nabla_\mu \delta^{[\mu}_\alpha \mathfrak{Q}^{\nu]} & = & \frac{1}{4}\lp \delta^\nu_\alpha \mathfrak{Q}_\mu Q^\mu + \mathfrak{Q}_\alpha Q^\nu\rp \nn \\ &-& \frac{1}{2}\sqrt{-\mathfrak{g}}\delta^\nu_\alpha\nabla_\mu Q^\mu \quad\,, \\
\nabla_\mu \delta^{[\mu}_\alpha \tilde{\mathfrak{Q}}^{\nu]} & = & \frac{1}{4}\lp \delta^\nu_\alpha \mathfrak{Q}_\mu \tilde{Q}^\mu + \mathfrak{Q}_\alpha \tilde{Q}^\nu\rp \nn \\ &-& \frac{1}{2}\sqrt{-\mathfrak{g}}\delta^\nu_\alpha\nabla_\mu \tilde{Q}^\mu \quad\,.
\ea
\end{subequations}
and therefore the correspondence is unique. 
Explicitly, we see that for a chosen superpotential $\tilde{\mathfrak{h}}^{\alpha\mu}{}_\nu$, the  corresponding effective inertial energy-momentum current $\tilde{\mathfrak{G}}^\mu{}_\nu$ is given as
\be \label{tildeg}
\tilde{\mathfrak{G}}^\mu{}_\nu = {\mathfrak{G}}^\mu{}_\nu + \nabla_\alpha\lp 
\tilde{\mathfrak{h}}^{\alpha\mu}{}_\nu - \mathfrak{h}^{\alpha\mu}{}_\nu\rp\,.
\ee
One reason justifying calling the ${h}^{\alpha\mu}{}_\nu$ in (\ref{complex2}) the canonical excitation tensor for gravity is that it corresponds to the canonical energy-momentum tensor  $G^\mu{}_\nu$ which is the unique Noether current for the metric tensor. For example, the M{\o}ller superpotential (\ref{moeller}) does not share this property.

We shall now review a variety of other alternative superpotentials proposed in the course of the previous century, and see that they can all be understood as particular limits of the canonical excitation tensor (\ref{complex2}). All these superpotentials are pseudotensors, and can be understood to be defined wrt a reference (Minkowski or otherwise) metric. The von Freud superpotential was probably the \1 proposal, and it was constructed in order to replace the Tolman pseudotensor with an object that has the antisymmetry property of a proper excitation (pseudo)tensor \cite{10.2307/1968929}. The von Freud superpotential can be written as
\bs
\ba
\underline{\mathfrak{h}}_{vF}^{\mu\nu}{}_\alpha & = & -\frac{1}{2}\sqrt{-\mathfrak{g}}\delta^{\mu\nu\gamma}_{\lambda\sigma\alpha} g^{\beta\lambda}
\lp\mathring{\Gamma}^\lambda{}_{\beta\gamma} - \underline{\mathring{\Gamma}}{}^\lambda{}_{\beta\gamma}\rp  \nn \\
& = & -\frac{1}{2}\sqrt{-\mathfrak{g}}\delta^{\mu\nu\gamma}_{\lambda\sigma\alpha}g^{\beta\sigma}g^{\lambda\delta}\mathring{\underline{\nabla}}_\beta g_{\delta\gamma}\,. 
\ea
The underlined quantities are evaluated wrt the reference metric. In the original form,
$\underline{\nabla}_\beta = \partial_\beta$, corresponding to some flat reference metric. 

An influential treatment of gravitational energy-momentum was due to Landau and Lifshitz \cite{Landau:1975pou}, who constructed a symmetric $G_{LL}^{\mu\nu}=G_{LL}^{\nu\mu}$, which corresponds to the superpotential 
\ba
\underline{\mathfrak{h}}_{LL}^{\mu\nu\alpha}
& = & \frac{1}{2}(-\underline{\mathfrak{g}})^{-\frac{1}{2}}\delta^{\mu\nu}_{\gamma\rho}\mathring{\underline{\nabla}}_\beta\lp -\mathfrak{g}g^{\rho\alpha}g^{\gamma^\beta}\rp \nn \\
& = & \frac{1}{2}\sqrt{\mathfrak{g}/\underline{\mathfrak{g}}}\delta^{\mu\nu}_{\gamma\rho}\delta^{\alpha\beta}_{\sigma\delta}g^{\rho\sigma}\mathring{\underline{\nabla}}_\beta\lp \sqrt{-\mathfrak{g}}g^{\gamma\delta}\rp\,. 
\ea
Considering the reference metric in the $G_{LL}^{\mu\nu}$ as an asymptotically Minkowski spacetime leads almost directly to the ADM expression, and this has been further generalised by Abbot and Deser \cite{Abbott:1981ff} to apply in spacetimes with more arbitrary asymptotic behavior, such that the conserved charges can be associated with isometries of the asymptotic geometry (see Ref.\cite{Ortin:2015hya}, sections 6.1.1 and 6.1.2 respectively).  
In their study of spin and angular momentum in GR, for whose conservation the property $G_{LL}^{\mu\nu}=G_{LL}^{\nu\mu}$ is crucial, Bergmann and Thomson \cite{Bergmann:1953jz} employed the Landau-Lifshitz prescription up a normalisation of the reference volume element,
\be
\underline{\mathfrak{h}}_{BT}^{\mu\nu\alpha} = \sqrt{\underline{g}/g}\,\,\underline{\mathfrak{h}}_{LL}^{\mu\nu\alpha}\,.
\ee
Another well-known superpotential was introduced by Papapetrou \cite{Papapetrou:1948jw}
\ba
\underline{\mathfrak{h}}_{P}^{\mu\nu\alpha} 
& = & \delta^{\mu\nu}_{\gamma\lambda}\delta^{\alpha\rho}_{\beta\sigma}\underline{g}^{\lambda\beta}\sqrt{-\mathfrak{g}}\lp\frac{1}{4}g^{\gamma\sigma}g^{\tau\delta}-\frac{1}{2}g^{\gamma\tau}g^{\sigma\delta}\rp\mathring{\underline{\nabla}}_\rho g_{\tau\sigma}  \nn \\
& = &
\frac{1}{2}\delta^{\mu\nu}_{\gamma\lambda}\delta^{\alpha\rho}_{\beta\sigma} g^{\gamma\beta}\mathring{\underline{\nabla}}_\rho \lp \sqrt{-\mathfrak{g}}g^{\gamma\sigma}\rp\,.  
\ea
Finally, we quote the form of the Weinberg superpotential \cite{Weinberg:1972kfs}
\ba
\underline{\mathfrak{h}}_{W}^{\mu\nu\alpha} =  
\delta^{\mu\nu}_{\gamma\lambda}\delta^{\alpha\rho}_{\beta\sigma}\underline{g}^{\lambda\beta}\sqrt{-\underline{\mathfrak{g}}}\lp\frac{1}{4}\underline{g}^{\gamma\sigma}\underline{g}^{\tau\delta}-\frac{1}{2}\underline{g}^{\gamma\tau}\underline{g}^{\sigma\delta}\rp\mathring{\underline{\nabla}}_\rho g_{\tau\sigma}\,.\quad
\ea
\es
We have written these superpotentials in alternative forms from which it is transparent that they are quite closely related, differing only in the way they incorporate the dependence upon the reference metric. It has been shown that the quasilocal energy expressions obtained from these five in principle distinct expressions, all yield the equivalent values of energy when a certain matching algorithm is exploited to fix the reference metric \cite{Chen:2018kcs} (using a so called 4-dimensional isometric matching or an energy optimisation as the criteria for the best match of the reference on the region’s boundary \cite{Nester:2012zi,Sun:2013ofz}). 

Our approach does not require a reference metric or other background structures. We can generalise the above five pseudotensors which rely upon a reference metric for their definition,
into proper covariant tensors, by 1) replacing the reference metric in the expressions by the dynamical metric $\underline{g}{}_{\mu\nu} \rightarrow g_{\mu\nu}$ and 2) replacing the reference metric-covariant derivative with the dynamical, independent covariant derivative 
$\mathring{\underline{\nabla}}{}_\alpha \rightarrow \nabla_\mu$. {This procedure closely resembles a} minimal coupling of matter to gravity, and now we are indeed minimally coupling various alternatives of energy-momentum as if the metric was as any other field. We obtain the expressions
\bs
\label{minimalcoupling}
\ba
{h}_{vF}^{\mu\nu}{}_\alpha & = & -\frac{1}{2}\delta^{\mu\nu\gamma}_{\lambda\sigma\alpha}g^{\beta\sigma}g^{\lambda\delta}Q_{\beta\delta\gamma}\,, \\
{h}_{LL}^{\mu\nu\alpha}
& = & \frac{1}{2}{\mathfrak{g}^{-1}}\delta^{\mu\nu}_{\gamma\rho}\nabla_\beta\lp \mathfrak{g}g^{\rho\alpha}g^{\gamma^\beta}\rp\,, \\ 
{h}_{BT}^{\mu\nu\alpha} & = & {h}_{LL}^{\mu\nu\alpha}\,, \\
{h}_{P}^{\mu\nu\alpha} & = &
\frac{1}{2}(-\mathfrak{g})^{-\frac{1}{2}}\delta^{\mu\nu}_{\gamma\lambda}\delta^{\alpha\rho}_{\beta\sigma} g^{\gamma\beta}\nabla_\rho \lp \sqrt{-\mathfrak{g}}g^{\gamma\sigma}\rp\,, \\
{h}_{W}^{\mu\nu\alpha} & = &  
\delta^{\mu\nu}_{\gamma\lambda}\delta^{\alpha\rho}_{\beta\sigma}{g}^{\lambda\beta}\lp\frac{1}{4}{g}^{\gamma\sigma}{g}^{\tau\delta}-\frac{1}{2}{g}^{\gamma\tau}{g}^{\sigma\delta}\rp Q_{\rho\tau\sigma}\,.\quad 
\ea
\es
With some algebra, we can confirm that all these expressions are equivalent, and identical with the canonical excitation tensor! {More precisely, we find the satisfactory coincidences}
\be \label{unification}
{h}_{vF}^{\mu\nu}{}_\alpha = {h}_{LL}^{\mu\nu}{}_\alpha
= {h}_{BT}^{\mu\nu}{}_\alpha = {h}_P^{\mu\nu}{}_\alpha = {h}_W^{\mu\nu}{}_\alpha = {h}^{\mu\nu}{}_\alpha\,.
\ee
The canonical resolution of the energy-momentum problem in GR is thus the background-independent unification of many previous prescriptions. 

On the other hand, not all previous proposals can be straightforwardly fitted into this unification. We will give 3 examples. The (generalised) Tolman superpotential (\ref{steom2}) does not have the right symmetry. The Komar superpotential (\ref{komar}), as discussed earlier, is non-canonical due to its dependence upon the derivative of the transformation parameter. The M{\o}ller prescription does not share this non-canonical property, but the superpotential (\ref{moeller}) does not coincide with (\ref{minimalcoupling}). Though obtained in a direct fashion in the tetrad formulation of GR, the M{\o}ller superpotential corresponds to the Neumann and not the Dirichlet boundary conditions for the metric \cite{Chang:1998wj}. Also, according to (\ref{tildeg}), the M{\o}ller superpotential corresponds to a non-canonical inertial energy-momentum current. 
 
\section{Energy}
\label{energy-momentum}

In this section, we will illustrate the workings of the canonical resolution by computing the 
energy of a black hole. We will consider the Schwarzschild-Reissner-Nordstr\"om-de Sitter case which is general enough to capture various different features, including those of non-vacuum spacetime.

Let us remark in passing that the canonical resolution of energy clarifies the properties of black holes beginning from the definition of what a black hole is. A physical definition (in contrast to alternative mathematical definitions that refer to global properties of spacetime that cannot be detected by any observer) \cite{Curiel:2018cbt} is based on the existence of a future trapped horizon. This property is determined from the expansion scalars, which however are foliation-dependent \cite{Faraoni:2016xgy}. The ambiguity can now be eliminated since the preferred foliation is uniquely given in an inertial frame. This is, in fact, highly relevant to currently ongoing discussions in the literature,  concerning e.g. the possible physical application of the Thakurta solution \cite{1981InJPh..55..304G} (in the context of primordial holes and their possible contribution to the dark matter density \cite{Boehm:2020jwd,Hutsi:2021nvs,Boehm:2021kzq,Hutsi:2021vha}). 
A formal definition seems to show that the Thakurta solution does not describe a black hole at all, since the solution does not possess a future trapped horizon \cite{Harada:2021xze}. However, this argument is devoid of physical content, since if one uses the Kodama time \cite{Kodama:1979vn} instead of the conformally-Schwarzschild time (as in Ref. \cite{Harada:2021xze}) to foliate the spacetime, the conclusion is completely different \cite{Kobakhidze:2021rsh}. The resolution is that in an inertial frame, the Misner-Sharp mass \cite{Misner:1964je} turns out to be the 
physical energy, and it is {\it not} the Noether charge of time translations wrt the conformally-Schwarzschild time, but rather wrt the Kodama time\footnote{The explicit and general proof of this is outside the scope of the present article, since it requires the generic Schwarzchild-Friedmann-Lema\^{i}tre(-Robertson-Walker) solution in an inertial frame. However, the following computations already confirm the argument in the limit of Schwarzchild-de Sitter solution.}. Thus, despite the claims of Ref.\cite{Harada:2021xze}, and irregardless of their possible relevance to the dark matter problem which is a separate issue \cite{Boehm:2020jwd,Hutsi:2021nvs,Boehm:2021kzq,Hutsi:2021vha}, Thakurta black holes {\it are} black holes. 

\subsection{Coordinates}
\label{Coordinates}

The relevant class of Kerr-Schild metrics are given by \cite{kerr1965atti} 
\be
g_{\mu\nu} = \eta_{\mu\nu} + V(r)\ell_\mu\ell_\nu\,, \label{kerr}
\ee
where $V$ is a scalar function of $r$ with $r^2=\delta_{ij}x^i x^j$ and $\ell_\mu$ is the null geodesic vector, 
\be
\ell_\mu\diff x^\mu = \diff t + \delta_{ij}\frac{x^i}{r}\diff x^j\,,
\ee
thus satisfying the properties
\begin{subequations}
\ba
g_{\mu\nu}\ell^\mu\ell^\nu & = & 0\,, \\
\eta_{\mu\nu}\ell^\mu\ell^\nu & = & 0\,, \\
\ell^\mu\mathring{\nabla}_\mu\ell_{\alpha} & = & 0\,, \\
\ell^\mu\partial_\mu\ell_{\alpha} & = & 0\,. 
\ea
\end{subequations}
We work in the coincident gauge which is the unitary gauge $\Gamma^\alpha{}_{\mu\nu}=0$. The coincident gauge for the connection, and the coordinates (\ref{kerr}) for the metric is a solution in the inertial frame
$G^\mu{}_\nu=0$. {To avoid any confusion that seems to exist in the literature, the coincident gauge is always a legitimate choice, but then making an Ansatz for the metric intends to find certain solutions that may or may not exist. In our case, the metric (\ref{kerr}) is a solution in the coincident gauge.} If we transformed this solution to a different coordinate system, we would generate a non-vanishing connection\footnote{Non-vanishing connections have been taken into account in several recent studies of modified symmetric teleparallel gravity \cite{Lin:2021uqa,DAmbrosio:2021zpm,Hohmann:2021rmp,Flathmann:2021itp,Hohmann:2021ast,Dimakis:2022rkd}. {See also \cite{BeltranJimenez:2022azb} for a clarification on the use of the coincident gauge.}}. 

It is then a simple calculation to obtain
\begin{subequations}
\ba
Q_{0\mu\nu} & = & 0\,, \\
Q_{i00} & = & V'\ell_i\,, \\
Q_{i0j} & = & Q_{ij0} = \frac{V}{r}\delta_{ij} + \lp V' -\frac{V}{r}\rp\ell_i\ell_j\,, \\
Q_{ijk} & = & \frac{2V}{r} \delta_{i(j}\ell_{k)} + \lp V' -\frac{2V}{r}\rp\ell_i\ell_j\ell_k\,.\quad 
\ea
\end{subequations}
Using $g^{\mu\nu} = \eta^{\mu\nu} - V\ell^\mu\ell^\nu$ we get further 
\begin{subequations}
\ba
Q^{0}{}_{\mu\nu} & = & VV'\ell_\mu\ell_\nu\,, \\
Q^{i}{}_{00} & = & \lp 1- V\rp V'\ell^i\,, \\
Q^{i}{}_{0j} & = & Q^{i}{}_{j0} = \frac{V}{r}\lp \delta^i_j-\ell^i \ell_j\rp + \lp 1-V\rp V'\ell^i\ell_j\,,\,\,\,\,\,\,\,\,\, \\
Q^i{}_{jk} & = & \frac{2V}{r}\lp \delta^i_{(j}\ell_{k)} - \ell^i\ell_j\ell_k\rp \nn \\
& + &  \lp 1 - V\rp V'\ell^i\ell_j\ell_k\,,\,\, 
\ea
\end{subequations}
and raising yet 1 more index,
\begin{subequations}
\label{raise2}
\ba
Q^{0\mu}{}_{\nu} & = & VV'\ell^\mu\ell_\nu\,, \\
Q^{i0}{}_0 & = & -\lp 1- V\rp V'\ell^i\,, \\
Q^{i0}{}_j & = & -\frac{V}{r}\lp\delta^i_j-\ell^i\ell_j\rp-\lp 1- V\rp V'\ell^i\ell_j\,, \\
Q^{ij}{}_0 & = & \frac{V}{r}\lp\delta^{ij}-\ell^i\ell^j\rp + \lp 1-V\rp V'\ell^i\ell^j\,, \\
Q^{ij}{}_{k} & = & \frac{2V}{r}\lp\delta^{i(j}\ell_{k)} - \ell^i\ell^j\ell_k\rp  \nn \\ & + & \lp 1-V\rp V' \ell^i\ell^j\ell_k\,.
\ea
\end{subequations}
The traces are  
\begin{subequations}
\label{traces}
\ba
{Q}_\mu & = & 0\,, \label{trace1} \\
\tilde{Q}_\mu & = & \lp \frac{2V}{r} + V'\rp \ell_\mu \,.
\ea
\end{subequations}
Thus, we can see from (\ref{raise2}) that
\begin{subequations}
\label{raise2b}
\ba
Q^{[0i]}{}_0 & = & \frac{1}{2}V'\ell^i\,, \\
Q^{[0i]}{}_j & = & \frac{1}{2}\lb \frac{V}{r}\lp\delta^i_j - \ell^i\ell_j\rp + V'\ell^i\ell_j\rb\,, \\
Q^{[ij]}{}_0 & = & 0\,, \\
Q^{[ij]}{}_k & = &  \frac{V}{r}\delta^{[i}_k\ell^{j]}\,, 
\ea
\end{subequations}
and from (\ref{traces}) that
\begin{subequations}
\label{trace2b}
\ba
 \delta^{[\mu}_\alpha {Q}^{\nu]} & = & 0 \,, \\
 \delta^{[\mu}_\alpha \tilde{Q}^{\nu]} & = & \lp \frac{2V}{r} + V'\rp\delta^{[\mu}_\alpha\ell^{\nu]}\,.
\ea 
\end{subequations}
We can compute from (\ref{raise2b}) that
\begin{subequations}
\label{raise3}
\be
\frac{1}{2}\int Q^{[0 i]}{}_\mu\ell_i  \diff S = \pi r^2V'\ell_\mu\,,
\ee
and from (\ref{trace2b}) that 
\be 
\frac{1}{2}\int \delta^{[0}_\mu Q^{i]}\ell_i  \diff S = \pi r^2\lp \frac{2V}{r} + V' \rp\ell_\mu\,.\label{Qell}
\ee
\end{subequations}
Interestingly, the charges $C_\mu \sim \ell_\mu$ will be proportional to the null vector $\ell_\mu$. This suggests that a photon on a radial trajectory sees no gravitational energy in the black hole, {what might in turn be telling us about its masslessness.

\subsection{Charges}

Let us now compare the energy charges of a black hole obtained by using the different potentials. For this, it is convenient to parameterise the superpotentials with 3 constants $a$, $b$, $c$ as follows,
\be
h^{\mu\nu}{}_\nu  = a Q^{[\mu\nu]}{}_\alpha + b\delta^{[\mu}_\alpha Q^{\nu]} 
+ c\delta^{[\mu}_\alpha \tilde{Q}^{\nu]}\,. \label{parameterisation}
\ee
Considering a black hole with mass $m_S$ and charge $q$ due to a possible non-vanishing electric field, in the presence of a cosmological constant $\Lambda$, we have
\be
V(r) = \frac{m_S}{4\pi m_P^2 r} - \frac{q^2}{8\pi m_P^2 r^2} + \frac{\Lambda}{3}r^2\,. \label{srnds}
\ee
Using (\ref{raise3}) we then obtain the general expression for the energy enclosed within a radius $r$ according to a static observer,
\ba
C_0(a,b,c) & = & \frac{m_S}{2}\lp c - a \rp +  \frac{q^2}{2r}a \nn \\
& + & \frac{4\pi m_P^2\Lambda r^3}{3}\lp a + 2c\rp\,.\,\, \label{srndsparameterisation}
\ea
The electric charge due to the field configuration $A^{\mu\nu}$ is computed from (\ref{qlocalcharge}), and the result is $q$. (In the conventions implicit in (\ref{srnds}) the charge $q$ is dimensionless and $\Lambda$ has the dimension energy squared.) We note that the Reissner-Nordstr{\"o}m black hole is an example where the formula (\ref{localcharge}) is not valid but only the formula (\ref{qlocalcharge})
yields the correct answer. Similarly, for the energy charge the quasilocal formula yields (\ref{srndsparameterisation}) but the \1 term there would not be seen in the local formula. 

\subsubsection{The canonical prescription}

Consider \1 the canonical excitation (\ref{complex0}) explicitly given by (\ref{complex2}), for which $a=-c=-1$. We obtain $C_\mu = 4\pi m_P^2Vr\ell_\mu$, and thus
\be
C_0 = m_S - \frac{q^2}{{2}r} + \frac{4}{3}\pi r^3 \rho_\Lambda\,, \label{einsteincharge}
\ee
where $\rho_\Lambda = m_P^2\Lambda$ is the energy density due to the cosmological constant. Interpreting
$C^E_0$ as the gravitational energy makes perfect sense for each of the contributions: 
\begin{itemize}
\item The mass energy. In the pure Schwarzschild case, the energy is simply $m_S$, regardless of the radius at which the flux is computed. This is the canonical energy of a gravitational monopole. Analogously, the canonical charge of an electron is $q=e$, though one can consider infinite number of alternative definitions $q=q(r)$.
\item The electric energy. The electric charge contributes a Coulombian potential energy term. It is precisely the electrostatic energy of a thin spherical shell of radius $r$ and total charge $q$, which is of course equal to the energy of a condenser with the same radius and total charge (recall section 8.2 of Ref.\cite{feynman2011feynman}). 
\item The vacuum energy. The cosmological constant contributes energy with a constant density $\rho_\Lambda$, the expression that is familiar from the current standard model of cosmology.
\end{itemize}
The \1 result was already obtained in an inertial frame in metric teleparallel gravity (and is considerably neater than the results in the different frames suggested by various alternative regularisation schemes), as we recall from section \ref{metrictele}. It of course follows from the result (\ref{unification}) that several well-known superpotentials,
when ``covariantised'' by the minimal coupling and evaluated in an inertial frame, all unambiguously agree that
$C^{vF}_0 = C^{LL}_0 = C^{BT}_0 = C^{P}_0 = C^{W}_0 = C_0$. 


\subsubsection{M{\o}ller prescription}

The M{\o}ller potential
(\ref{moeller}), for which $a=-2$ and $b=c=0$, yields a slightly different result,  $C^M_\mu = - 4\pi m_P^2V' r^2\ell_\mu$, which for the (\ref{srnds}) gives
\be \label{mcharge}
C^M_0 =  m_S - \frac{q^2}{r} - \frac{8}{3}\pi r^3 \rho_\Lambda\,, 
\ee
The pure Schwarzschild contribution is again the correct one, but the numerical factors in the 2 other pieces suggest that the superpotential does not directly correspond to any standard concepts energy. The vacuum energy has even the wrong sign. Our computation agrees with the previous results in the literature, see e.g. the summary presented in \cite{Xulu:2002ix}. It has been found that the von Freud and the M{\o}ller prescriptions agree for the Schwarzschild, Vaidya and Janis-Newman-Winicour spacetimes, but there is a factor of 2 difference in the electric energy term in the Reissner-Nordstr{\"o}m
spacetimes \cite{Xulu:2000jf}, as shown in (\ref{mcharge}). 

\subsubsection{Komar prescription}

The same discrepancy, due to which the result (\ref{mcharge}) does not agree with the weak field limit, has been arrived at also from Komar's definition of energy \cite{doi:10.1063/1.526217} (again, one needs to renormalise $J^{\mu\nu}_K$ by a factor of 2). It had been pointed out by Tod \cite{Tod:1983waa} who, using Penrose's definition of the quasilocal energy \cite{Penrose:1982wp}, obtained the result which for the Reissner-Nordstr{\"o}m agrees with the result from
our canonical prescription (\ref{einstein}) but disagrees with the result from the Komar and the M{\"o}ller prescription (\ref{mcharge}), that the latter is not consistent with the linear theory.

However, there is a special property of the Komar energy that seems to suggest some possible physical interpretation. The black hole horizon is computed from $V(r)=1$, and has the more tractable expression when we set $\Lambda=0$,
\be
8\pi m_P^2 r_{\pm} = m_S \pm \sqrt{m_S^2 - 8\pi m_P^2 q^2}\,.
\ee
The black hole is called extremal when the 2 horizons are degenerate and 
thus $q^2 = m_S^2/(8\pi m_P^2)$. It follows that the Komar (and thus, also the M{\o}ller) energy enclosed within the horizon neatly disappears for the extremal Reissner-Nordstr{\"o}m black hole \cite{Deruelle:2005qk}, whereas the canonical energy is $C_0=m_S/2$. In either case, according to the standard definition of temperature we obtain
\be
T =-\frac{V'(r_+)}{4\pi}= \frac{1}{16\pi^2}\lp m_S - \frac{q^2}{r_+}\rp\,,
\ee
and thus, the extremal black hole is at zero temperature.  
(Perhaps, the vanishing of the M{\"o}ller and the Komar charges at the horizon in the extremal limit hints that they could be related to some different thermodynamical energy. As mentioned earlier, both the M{\"o}ller and the Komar prescriptions can be seen imply different boundary conditions for the Hamiltonian than the canonical prescription, and these different boundary conditions might reflect different control versus response parameters prescription in a thermodynamical interpretation.)    

\subsubsection{Tolman prescription}

Now we check the result from using the symmetric potential of Tolman \cite{Tolman1934-TOLRTA}, whose covariant version is given by (\ref{ptensor}). Since in the coordinates (\ref{kerr}),
\be 
q^{\mu\nu}{}_\alpha = -\frac{m_P^2}{2}L^{\mu\nu}{}_\alpha - \frac{m_P^2}{4}\tilde{Q}^\mu\delta^\nu_\alpha\,,
\ee
in order to calculate the conserved energy-momentum current we only need the components
\begin{subequations}
\ba
L^{0i}{}_0 & = & \frac{1}{2}V'\ell^i\,, \\ 
L^{0i}{}_j & = &  \frac{V}{r}\lp\delta^i_j - \ell^i\ell_j\rp + V'\ell^i\ell_j\,. 
\ea
\end{subequations}
We then obtain the relevant parts,
\begin{subequations}
\ba
q^{0i}{}_0 & = & -\frac{m_P^2}{2}V'\ell^i\,, \\ 
q^{0i}{}_j & = & m_P^2\frac{V}{r}\lp\delta^i_j - \ell^i\ell_j\rp + m_P^2V'\ell^i\ell_j \,. 
\ea
\end{subequations}
From this we obtain that $C^T_\mu =  C^M_\mu$, the Tolman superpotential gives the same Noether energy charge (\ref{mcharge}) as the M{\o}ller superpotential. 

\subsubsection{G{\"u}rses-G{\"u}rsey prescription}

We should review some previous calculations in the literature.
We find a slight disagreement with the rather often-cited results of Ref. \cite{Aguirregabiria:1995qz}. Most importantly though, our result (\ref{einstein}) for the time translation charge $C_0$ completely agrees with theirs in the Reissner-Nordstr{\"o}m case. However, our conclusion that the charges can be used to distinguish between the different prescriptions, does not agree with their claim that the various different prescriptions they studied, which included the Tolman and the so called Einstein energy-momentum complex\footnote{Which is actually the von Freud's superpotential \cite{10.2307/1968929,Bohmer:2017zid}, though often attributed to Einstein's original paper \cite{1916SPAW......1111E} (this error was repeated in e.g. \cite{Koivisto:2021jdy}). Only the $\underline{q}^{\alpha\mu}{}_\nu$, later known as Tolman pseudotensor, featured in Ref.\cite{1916SPAW......1111E}, and its quasilocal integral was not taken into consideration.} 
would yield identical charges. 
The different conclusion might be due to assuming that when 2 superpotentials yield the same energy momentum (pseudo)tensor, they would yield also the same charges. Assuming this, the same superpotential might have been used in all of the cases, which would explain the different conclusion from ours. The superpotential adopted in Ref.\cite{Aguirregabiria:1995qz} was \1 presented by G{\"u}rses and G{\"u}rsey  \cite{doi:10.1063/1.522480}, who considered the Landau-Lifshitz and the so called Einstein energy-momentum complex. From our discussion in section \ref{boundary} we see that the conclusion of G{\"u}rses and G{\"u}rsey generalises from the class of Kerr-Schild metrics to arbitrary geometries, and that it applies further to the Papapetrou, Weinberg and Bergmann-Thomson superpotentials. The unification (\ref{unification}) of these superpotentials was thus anticipated by the result of
G{\"u}rses and G{\"u}rsey, extended by the ``empirical'' investigation of Aguirregabiria {\it et al} \cite{Aguirregabiria:1995qz} and further generalised by the formal arguments of Chen {\it et al} \cite{Chen:2018kcs} (recall discussion in section \ref{boundary}). 

\subsubsection{Alternative prescriptions}

Finally, we shall briefly comment on the Noether currents obtained from alternative formulations of the symmetric teleparallel equivalent of GR in section \ref{alternatives}. The Lagrange multiplier formulation leads to the 
``improved'' Komar superpotential (\ref{komar_improved}) that was also found 
in Ref.\cite{Heisenberg:2022nvs}. As it differs from the canonical result only by a covariant derivative of the velocity, there is no difference for a static observer since our solution is in the coincident gauge. Thus, one should consider a more general set-up to distinguish the effect of the non-canonical term.

The superpotential (\ref{theresult}) derived in the inertial connection approach, however, does not yield a viable result. 
Since in the parameterisation (\ref{parameterisation}) now $a=-1$, $b=-3/4$ and $c=1/2$, we now obtain from (\ref{srndsparameterisation}) the energy of the Reissner-Nordstr{\"o}m-de Sitter black hole (\ref{srnds}) as
\be
C^A_\mu = \lp\frac{3}{4}m_S - \frac{q^2}{8r}\rp\ell_\mu\,. 
\ee
That the cosmological constant contribution drops away would suggest the interpretation that the charge is (proportional to) the enthalpy $H=E+p\mathcal{V}$ rather than the internal energy $E$. It does make sense to interpret the mass $m_S$ as enthalpy \cite{Dolan:2010ha}. However, we would have to invoke the radiation equation of state $m_S=H=\frac{4}{3}E$ in order to recover the energy charge $C^A_0=E$, but that seems too contrived.


\section{Gravitational waves}
\label{gravitationalwaves}

Gravitational waves have been a subject of great controversies since their conception \cite{article}.
Nowadays, due to recent fantastic progress, we have even observational evidence for their existence.
Nevertheless, the energy carried by gravitational waves remains a theoretical conundrum \cite{cite-key,cooperstock,Cooperstock:2015lga,Cooperstock:2018udm,Cai:2021jbi}. There are many false common wisdoms about the energy-momentum of gravitational waves; for a recent critical assessment of 4 such arguments, see Ref. \cite{Duerr:2019fsv}. The canonical resolution of energy-momentum reveals that the energy-momentum of gravitational waves in an absolute vacuum (i.e. in the strict absence of any non-gravitational fields, including a detector of any kind) is null. This does not imply that gravitational waves would be unphysical, but that an absolute vacuum is an unphysical idealisation. 

The property of the gravitational waves might again be considered in analogy with electromagnetism (a recurring theme of this article). Electromagnetic waves can very well propagate without ``carrying'' electromagnetic charge, yet these waves also can be detected through their interaction with charged matter. Similarly, gravitational waves can propagate without inherent gravitational charge (= energy), but these waves are detectable via their gravitational (=energetic) interaction with matter. This follows consistently from our identification of energy as the canonical Noether charge, and of course the energy according to this definition is conserved, by construction.  

To clarify the detectability of gravitational wave energy, we recall the local energy-momentum current, obtained from the divergence of the canonical excitation as
\be
\mathfrak{J}^\mu=\nabla_\alpha \mathfrak{J}^{\alpha\nu}
= \lp \mathfrak{T}^\mu{}_\nu +\mathfrak{t}^\mu{}_\nu\rp v^\nu
+ \frac{1}{2}\mathfrak{h}^{\alpha\mu}{}_\nu\nabla_\alpha v^\nu\,.   
\ee
We can recognise 5 cases wherein gravitational waves may in principle result in nontrivial energy charge:
\begin{enumerate}[label={\it \roman*})]
\item In the presence of matter. Since $\mathfrak{T}^\mu{}_\nu = \sqrt{-g}T^\mu{}_\nu$, fluctuations in the metric influence the material current.
\item In the presence of background gravitational field. In physical (as opposed to absolute) vacuum, the charge $C_0$ can quantify the nonzero energy of gravitational waves\footnote{The gravitational field of e.g. a binary system extends to infinity. It is associated with some total mass, which decreases as the binary emits a pulse of gravitational waves. The change in the mass before and after the pulse is reflected in the charge $C_0$, as should be obvious from the derivations of the previous section \ref{energy-momentum} (even though of course for a binary one should in principle take into account a more complicated configuration exhibiting a quadrupole moment at the emission etc.).}.
\item In a non-inertial frame. Wrt a non-inertial frame, wherein $\mathfrak{t}^\mu{}_\nu \neq 0$, there obviously can appear energies even into an absolute vacuum.
\item Wrt (covariantly) non-static clock. An accelerated observer, for whom $\nabla_\alpha v^\mu \neq 0$ , can probe the $\mathfrak{h}^{\alpha\mu}{}_\nu$ generated by gravitational waves.
\item In nontrivial topology. Gravitational waves with nontrivial topology could generate energy charges without any local current ($\mathfrak{J}^\mu=0$).  
\end{enumerate}
It could be interesting to investigate some of these cases (we do not claim that all of them would be realistic or relevant). As the \1 exploration, in the following we will only confirm that, as expected, in the case that none of the above 5 conditions holds, the energy-momentum of gravitational waves consistently vanishes. 


\subsection{Linear perturbations}
\label{linear}

Consider the perturbation $\delta g_{\mu\nu}$ of the flat metric $\eta_{\mu\nu}$,
\be
g_{\mu\nu} = \eta_{\mu\nu} + \delta g_{\mu\nu}\,.
\ee
Using the 1+3 decomposition familiar from cosmological perturbation theory, we decompose the perturbation $\delta g_{\mu\nu}$ into scalars $\phi$, $\psi$, $\beta$, $\sigma$, transverse vectors $B_i$, $E_i$, and transverse and traceless tensors $h_{ij}$ as follows:
\begin{subequations}
\ba
\delta g_{00} & = & -2\phi\,, \\ 
\delta g_{0i} & = & -\beta_{,i} + B_i\,, \\ 
\delta g_{ij} & = & -2\psi\delta_{ij} + \sigma_{,ij}-\frac{1}{3}\nabla^2\sigma\delta_{ij} \nn \\ & + & 2 E_{(i,j)} + 2h_{ij}\,.
\ea
\end{subequations}
Then we find that at the linear order in perturbations, the gravitational energy-momentum vanishes identically,
\be
G^\mu{}_\nu =0\,,
\ee
and the field equations become
\begin{subequations}
\ba
T^0{}_0 &  = & -2\nabla^2\varphi\,, \label{pert_energy}\\
T^0{}_i & = &   -2\dot{\varphi}_{,i} + \frac{1}{2}\nabla^2 V_i\,,\\
T^i{}_j  & = & \lb 3\ddot{\psi}-\ddot{\varphi}+\nabla^2\lp \phi - \varphi - \dot{\beta}\rp\rb\delta^i{}_j \nn \\
& - & \lp\phi - \varphi - \dot{\beta}\rp{}^{,i}{}_{,j} \nn \\
& - & \dot{V}^{(i}{}_{,j)} - \frac{1}{2}\dot{B}^i{}_{,j} + \ddot{h}^i{}_j - \nabla^2h^i{}_j\,.
\ea
\end{subequations}
where we have defined $\varphi=\psi+\frac{1}{6}\nabla^2\sigma$ and $V_i = B_i-\dot{E}_i$. One can check that consistently
$\nabla_\mu T^\mu{}_\nu = 0$.
We have
\begin{subequations}
\ba
m_P^{-2}\mathfrak{q}^{i0}{}_0 & = & \varphi^{,i} + \frac{1}{4}\dot{\beta}^{,i} + \frac{1}{4}\dot{B}^i + \frac{1}{4}\nabla^2 E^i\,, \\
m_P^{-2}\mathfrak{q}^{i0}{}_j & = &  \frac{1}{4}\lp -\dot{\phi}+\dot{\varphi} - \frac{1}{6}\nabla^2\dot{\sigma}\rp\delta^i_j + \frac{1}{4}\dot{\sigma}^{,i}{}_{,j} \nn \\ & + & \frac{1}{2}B^{[i}{}_{,j]} + \frac{1}{2}\dot{E}^{(i}{}_{,j)} + \dot{h}^i{}_j \,.
\ea
\end{subequations}
We are most interested in the components of the tensor density\footnote{These were reported in Ref. \cite{Koivisto:2021jdy} for the generic 13-parameter theory with a local and linear constitutive law, though unfortunately with some typos.} $\mathfrak{h}^{\mu\nu}{}_\alpha$. For these components we obtain the expressions
\bs
\ba
m_P^{-2}\mathfrak{h}^{i0}{}_0 & = & -2\varphi^{,i} - \frac{1}{2}\nabla^2 E^i\,, \label{pert_h00} \\
m_P^{-2}\mathfrak{h}^{ij}{}_0 & = & B^{[i,j]}\,, \\
m_P^{-2}\mathfrak{h}^{0i}{}_j & = & \lp 3\dot{\psi} - \dot{\varphi} - \frac{1}{2}\nabla^2\beta\rp\delta^i_j \nn \\
& + & \frac{1}{2}\lp \beta + \dot{\sigma}\rp{}^{,i}{}_{,j}
- V^{(i}{}_{,j)} + \dot{h}^i{}_j\,,  \\
m_P^{-2}\mathfrak{h}^{ij}{}_k & = & -\lp 2\varphi - 2\phi + \dot{\beta}\rp{}^{[,i}\delta^{j]}_k + E^{[i,j]}{}_{,k}\nn \\
& - & \lp \nabla^2 E - \dot{B}\rp{}^{[i}\delta^{j]}_k 
 + 2h_k{}^{[i,j]}\,. 
\ea
\es
Again we can make a consistency check by checking that $\nabla_\mu H^{\mu\nu}{}_\alpha = m_P^2 T^\nu{}_\alpha$. 
The component (\ref{pert_h00}) determines the energy as
\ba
C_0 & = & m_P^2\oint \diff^2 x\lp -2\psi_{,i} - \frac{1}{3}\nabla^2\sigma_{,i} - \frac{1}{2}E_i\rp n^i \nn \\
& = & m_P^2\oint \diff^2 x g^{jk}\partial_{[i}g_{j]k}n^i 
 =  E_{\text{ADM}}\,. \label{C_adm}
\ea
We have recognised the energy integral as precisely the ADM energy expression. Note that the flux of the possible vector modes $E^i$ can contribute energy and this part one could not have deduced from the local energy density (\ref{pert_energy}).
It is thus non-trivial that we recover precisely the canonical
ADM energy at the linear order in perturbations, which could be 
regarded as the minimal requirement for a successful definition 
of gravitational energy. For example, the Komar integral does not agree with the ADM expression,
\be
C_{\text{Komar}} = \int \diff^2 x\lp {\phi}_{,i} -  \frac{1}{2}\dot{\beta}_{,i} + \frac{1}{2}B_i\rp n^i\,.  \label{C_komar}
\ee
With some assumptions, however, an agreement can be established \cite{Jaramillo:2010ay}. In the present set-up, we see that we need to assume 0) renormalisation by the factor of 2, 1) the absence of shift and vector perturbations, 3) vanishing effective anisotropic stress. $\phi=-\psi$. Only scalars and vectors contribute to either (\ref{C_adm}) or
(\ref{C_komar}) and there is no energy due to tensor modes. 

We will have a closer look at the transverse-traceless components $h_{ij}$, which are the only components that may propagate in vacuum. Taking this into account, the relevant components of the tensor density $\mathfrak{h}^{\mu\nu}{}_\alpha$ are simplified to 
\begin{subequations}
\label{gwh}
\ba
\mathfrak{h}^{i0}{}_0 & = & 0\,, \\
\mathfrak{h}^{i0}{}_j & = & -m_P^2\dot{h}^i{}_j\,.
\ea
\end{subequations}
We can also check that the field equations, which now reduce to $\nabla_\alpha\mathfrak{h}^{\alpha\mu}{}_{\nu}=0$ since $\mathfrak{G}^\mu{}_\nu=0$, are given by 
\begin{subequations}
\ba
\nabla_\mu\mathfrak{h}^{\mu 0}{}_0  = 
\nabla_\mu\mathfrak{h}^{\mu 0}{}_i & = & 0\,, \\ 
\nabla_\mu\mathfrak{h}^{\mu i}{}_j & = & m_P^2\Box h^i{}_j\,,
\ea
\end{subequations}
and thus imply the wave equation $\Box h^i{}_j  = 0$ as expected. From (\ref{gwh}) we can determine the conserved charges
\bs
\ba
C^0 & = & 0\,, \\
C^i & = & m_P^2\oint \diff^2 x \dot{h}^i{}_j n^j\,. \label{gwmom}
\ea
\es
This confirms that the gravitational waves do not have energy associated to them in the linearised approximation. It is easy to see that despite the non-zero integrand (\ref{gwmom}), also the momentum charge vanishes. Without loss of generality, we may consider the standard
parameterisation of gravitational wave polarisations $h_+$ and $h_\times$ travelling, for concreteness, in the $z$-direction,
\be
h^i{}_j =
\lp
\begin{matrix}
h_+ & h_\times & 0\\
h_\times & -h_+ & 0 \\
0 & 0 & 0
\end{matrix}
\rp\,. 
\ee
The surface integral can then be performed over a spherical surface at a radius $r=\sqrt{x^2 + y^2 + z^2}$ using the standard spherical coordinates,
\bs
\ba
n^x & = & \cos\phi\sin\theta\,, \\
n^y & = & \sin\phi\sin\theta\,, \\
n^z & = & \cos\theta\,. 
\ea
\es
We obtain
\bs
\ba
C^x & = & -\frac{m_P^2}{2}\oint r^2 \sin\theta\lp h_+ n^x + h_\times n^y\rp\diff\theta\diff\phi \nn \\
& = & -\frac{\pi m_P^2 r^2}{4}\int \lp h_+ \cos\phi + h_\times\sin\phi\rp\diff\phi\nn \\ & = & 0\,,\quad  \\
C^y & = & \frac{m_P^2}{2}\oint r^2 \sin\theta\lp h_+ n^y - h_\times n^x\rp\diff\theta\diff\phi \nn \\
& = & \frac{\pi m_P^2 r^2}{4}\int \lp h_+ \sin\phi - h_\times\cos\phi\rp\diff\phi \nn \\  & = & 0\,,  \\
C^z & = & 0\,.
\ea 
\es
For the transverse momenta, the averages over the azimuthal angles cancel, and therefore the net charges vanish. {The vanishing of the charges is in agreement with our remark below \eqref{Qell} since gravitational waves correspond to massless particles very much like photons.}

This result supports the intuition that in an absolute vacuum, gravitational waves cannot be associated with physical energy-momentum (according to static observer in an inertial frame). However, it is reasonable to suspect that energy-momentum should be quadratic order in perturbations, but in this article we shall not proceed to the study of higher order perturbation theory. Instead, we can find further support to the conclusion by studying a class of exact, non-perturbative gravitational wave solutions.

\subsection{Exact solutions}

A spacetime filled with plane-fronted waves with parallel propagation, i.e. a PP-wave spacetime describing gravitational waves, can be parameterised by the line element
\be \label{ppwave}
\diff s^2 = H(u,x,y)\diff u^2 + 2\diff u\diff v + \diff x^2 + \diff y^2\,.
\ee
It turns out that in the coincident gauge this solution is in the inertial frame, $t^\mu{}_\nu=0$. 
The same result was found in \cite{doi:10.1063/1.522480}, see also \cite{Fabbri:2008vp,Fabbri:2011zz}, and in metric teleparallelism in \cite{Obukhov:2009gv}.

There is only 1 non-trivial component of
\be
\nabla_\alpha \mathfrak{q}^{\alpha\mu}{}_{\nu} = -\frac{m_P^2}{2}\delta^\mu_v\delta_\nu^v\lp H_{,xx} + H_{,yy}\rp\,, 
\ee
which also vanishes on shell.
If we restrict to quadratic solutions of the Einstein equation, we obtain the well-known plane gravitational waves $H = A(u)\lp x^2 - y^2\rp + 2 B(u)xy$ described by the 2 functions $A(u)$ and $B(u)$.  There are nonzero
components of 
\ba
\mathfrak{q}^{\mu\alpha}{}_\nu & = & -\frac{m_P^2}{4}\delta^\mu_v\Big[\delta^\alpha_v \lp \delta_\nu^u H_{,u} + \delta_\nu^x H_{,x} + \delta_\nu^y H_{,y}\rp \nn \\ & + & \delta^u_\nu\lp\delta^\alpha_x  H_{,x} + \delta^\alpha_y H_{,y}\rp\Big] \nn \\
 & + &  \frac{1}{4}\delta^\alpha_v\delta_\nu^u\lp \delta^\alpha_x H_{,x} + \delta^\alpha_y H_{,y}\rp\,.
\ea
However, for the calculation of energy and entropy we need the superpotential $\mathfrak{h}^{\mu\alpha}{}_\nu$ instead. Also, it is more convenient to employ the Cartesian coordinates
\bs
\ba
u & = & z-t\,, \\
2v & = & z +t\,.
\ea
\es
It is easy to see that this linear change in coordinates does not take us away from the inertial frame but retains $t^\mu{}_\nu = 0$ in the coincident gauge. The line element (\ref{ppwave}) reads
\be
\diff s^2 = \lp H-1\rp\diff t^2 - 2H\diff t\diff z + \lp 1+H\rp\diff z^2 + \diff x^2 + \diff y^2\,,
\ee
where $H=H(z-t,x,y)$. The conserved currents are determined from the components
\be
\mathfrak{h}^{0\mu}{}_\nu = \frac{m_P^2}{2}\delta^\mu_z\lp \delta^x_\nu H_{,x} +  \delta^y_\nu H_{,y}\rp\,. 
\ee
Again, we see immediately that the energy is zero,
\bs
\ba
C_0 & = & 0\,, \\
C_i & = & \frac{m_P^2}{4}\oint \lp\delta_i^x H_{,x} + \delta_i^y H_{,y}\rp n_z\,.
\ea
\es
Inserting the quadratic solution for $H(u,x,y)$ and using $x=r n^x$, $y=r n^y$, gives
\bs
\ba
C_x & = & \frac{m_P^2}{2}\oint\lb A(u)x + B(u)y\rb n^z\diff\theta\diff\phi  \nn \\
& = & \frac{m_P^2}{2}\oint r^3\sin^2\theta \lb A(u)\cos\phi + B(u)\sin\phi\rb \cos\theta \diff\theta\diff\phi \nn \\
& = & 0\,, \\
C_y & = & -\frac{m_P^2}{2}\oint \lb A(u)y-B(u)x\rb n^z\diff\theta\diff\phi  \nn \\
& = & -\frac{m_P^2}{2}\oint r^3\sin^2\theta \lb A(u)\sin\phi - B(u)\cos\phi\rb \cos\theta \diff\theta\diff\phi \nn \\
& = & 0\,, \\
C_z & = & 0\,.
\ea
\es
Again, it is the integration over the azimuthal angle that averages to zero 
(since $u$ depends upon $z$ and thus the polar angle, the integral over $\theta$ would not necessarily cancel, as it did not in the case of \ref{linear}). Thus, the non-perturbative calculation confirms the conclusions about plane gravitational waves in absolute vacuum suggested by the calculation of linear perturbations.

\section{Entropy}
\label{entropysection}

In section \ref{energy-momentum} we discussed physical definitions of a black hole.
In purely formal terms, a black hole $\mathcal{B}$ can be defined as a region of a manifold $\mathcal{M}$ where the future null infinity $\mathcal{I}^+$ has no chronological past $I^{-}$, $\mathcal{B}= \mathcal{M} - I^{-}(\mathcal{I}^+)$.
The horizon $\mathcal{H}$ of $\mathcal{B}$ is then the boundary $\mathcal{H} = \partial I^{-}(\mathcal{I}^+)$, which is a null surface. $\mathcal{B}$ is stationary if there exists a Killing field $\mathring{\nabla}_{(\mu} t_{\nu)}=0$ which at infinity $t^\mu t_\mu = -1$. In such a (rigid) case, $\mathcal{H}$ is a Killing horizon i.e. its null generators coincide with the orbits of a 1-parameter group of isometries and so there
is a Killing vector normal to $\mathcal{H}$. $\mathcal{B}$ is called static if $t^\mu$ is a hypersurface orthogonal, and then it follows that $t^\mu \in \mathcal{H}^\perp$. Under more general circumstances in which $\mathcal{H}$ could also have angular velocity $\Omega$, there exists a Killing field $\xi^\mu = t^\mu + \Omega\phi^\mu$. Then there is always defined a function $\kappa$ such that 
$\mathring{\nabla}^\mu(\xi^\nu\xi_\nu)=-2\kappa\xi^\mu$. It must be a constant along the null geodesic generator of a Killing horizon and therefore is constant over the $\mathcal{H}$: that is the 0$^{\text{th}}$ law of black hole thermodynamics. 

A bifurcate Killing horizon is a pair of Killing horizons, with respect to the same Killing field $\xi^\mu$, and intersecting on a spacelike hypersurface $\mathcal{C}$. Then $\xi^\mu(x) = 0$ when $x \in \mathcal{C}$. A consequence of the 0$^{\text{th}}$ law is that unless the $\mathcal{B}$ is extremal i.e. $\kappa=0$, then $\mathcal{H}$ comprises a branch of bifurcate Killing horizon. The 
1$^{\text{st}}$ law of black hole thermodynamics is $\delta M = m_P^2\kappa \delta A + \Omega \delta J$, where $A$ is the area of $\mathcal{H}$ {and $J$ the angular momentum}. The area-entropy law then results in 
\be
\delta M = \frac{m_P^2\kappa}{4}S + \Omega\delta J\,. \label{1stlaw}
\ee
Wald has found a famous expression for the entropy $S$ that can be interpreted as a Noether charge \cite{Wald:1993nt,Iyer:1994ys}. It was derived using the integral of the form (\ref{integral1}) with the Noether potential $J^{\mu\nu}_\xi$ generated by a Killing vector $\xi^\mu$ which asymptotically corresponds to time translations and thus yields the correct (equal to the ADM) energy charge and can be directly related to the Hamiltonian in the symplectic formalism. We will not review the whole derivation here, which has also recently been translated to the symmetric teleparallel geometry \cite{Heisenberg:2022nvs}. 

The essential point is that the energy charge (diffeomorphism charge wrt time translations) can be recasted into an expression for entropy by using the \1 law that relates energy change to entropy change. What required some ingeniosity and elaborate differential-geometric constructions was the fact that the energy charge was not known, or rather it was known only under certain assumptions at the infinity. However, we have claimed the possible resolution of the energy problem. As we know the energy charge, we can almost trivially express the entropy $S$ in terms of this energy charge, using the \1 law (\ref{1stlaw}). The surface gravity $\kappa$ appears in this formula,     
\be
\label{integral3}
S = \frac{2 \pi}{\kappa}\oint_{\mathcal{H}} \diff^{n-2}\sigma_{\mu\nu} \mathfrak{h}^{\mu\nu}{}_\alpha v^\alpha\,. 
\ee
This expression is valid for any gravity theory formulated in any geometric setting, if one replaces $\mathfrak{h}^{\mu\nu}{}_\alpha v^\alpha = \mathfrak{J}^{\mu\nu}_v$ correspondingly. 
To check the formula, we recall that the temperature of a Schwarzschild black hole is $m_P^2/m_S$, and thus the surface gravity $\kappa = 2\pi m_P^2/m_S$, adapting these quantities to our mass-based units. Then, as we compute the surface integral (\ref{integral3}) we recognise that it is proportional to the energy integral we have already computed,
\be \label{result1}
S = \frac{m_S}{2m_P^2}C_0 = \frac{1}{2}\lp \frac{m_S}{m_P}\rp^2\,. 
\ee
Since the area of the black hole is $ A = m_S^2/4\pi m_P^4$, (\ref{result1}) is precisely the usual entropy-area law. 

Let us now consider another approach, based on the ``\1 new insight'' of Ref. \cite{Jimenez:2021nup}. There we noticed that the black hole entropy is the
Noether charge, corresponding to the symmetry of radially boosting the horizon. Since Lorentz transformations operate in the frame bundle, we do not repeat the (very simple) derivation here, but state the result:
\be \label{wald1}
{S} = 
-2\pi\int_\mathcal{C} {\diff^2 x} \mathfrak{r}_{\alpha}{}^{\beta\mu\nu} n^\alpha{}_{\beta}n_{\mu\nu}\,,
\ee
where $n_{\alpha\beta}$ is the binormal to the $\mathcal{C}$ normalised such that $n_{\alpha\beta}n^{\alpha\beta}=-2$. 1 of these terms stems from just writing differently $\diff^2 \sigma_{\mu\nu} \sim {\diff^2 x} n_{\mu\nu}$, and the other stems from the choice of the Lorentz boost. The conserved charge is called the center of mass momentum. In non-relativistic physics, the conservation of this charge is the consequence of the energy-momentum conservation, but in special relativity it is an independent quantity. It is very intriguing that its physical interpretation (which, in comparison to the energy-momentum and the angular momentum is quite obscure) turns out to be related to entropy, at least for black holes. (In the context of black holes, fluctuations of the center of mass momentum have been considered potentially relevant for the unitary of the evaporation, e.g \cite{Flanagan:2021svq}.) 

It is not difficult to see that the center of mass momentum charge (\ref{wald1}) is a viable generalisation of the Wald entropy. In the special case of coincident GR, we obtain from the solution (\ref{rsolution}) 
\be \label{cgrentropy}
S = \pi m_P^2 \int_{\mathcal{H}} \diff x^2 g^{\beta[\mu}\delta^{\nu]}_\alpha n^\alpha{}_\beta n_{\mu\nu}\,.
\ee
This could be easily evaluated e.g. using the geometry of section \ref{energy-momentum} and the binormal $n_{\alpha\beta} = 2\ell_0\ell_r \eta_{0[\alpha}\eta_{\beta]r}$.  
Instead, we recall the original expression for Wald entropy \cite{Wald:1993nt}
\be \label{wald0}
\mathring{S} = -2\pi\int_\mathcal{C}\diff^2 x \frac{\partial\mathfrak{L}}{\partial {\mathring{R}}_{\alpha\beta\mu\nu}}n_\alpha{}_{\beta}n_{\mu\nu}\,,
\ee
where the hats remind that the derivation was based on considering
the metric as the only gravitational variable. In the case of GR, we obtain from (\ref{grconst}) precisely the same expression as (\ref{cgrentropy}). 
In our 1$^{\text{st}}$ order framework, the Riemann curvature ${\mathring{R}}^{\alpha}{}_{\beta\mu\nu}$ cannot enter into the
action $I$ explicitly, but from (\ref{riemann}) one can see
\be 
{R}^\alpha{}_{\beta\mu\nu} =  \mathring{R}^\alpha{}_{\beta\mu\nu}   
   +   2\mathring{\nabla}_{[\mu} N^\alpha_{\phantom{\alpha}\nu]\beta}
+ 2N^\alpha_{\phantom{\alpha}[\mu\lvert\lambda\rvert}N^\lambda_{\phantom{\lambda}\nu]\beta}\,,
\ee 
and thus (\ref{wald1}) is the straightforward generalisation of (\ref{wald0}). 


An entropy formula for metric teleparallelism had been derived earlier by Hammad {\it et al} \cite{Hammad:2019oyb}, and is the correct limit (\ref{integral3}) when adapted to the metric teleparallel geometry. Hammad {\it et al} noted that the result can be rewritten as a volume integral 
\be \label{volume}
S = \int_{\mathcal{B}}\diff^3 x\mathring{\nabla}_i \lp \mathfrak{h}^{0i}{}_\alpha v^\alpha\rp\,,
\ee
and thus it clarifies a possible conformal problem with the standard area-entropy law in GR. Though, we do not know what forbids using the Stokes law in GR exactly the same way. Also, we note technical issues with the computations in \cite{Hammad:2019oyb}, as their energy formula would give a divergent result (the regularisations discussed in \ref{metrictele} were not used explicitly), and their entropy formula would not either give the correct result (since an inertial frame was not identified), but the result was forced by implementing additional constraints on the Killing vectors which we will briefly discuss below. Despite these details, the entropy formula in Ref.\cite{Hammad:2019oyb} agrees with the expression (\ref{integral3}), and they also noted the benefit of teleparallel formulation and other \1 order formulations \cite{Corichi:2013zza}, that one does not have to invoke a bifurcation surface and resort to a Killing vector that vanishes on that surface.

The Killing vectors of the metric are parallel transported by the metric Levi-Civita connection. 
One could also consider the different transport of the vector along itself, $\xi^\mu \nabla_\mu \xi^\nu = \tilde{\kappa}\xi^\nu$, but in general $\tilde{\kappa} \neq \kappa$. If the vector is autoparallel, $\tilde{\kappa}=0$. From (\ref{gmorf}) we see that for a Killing vector $\xi^\mu$,
\begin{subequations}
\be
\nabla_{(\mu}\xi_{\nu)}=-\lp T_{(\mu\nu)\alpha} - L_{\alpha\mu\nu}\rp \xi^\alpha\,, 
\ee
implying that
\be
\partial_\mu \xi^2 + \tilde{\kappa}\xi_\mu = -\lp T_{\alpha\mu\nu} - 2Q_{[\alpha\mu]\nu}\rp \xi^\alpha\xi^\nu\,. 
\ee
\end{subequations}
In general there is no reason to expect the RHS of these equations to vanish. (However, this condition was imposed in  Ref.\cite{Hammad:2019oyb}, adapted from Ref. \cite{Dey:2017fld}.) The surface gravity in black hole thermodynamics is a purely metrical concept, and requiring $\tilde{\kappa} = \kappa$ would entail somehow totally different interpretation of temperature and thermodynamics in general. 

A generalisation of a Killing vector, which has a physical interpretation, could rather be defined by the requirement of the $\xi^\mu$ being also a symmetry of the independent connection, \cite{Hohmann:2019fvf,Hohmann:2020zre,Hohmann:2021ast} $\delta_\xi \Gamma^\alpha{}_{\mu\nu}=0$. Then $\xi^\mu$ can be called an affine Killing vector. From (\ref{cmorf}) we then get that this condition is a 2$^{\text{nd}}$ order differential equation for for $\xi^\mu$,
\ba
g^{\alpha\beta}\nabla_\mu\nabla_\nu \xi_\beta & - & 2Q_{(\mu}{}^{\alpha\beta}\nabla_{\nu)}\xi_\beta - \nabla_\mu Q_\nu{}^{\alpha\beta}\xi_\beta \nn \\
& = & \nabla_\mu\lp T^\alpha{}_{\beta\nu}\xi^\beta\rp + R^\alpha{}_{\nu\beta\mu}\xi^\beta\,.
\ea
If we restrict to symmetric teleparallelism, this is simply $\nabla_\mu\nabla_\nu \xi^\alpha=0$, and the generalised Killing condition leads to the 
identity $Q_{\alpha\lambda(\mu}\nabla_{\nu)}\xi^\lambda = \nabla_\alpha(Q_{\lambda\mu\nu}\xi^\lambda) = -\nabla_\xi Q_{\lambda\mu\nu} + Q_{\lambda\mu\nu}\nabla_\alpha \xi^\lambda$, where in the \2 equality we recalled from (\ref{bianchi3}) that now $\nabla_{[\alpha}Q_{\beta]\mu\nu}=0$, and one sees that the equation boils down to, consistently, nothing but the condition that $\delta_\xi Q_\alpha{}^{\mu\nu} = 0$. Note though that in general we have no reason to impose this constraint for the Killing vectors of black hole spacetimes. In fact, the formulas we have derived for energy and entropy do not require considering Killing vectors at all.

\begin{widetext}
\begin{center}
\begin{table}
\begin{tabular}{ |c|c|c|c| } 
 \hline
 formulation & constraints & superpotential & canonical frame \\ 
 \hline \hline
  symm. tele$_\parallel$  & $R^\alpha{}_{\beta\mu\nu}=T^\alpha{}_{\mu\nu}=0$ 
 & $m_P^{-2}h^{\mu\nu}{}_\alpha =   \delta^{[\mu}_\alpha\tilde{Q}^{\nu]} - \delta^{[\mu}_\alpha Q^{\nu]} -Q^{[\mu\nu]}{}_\alpha$ & $\phantom{\Bigg(}$ $t^\mu{}_\nu = q^\mu{}_{\alpha\beta}Q_\nu{}^{\alpha\beta} - \frac{1}{2}\delta^\mu_\nu q^\alpha{}_{\beta\gamma}Q_\alpha{}^{\beta\gamma} = 0$ $\phantom{\Bigg)}$ \\
 \hline
   metric tele$_\parallel$  & $R^\alpha{}_{\beta\mu\nu}=Q_\alpha{}^{\mu\nu}=0$ 
 & $m_P^{-2}t_\alpha{}^{\mu\nu} = \frac{1}{2}T^\mu{}_\alpha{}^\nu+T^{[\mu\nu]}{}_{\alpha}+2\delta_\alpha^{[\mu}T^{\nu]}$ &
 $\phantom{\Bigg(}$ $t^\mu{}_\nu = 2t_\alpha{}^{\beta\mu}T^\alpha{}_{\nu\beta}-\frac{1}{2}\delta^\mu_\nu t_\alpha{}^{\beta\gamma}T^\alpha{}_{\beta\gamma}=0$ $\phantom{\Bigg)}$ \\ 
 \hline
 $\phantom{\Bigg(}$ Palatini $\phantom{\Bigg)}$ &  $-$ &  $m_P^{-2}{h}_K^{\mu\nu}{}_\alpha = \mathring{\nabla}^{[\mu}\delta^{\nu]}_{\hat{\alpha}}$ & $t^\mu_{K\nu} \overset{?}{=} \frac{m_P^2}{2}\lb \lp \mathring{R}-\mathring{\Box}\rp\delta^\mu_{\hat{\nu}} + 
\lp 2\mathring{\nabla}^\mu\mathring{\nabla}_\alpha - \mathring{\nabla}_\alpha\mathring{\nabla}^\mu\rp\delta^\alpha_{\hat{\nu}}\rb$ \\
\hline
\end{tabular}
\caption{Summary of the superpotentials and the inertial energy-momentum currents in the Geometrical Trinity. \label{table}}
\end{table}
\end{center}
\end{widetext}

\section{Conclusions}
\label{conclusions}

We studied the diffeomorphism Noether currents in the context of general Palatini theories of gravity. The results were adapted to the Geometrical Trinity, in particular the metric teleparallel and the symmetric teleparallel versions of GR.  The \1 study of black holes and of gravitational waves in terms of the canonical Noether currents of teleparallel theories were carried out. Results to highlight was $C_0$ given by (\ref{einsteincharge}) for the black hole and $C_\mu=0$ for gravitational waves. Several previous calculations in the literature were reviewed, corrected and completed.

As far as we know, all previous considerations of Noether currents in either metric or symmetric teleparallelism have been compromised by gauge-dependence. In metric teleparallelism the results are sensitive to the choice of the Lorentz frame, and in symmetric teleparallelism the results are sensitive to the choice of the translational gauge. The latter is a direct reflection of the coordinate ambiguity in the conventional pseudotensor superpotential approaches. Reasonable results can of course be obtained, but in principle one can also always obtain a vanishing charge, or indeed anything between $\pm \infty$. We reviewed many different prescriptions that have been proposed to fix at least part of the ambiguities. With such prescriptions, the results have been made to converge at the asymptotic infinity to the correct charges, s.t. e.g. the energy converges to the ADM energy. The physical criterion we have advocated, fixing the gauge freedom by evaluating the results in an inertial frame, seems to be (we might say, an infinite) improvement to this. 

We arrived at unique expressions for the currents. In the standard curvature formulation of GR, the current is given by the famous Komar expression, which is gauge-invariant even though the connection has the projective gauge freedom. If the Komar charges always matched the physical energy of a given gravitational system, this study would not have been needed. It had been noticed that by a (rather ad hoc) modification (\ref{komar3}), one could fix the problems of the Komar superpotential. Gratifyingly, the current of this form (\ref{komar_improved}) can emerge as a Noether curren in the symmetric teleparallel equivalent of GR. Even better, the canonical current (\ref{Ccurrent}) in coincident GR is the simplification of the improved Komar obtained by erasing the non-standard and undesired derivative term. The expression for the canonical current was shown the background-independent unification of the  Bergmann-Thomson, von Freud, Landau-Lifshitz, Papapetrou and Weinberg superpotentials, which results from the unique minimal coupling prescription applied to each of those distinct
superpotentials. 

In metric teleparallelism, the form of the current (\ref{tcurrent}) was also expected from previous, in the metric sector vastly more numerous studies. 
The metric and the symmetric teleparallel formulations have analogous structure that differs from the Einstein-Palatini case. The Noether currents turn out to be encoded into the field equations s.t. the divergence of the superpotential equals the local energy-momentum current. The gauge-dependence of the currents now rather saves than compromises the theories, since it can be exploited to set the solution into an inertial frame, wherein the local current is solely due to the matter fields. We summarise the expressions for the canonical currents and the inertial energy-momentum tensors in the Geometrical Trinity in table \ref{table}. 

We have several expressions for entropy, derived from different considerations.
The values of these integral expressions, for a given gravitational system, are uniquely fixed in an inertial frame. Thus, the fact that these completely different kinds of expressions yielded the correct area-entropy law for a black provided a very non-trivial consistency check of our formulations. 
To summarise, one can consider the following entropy expressions.
\newline
\begin{itemize}
\item 2-dimensional surface integrals. The formula (\ref{integral3}) is the generalised Wald entropy that is equivalent to the center of mass momentum of the black hole.
\item 3-dimensional volume integral. The expression (\ref{volume}) is a rewriting of (\ref{wald1}), which was deduced from the \1 law as a Noether charge wrt time translation.
\item 4-dimensional spacetime integral. In Euclidean quantum gravity, the on-shell action can be related to the entropy. We had only a preliminary look at this Ref.\cite{Jimenez:2019yyx} and further investigation is called for. 
\end{itemize}
We emphasise that both the energy charge (\ref{wald1}) and the center of mass momentum charge
(\ref{integral3}) were deduced without any assumption about the theory or its geometrical setting. In particular, the center of mass momentum of a black hole is nonzero in standard GR and it equals the entropy of the black hole in standard GR. If one regards the latter fact as a geometrical coincidence, one then needs to explain what the former fact means.

The energy charge turned out to be gauge-dependent, the center of mass momentum charge not. Whether the former, fixed to an inertial frame, is equivalent to the latter, remains to be strictly proven. Also, we had previously argued that the metric and the symmetric formulations should give equivalent results in the respective inertial frames, and this is supported by the case study of a black hole in this article, but the relation of the formulations would be interesting to consider at more depth. Also, the tetrad formulation offers a more direct handle on the observer frame, and the interpretation of the inertial frame will be interesting to elaborate in the tetrad language. 

Another issue to be clarified could be the precise origin of the non-canonical piece $J^{\mu\nu}_{\text{non}}$ that is found in the (Lagrange multiplier -version of the) tensor realisation of the diffeomorphism transformation. This seems a more formal issue, but it could reveal some insight into the possible physical interpretation of non-canonical charges such as the Komar or the M{\o}ller charge. Also, it is a gravitational aspect of the long-standing problem of defining the proper canonical Noether currents (as opposed to pseudocanonical pseudotensors which require ad hoc improvements) for matter fields. The apparent ambiguities of course stem from that the theory is invariant under arbitrary diffeomorphisms. However, establishing the proper canonical transformation, would not only beautify the formalism but provide a tool to construct the proper canonical theory itself.

We believe that the clarification of energy, which was only possible at the edge between the metric and the symmetric teleparallel vertices of the Geometrical Trinity, could turn out to be a milestone on the road towards quantum gravity. The epitome of challenges on this road, the problem of time, the {\it conjugate} of energy, accentuates the conceptual conflicts between GR and quantum mechanics. According to the premetric programme (recall section \ref{gravi2}), based on the developments by F. Kottler, {\'E}. Cartan, D. van Dantzig etc, it is the conserved charges that are 
the fundamental primitives - electric charges rather than complex phases, energy-momentum rather than spacetime. The emergent nature of the latter seems to be suggested already by the fundamental quasilocality of the charges in gauge theories, including gravity. A promising, recently proposed theory, based on quantum energetic causal sets \cite{Cortes:2013pba}, postulates that energy is fundamental \cite{Cortes:2013uka}. Indeed, energy and momentum are not emergent from spacetime, but rather the opposite is the case.


\begin{acknowledgments}
This work was supported by the Estonian Research Council grants PRG356 “Gauge Gravity” and MOBTT86, and by the European Regional Development Fund CoE program TK133 “The Dark Side of the Universe”. J.B.J. was supported by the Project PGC2018-096038-B-100 funded by the Spanish "Ministerio de Ciencia e Innovaci\'on"
\end{acknowledgments}

\bibliography{Qrefs}

\end{document}